\documentclass[nojss]{jss}
\usepackage{thumbpdf, lmodern}

\usepackage{amsmath,amssymb,amsthm}
\usepackage{bm}
\usepackage{bbm}
\usepackage{caption}
\usepackage{multirow}
\usepackage{fancyvrb}

\usepackage[linesnumbered,ruled,vlined]{algorithm2e}
\SetKwInput{KwInput}{Input}
\SetKwInput{KwOutput}{Output}
\SetKwInput{KwInit}{Initialization}

\usepackage{tikz}
\usetikzlibrary{shapes.geometric, arrows}
\usepackage{graphicx}
\usepackage{xcolor}

\usepackage{listings}
\graphicspath{ {images/} }
\newcommand{\R}{\mathbb{R}}
\newtheorem{remark}{Remark}
\DeclareMathOperator*{\argmin}{arg\,min}

\author{Peiliang Bai \\ Department of Statistics \\ University of Florida \And
        Yue Bai \\ Department of Statistics \\ University of Florida \AND
        Abolfazl Safikhani \\ Department of Statistics \\ Informatics Institute \\ University of Florida \And
        George Michailidis \\ Department of Statistics \\ Informatics Institute \\ University of Florida}
\title{Multiple Change Point Detection in Structured VAR Models: the \pkg{VARDetect} \proglang{R} Package}

\Plainauthor{Peiliang Bai, Yue Bai, Abolfazl Safikhani, George Michailidis} 
\Plaintitle{Multiple Change Point Detection in Structured VAR Models} 
\Shorttitle{\pkg{VARDetect}: Multiple Change Point Detection in Structured VAR Models} 

\Abstract{
 Vector Auto-Regressive (VAR) models capture lead-lag temporal dynamics of multivariate time series data. They have been widely used in macroeconomics, financial econometrics, neuroscience and functional genomics. In many applications, the data exhibit structural changes in their autoregressive dynamics, which correspond to changes in the transition matrices of the VAR model that specify such dynamics. We present the \proglang{R} package \pkg{VARDetect} that implements two classes of algorithms to detect multiple change points in piecewise stationary VAR models. The first exhibits \textit{sublinear} computational complexity in the number of time points and is best suited for structured sparse models, while the second exhibits \textit{linear} time complexity and is designed for models whose transition matrices are assumed to have a low rank plus sparse decomposition. The package also has functions to generate data from the various variants of VAR models discussed, which is useful in simulation studies, as well as to visualize the results through network layouts.
}

\Plainkeywords{keywords, comma-separated, not capitalized, R} 

\Address{
  Peiliang Bai \\
  Department of Statistics \\
  University of Florida\\
  Gainesville, Florida, USA\\
  E-mail: \email{baipl92@ufl.edu} \\
  
  Yue Bai \\
  Department of Statistics \\
  University of Florida\\
  Gainesville, Florida, USA\\
  E-mail: \email{baiyue@ufl.edu} \\
  
  Abolfazl Safikhani \\
  Department of Statistics \\
  \& Informatics Institute \\
  University of Florida\\
  Gainesville, Florida, USA\\
  E-mail: \email{a.safikhani@ufl.edu} \\
  
  George Michailidis \\
  Department of Statistics \\
  \& Informatics Institute \\
  University of Florida\\
  Gainesville, Florida, USA\\
  E-mail: \email{gmichail@ufl.edu} \\
}

\begin{document}

\section{Introduction}

Vector Auto-Regressive (VAR) models aim to capture self and cross autocorrelation structure in multivariate time series data. They have been widely used in diverse fields including economics \citep{kilian2017structural,stock2016dynamic,lin2017regularized,Primiceri_2005}, finance and banking \citep{zhu2015change,basu2019low}, functional genomics \citep{basu2015network,michailidis2013autoregressive} and neuroscience \citep{friston2014granger,safikhani2020joint}.

However, in many application areas, the underlying data exhibit stationary behavior only within segments of the observations, but their overall autoregressive dynamics exhibit changes across these segments. To model such behavior, \textit{piecewise-stationary} VAR models have been developed that are also easy to interpret. Under such a setting, it is assumed that the autocovariance structure of the data changes at selected time points, henceforth called \textit{change points}, while it remains constant in the time segments between them.

The problem of \textit{offline} change point detection for time ordered data has a long history in statistics and signal processing \citep{basseville1988detecting,csorgo1997limit}. Numerous algorithms have been proposed for various statistical models, including univariate and multivariate mean shift models \citep{killick2012optimal,fryzlewicz2014wild,cho2015multiple}, exponential family models \citep{frick2014multiscale,matteson2014nonparametric}, regressions models \citep{bai1994least,bai1997estimation,harchaoui2010multiple},
covariance models \cite{aue2009break} and graphical models \citep{roy2017change}. A more expansive recent survey on this topic is \cite{truong2020selective}.

Detection algorithms include (i) optimizing cost functions implemented in \proglang{R} packages  \pkg{structchange} \citep{zeileis2002strucchange}, \pkg{changepoint} \citep{killick2016changepoint}, \pkg{changepoint.np} \citep{haynes2021changepointnp}, \pkg{fpop} \citep{rigaill2019fpop}, \pkg{ecp} \citep{james2020ecp}, \pkg{mosum} \citep{meier2021mosumpkg}; (ii) multiscale methods implemented in \proglang{R} packages \pkg{breakfast} \citep{anastasiou2020breakfast}, \pkg{stepR} \citep{florian2020stepR}, \pkg{FDRSeg} \citep{li2017FDRSeg}; (iii) regularized least squares functions \pkg{TSMCP} \citep{li2018tsmcp}.

Turning to the offline change point detection problem for
VAR models, some early work appeared in \cite{lavielle2006detection, gazeaux2011inferring} for settings involving a small number of time series. However, in many of the applications mentioned above, the number of time series is large, thus giving rise to high dimensional VAR models. Hence, there has been recent work of developing methods for detecting change points in such high dimensional VAR models \citep{wang2019localizing, cribben2013detecting, bai2020multiple}. Most of the work has focused on \textit{sparse} VAR models \citep{basu2015regularized}. Nevertheless, on certain applications the autoregressive dynamics exhibit low dimensional structure, which gives to rise to reduced rank VAR models \citep{velu1986reduced}. 

This paper presents the \pkg{VARDetect} package \citep{bai2021vardetect} in the \proglang{R} language that addresses (i) the problem of change point detection in piece-wise stationary high-diemensional VAR models and {(ii) estimating the model parameters of the underlying VAR models, under different settings regarding the structure of their transition matrices (autoregressive dynamics)}; specifically, the following cases are included: (i) sparse, (ii) structured sparse, and (iii) low rank plus sparse. It includes two classes of algorithms:
the first based on a regularized least squares objective function is most suitable for structured sparse transition matrices and exhibits \textit{sublinear} computational complexity in the number of observations; the second performs an exhaustive search over a rolling window that contains subsets of the data, is best suited for low rank plus sparse transition matrices and exhibits \textit{linear} computational complexity in the number of observations. Further, \pkg{VARDetect} contains a function to generate data from any of three types of transition matrices previously mentioned that is useful for simulation studies. Finally, at present, there is no package in \proglang{R} or any other programming language that implements change point detection methodology for VAR models.

The remainder of the paper is organized as follows. Section \ref{sec:2} introduces high-dimensional VAR models and their piecewise stationary counterparts, together with the three types of transition matrices under consideration. It also 
presents the change point detection algorithms. Section \ref{sec:3} summarizes the main structure of the developed package, and provides the details of the proposed algorithms TBSS and LSTSP, respectively. Finally, Section \ref{sec:4} illustrates the features of \pkg{VARDetect} through a series of examples for both synthetic data and real data. The \pkg{VARDetect} package is available from the Comprehensive \proglang{R} Archive Network (CRAN) at \url{https://CRAN.R-project.org/package=VARDetect}.

\section{Modeling Framework and Detection Algorithms}\label{sec:2}

We start by introducing the piecewise stationary VAR model and consider the different structures for the transition matrices included in the \pkg{VARDetect} package. Subsequently, we introduce two general algorithms for detection of change points: the first, coined \textit{\textbf{T}hresholded \textbf{B}lock \textbf{S}egmentation \textbf{S}cheme} (TBSS), is suited for (structured) sparse transition matrices, while the second, coined \textit{\textbf{L}ow-rank plus \textbf{S}parse \textbf{T}wo \textbf{S}tep \textbf{P}rocedure} (LSTSP), is suited for low rank transitions matrices. 

\subsection{High Dimensional VAR Models}\label{sec:2.1}
A $p$-dimensional \textit{stationary} VAR process $\{X_t\}$ with $q$ time lags is defined as
\begin{equation}
    \label{eq:1}
    y_t = \Phi^{(1)} y_{t-1} + \Phi^{(2)} y_{t-2} + \cdots + \Phi^{(q)} y_{t-q} + \epsilon_t,
\end{equation}
where $y_t$ is a $p$-dimensional vector of observations at time $t$, $\Phi^{(l)} \in \mathbb{R}^{p\times p}, l=1,\cdots,q$ is the transition matrix corresponding to the $l$-th lag of the VAR process, and $\epsilon_t$ is a multivariate Gaussian white noise term with independent components; i.e., $\epsilon_t \overset{i.i.d.}{\sim} \mathcal{N}_p(\mathbf{0}, \Sigma)$, with $\Sigma$ denoting the corresponding covariance matrix of the noise process. 

Next, we describe the different structures for the transition matrices $\Phi = \left( \Phi^{(1)}, \ldots, \Phi^{(q)} \right)$. 
\begin{itemize}
    \item[1.] {\it Sparse:} in this case, the number of non-zero elements in each $\Phi^{(l)}$ is $d_l\ll p^2$. 

    \item[2.] {\it Group sparse:} we consider a group sparse structure for the transition matrices. Specifically, let $\{G_1, G_2, \dots, G_L\}$ denote a partition of $\{1, 2, \dots , p^2q\}$ into $L$  column-wise, row-wise or lag groups, and each group is assumed to be dense.
    The group structure includes  (1) row-wise simultaneous across all lags, i.e. groups are of the form $ \{ \Phi^{(1)}(i,.), \ldots, \Phi^{(q)}(i,.) \} $ for $i=1, \ldots, p$; (2) column-wise simultaneous across all lags, i.e. $ \{ \Phi^{(1)}(.,i), \ldots, \Phi^{(q)}(.,i) \} $ for $i=1, \ldots, p$; (3) row-wise separate across all lags, i.e. groups are of the form $ \{ \Phi^{(j)}(i,.) \} $ for $i=1, \ldots, p$ and $j=1, \ldots, q$; (4) column-wise separate across all lags, i.e. groups are of the form $ \{ \Phi^{(j)}(.,i) \} $ for $i=1, \ldots, p$ and $j=1, \ldots, q$; (5) 
    hierarchical lag-based grouping, 
    this one is an over-lapping group structure as described in \cite{nicholson2020high} in which higher lag elements are penalized more compared to lower lag ones. Specifically, groups can be of the form  $ \{ \Phi^{(1:q)}(i,.), \Phi^{(2:q)}(i,.), \ldots, \Phi^{(q)}(i,.) \} $ for $i=1, \ldots, p$ where $\Phi^{(a:b)}(i,.) = \left( \Phi^{(a}(i,.), \ldots, \Phi^{(b)}(i,.) \right)$.
     
    \item[3.] {\it Low rank plus sparse:} we primarily focus on a lag $q=1$ VAR model due to both interpretation and technical challenges (see discussion in \cite{basu2019low}). Then, the model in \eqref{eq:1} becomes:
    \begin{equation*}
        y_t = \Phi y_{t-1} + \epsilon_t,
    \end{equation*}
    and the transition matrix $\Phi$ satisfies: $\Phi = L + S$, where $L$ is a low rank component with rank $r \ll p$, and $S$ is a sparse component with sparsity level $d \ll p^2$. 
\end{itemize}

\subsection{Change Points in High-Dimensional VAR Models}\label{sec:2.2}
We first define a piecewise stationary VAR model. Suppose there exist $m_0$ break points $0=t_0 < t_1 < \cdots t_{m_0} < t_{m_0+1} = T+1$ such that for segment $t_{j-1} \leq t < t_j$, $j=1,2,\dots, m_0+1$, a stationary VAR process $\{X_t\}$ with $q$ lags exists:
\begin{equation}
    \label{eq:2}
    y_t = \Phi^{(1,j)}y_{t-1} + \Phi^{(2,j)}y_{t-2} + \cdots + \Phi^{(q_j,j)}y_{t-q} + \epsilon_t^j,
\end{equation}
where $y_t$ is a $p$-dimensional vector of observation at time $t$, $\Phi^{(l,j)} \in \mathbb{R}^{p\times p}$ is the transition matrix corresponding to the $l$-th lag of VAR($q_j$) process for the $j$-th segment, where $j=1,2,\dots, m_0+1$, and noise term $\epsilon_t$ is multivariate Gaussian distributed with zero mean and covariance matrix $\Sigma_j$. To avoid any possible identifiability issues of the model parameters, we assume a simple structure for the covariance of the error terms, i.e., we assume $\Sigma_j = \sigma_j^2 \text{I}_p$ where $\text{I}_p$ is the $p \times p$ identity matrix. Note that the change points are mainly induced by changes in the transition matrices $\Phi^{(l,j)}$. 

The algorithms included in \pkg{VARDetect} aim to detect the break points $t_j$ in a computationally highly scalable as a function of $T$, as well as accurately estimate the model parameters $\Phi^{(l,j)}$ under high dimensional scaling $(p^2 \gg T)$. 

\subsubsection{Structured Sparse VAR Models}\label{sec:structure-sparse} 
Next, we introduce a \textit{reparametrization} of the transitions matrices of these models, that proves beneficial for both detection and computational purposes. We first define $n = T-q+1$, and will use the suppressed $n$-index throughout the paper. Define a sequence of time points $q = r_0 < r_1 < ... < r_{k_n} = T+1$ which play the role of end points for blocks of observations; i.e., $ r_{i+1} - r_i = b_n $ is the block size for $ i = 0, ..., k_n-2 $, and $k_n = \lceil \frac{n}{b_n}  \rceil$ is the total number of blocks, where $n=T-q+1$.

Denote by $\Phi^{(\cdot, j)} = (\Phi^{(1,j)}, \cdots, \Phi^{(q,j)}) \in \mathbb{R}^{p \times pq}$, set $\theta_1 = \Phi^{(\cdot, 1)}$, for $i=2,3,\cdots, n$, and define the remaining parameters $\theta$ as follows:
\begin{equation}
    \theta_i = 
    \begin{cases}
        \Phi^{(\cdot, j+1)} - \Phi^{(\cdot, j)},&\quad i=t_j\ \text{for some }j \\
        0, &\quad \text{otherwise}.
    \end{cases}
\end{equation}
Note that by using this parameterization, $\theta_i \neq 0$ for some $i \geq 2$ implies a change in the elements of the transition matrices. Therefore, we obtain that:
\begin{equation}
\label{eq:regression_block}
\underbrace{\begin{pmatrix} y_q^\prime \\ \vdots  \\  y_{r_{1}-1}^\prime \\  y_{r_{1}}^\prime \\ \vdots \\ y_{r_{2}-1}^\prime \\ \\ \vdots \\ \\ y_{r_{k_n-1}}^\prime \\ \vdots \\ y_T^\prime \end{pmatrix}}_{\mathcal{Y} } 
= 
\underbrace{\begin{pmatrix} Y_{q-1}^\prime  \\ \vdots & 0 & \ldots & 0 \\ Y_{r_{1}-2}^\prime \\ Y_{r_{1}-1}^\prime & Y_{r_{1}-1}^\prime \\ \vdots & \vdots & \ldots & 0 \\ Y_{r_{2}-2}^\prime & Y_{r_{2}-2}^\prime \\ &&& \\ \vdots & \vdots & \ddots & \vdots \\ &&& \\  Y_{r_{k_n-1}-1}^\prime & Y_{r_{k_n-1}-1}^\prime & & Y_{r_{k_n-1}-1}^\prime \\ \vdots & \vdots & \ldots & \vdots \\ Y_{T-1}^\prime & Y_{T-1}^\prime & & Y_{T-1}^\prime \end{pmatrix}}_{\mathcal{X}} \underbrace{\begin{pmatrix} \theta_1^\prime \\  \theta_{2}^\prime \\ \vdots  \\ \theta_{k_n}^\prime \end{pmatrix} }_{ \Theta}+ \underbrace{\begin{pmatrix}  \varepsilon_q^\prime \\ \vdots  \\  \varepsilon_{r_{1}-1}^\prime \\  \varepsilon_{r_{1}}^\prime \\ \vdots \\ \varepsilon_{r_{2}-1}^\prime \\ \\ \vdots \\ \\ \varepsilon_{r_{k_n-1}}^\prime \\ \vdots \\ \varepsilon_T^\prime \end{pmatrix}}_{E},
\end{equation}
where $ Y_l^\prime = \left( y_l^\prime \ldots y_{l-q+1}^\prime  \right)_{1 \times pq} $, 
$\mathcal{Y} \in \R^{n \times p} $, $\mathcal{X} \in \R^{n \times k_n pq} $, ${\Theta} \in \R^{k_npq \times  p}$ and $E \in \R^{n \times p} $. Therefore, estimates of the underlying change points $t_j, j=1,\cdots,m_0$ 
correspond to block-end time points $r_{i-1}$, with $i \geq 2$ and $\theta_i \neq 0$. 

We can rewrite the linear regression model \eqref{eq:regression_block} in vector form as 
\begin{equation}
    \label{eq:vectorform_block}
    \textbf{Y} = \textbf{Z} \mathbf{\Theta} + \textbf{E}, 
\end{equation}
where $ \textbf{Y} = \mbox{vec}(\mathcal{Y})\in \mathbb{R}^{np \times 1} $, $ \textbf{Z} = \mathbf{I}_p \otimes \mathcal{X} \in \mathbb{R}^{np \times \pi_b}$, $ \mathbf{\Theta} = \mbox{vec}(\Theta) \in \mathbb{R}^{\pi_b \times 1}  $ and $ \textbf{E} = \mbox{vec}(E)\in \mathbb{R}^{np \times 1} $, with $ \otimes $ denoting the tensor product of two matrices and $ \pi_b = k_n p^2 q $. 

The linear regression model presented in \eqref{eq:vectorform_block} suggests that the model parameters $\bm{\Theta}$ can be estimated by employing a regularization function. There are two regularization options under the posited model reparameterization:
\begin{equation}
    \label{eq:general-penalty}
    \widehat{\bm{\Theta}} = \argmin_{\bm{\Theta}} \left\{ \frac{1}{n}\|\mathbf{Y} - \mathbf{Z}\bm{\Theta}\|_2^2 + \lambda\mathcal{R}(\bm{\Theta}) \right\},
\end{equation}
where $\lambda > 0$ is a user-defined regularization penalty and $\mathcal{R}$ denotes a generic penalty function. Under the current model setup \eqref{eq:regression_block}, the code supports the following two penalties:
\begin{itemize}
    \item[(A)] {\it Fused lasso penalty} given by:
    \begin{equation}
        \label{eq:fused-lasso}
        \lambda_{1,n}\mathcal{R}(\bm{\Theta}) \overset{\text{def}}{=} \lambda_{1,n}\|\bm{\Theta}\|_1 + \lambda_{2,n}\sum_{i=1}^{k_n}\left\| \sum_{j=1}^{i}\theta_j \right\|_1
    \end{equation}
    
    \item[(B)] {\it Group lasso penalty:} Under this setting, let $\{G_1, G_2, \dots, G_L\}$ denote a partition of $\{1, 2, \dots , p^2q\}$ into $L$  column-wise, row-wise or lag groups. Let the norm $\Vert A \Vert_{2,1}$ denote $\sum_{l = 1}^L\Vert (A)_{G_l} \Vert_{F}$ for some generic matrix $A$. Then, the corresponding penalty function is given by:
    \begin{equation}
        \label{eq:group-lasso}
        \lambda_{1,n}\mathcal{R}(\bm{\Theta}) \overset{\text{def}}{=} \lambda_{1,n}\|\bm{\Theta}\|_1 + \lambda_{2,n}\sum_{i=1}^{k_n}\left\|\sum_{j=1}^i \theta_j \right\|_{2,1},
    \end{equation}
    where we use $\ell_{2,1}$-norm to denote $\sum_{l=1}^L \|(\theta)_{G_l}\|_F$.
    
    \item[(C)] {\it Weighted penalty for time-varying sparse and fixed low rank components:} In this setting, we consider a generalized model whose transition matrices $\Phi^{(\cdot, j)}$'s are decomposed into a \textit{fixed} low rank component $L$ and a \textit{time-varying} sparse component $S_j$, i.e., $\Phi^{(\cdot, j)} = L + S_j$. The corresponding penalty function $\mathcal{R}$ becomes:
    \begin{equation}
        \label{eq:lps-fixed}
        \lambda_{1,n}\mathcal{R}(\bm{\Theta}) \overset{\text{def}}{=} \lambda_{1,n}\|\bm{\Theta}\|_1 + \lambda_{2,n}\sum_{i=1}^{k_n}\left\| \sum_{j=1}^i \theta_j \right\|_1 + \mu\|L\|_*,
    \end{equation}
    wherein the $\|\cdot\|_*$ denotes the nuclear norm for regularization of the low rank component $L$, and $\mu > 0$ is the corresponding tuning parameter. The parameter $\theta_i$ is now defined as the difference between sparse components $S_j$ instead of $\Phi^{(\cdot, j)}$:
    \begin{equation*}
        \theta_i = 
        \begin{cases}
            S_{j+1} - S_j,\ &\text{if } t_j \in [r_{i-1}, r_i)\ \text{for some }j, \\
            0,\ &\text{otherwise}
        \end{cases}
    \end{equation*}
\end{itemize}

\subsubsection{Reduced Rank VAR Model}\label{sec:reduced-rank}
The reduced rank VAR model with $m_0$ change points $0 = t_0 < t_1 < \cdots < t_{m_0} < t_{m_0+1} = n$ is given by:
\begin{equation}
    \label{eq:reducedrank}
    X_t = \sum_{j=1}^{m_0+1}(A_jX_{t-1} + \epsilon_t^j)\mathbf{I}(t_{j-1} \leq t < t_j),\quad t=1,2,\dots, n, 
\end{equation}
where $A_j$ is a $p\times p$ coefficient matrix for the $j$-th segment, $j=1,2,\dots, m_0+1$, $\mathbf{I}(t_{j-1} \leq t < t_j)$ presents the indicator function of the $j$-th segment, and $\epsilon_t^j$'s are $m_0+1$ independent zero mean Gaussian noise processes. It's assumed that the coefficient matrix $A_j$ can be decomposed into a low rank component plus a sparse component: $A_j = L_j + S_j$, where $L_j$ is a low-rank component with rank $r_j$ ($r_j \ll p$), and $S_j$ is a sparse component with $d_j$ ($d_j \ll p^2$) non-zero entries.

Note that there exists an identifiability issue of model parameters due to the structure of the transition matrices. To resolve this problem, it is important to introduce a quantity, named \textit{information ratio}, for leveraging the strength of signals coming from the low rank and the sparse components. The definition of information ratio is given by: for the $j$-th segment,
\begin{equation*}
    \gamma_j \overset{\text{def}}{=} \frac{\|L_j\|_\infty}{\|S_j\|_\infty},\quad j=1,2,\dots, m_0+1.
\end{equation*}
The information ratio is used to generate synthetic data in the data generation function in Section \ref{sec:3}.

\subsection{Detection Algorithms for Proposed VAR Models}\label{sec:2.3}
In this section, we discuss the main ideas of two proposed algorithms: TBSS and LSTSP, respectively, and describe the procedures for detecting multiple change points and estimating model parameters, which are implemented in the \pkg{VARDetect} package. 

\subsubsection{Thresholded Block Segmentation Scheme (TBSS)}
The main idea of the proposed algorithm is to use block fused lasso with block size $b_n$ and fix the VAR model parameters within each block. The key steps are summarized next. 

\noindent
{\bf Step 1. Identification of candidate change points:} According to the linear regression model in \eqref{eq:vectorform_block}, the model parameters $\bm{\Theta}$ (and $L$) will be estimated by \eqref{eq:general-penalty} with different penalties. The corresponding optimization problem is convex and hence can be solved efficiently through standard algorithms \citep{hastie2015statistical}. Denote the set of estimated candidate change points by $\widehat{\mathcal{A}}_n = \left\{ i \geq 2: \widehat{\theta}_i \neq 0 \right\}$.

The cardinality of this set corresponds to the estimated number of candidate change points; i.e., $\widehat{m} = |\widehat{\mathcal{A}}|$. Further, let $\widehat{t}_j$, $j= 1, \ldots, \widehat{m}$ denote their estimated locations. Then, the relationship between $\widehat{\theta}_j$ and $\widehat{\Phi}^{(.,j)}$ in each of the estimated segments is given by:
\begin{equation}
    \label{eq:psi}
    \widehat{\Phi}^{(.,1)} = \widehat{\theta}_1\  \mbox{and}\ \widehat{\Phi}^{(.,j)} = \sum_{i=1}^{\widehat{t}_j} \widehat{\theta}_i,\quad j = 1, 2, \ldots, \widehat{m}. 
\end{equation}

\noindent
{\bf Step 2. Local screening:} The local screening step is to remove the redundant candidate change points obtained from the previous step. The main idea is to estimate VAR model parameters \emph{locally} on the left and right hand sides of each selected candidate change point and compare them to one VAR model parameter estimated from combining the left and right segments of selected candidate change points as one large stationary segment. Therefore, we construct the following \emph{localized information criterion} (LIC) as follows.

Recall that the candidate change points set is denoted as $\widehat{\mathcal{A}}_n = \left\{ \widehat{t}_1, \dots, \widehat{t}_{\widehat{m}} \right\}$, then for each subset $A \subseteq \widehat{\mathcal{A}}_n$, we define the following local VAR parameter estimates:
if $\widehat{t}_i \in A$, then 
\begin{eqnarray}\label{eq:lic:est:1}
\widehat{\psi}_{\widehat{t}_i,1} &=& \mbox{argmin}_{\psi_{\widehat{t}_i,1}} \Bigg\lbrace \frac{1}{a_n}\sum_{t= \widehat{t}_i-a_n}^{\widehat{t}_i-1} \left \| y_t - \psi_{\widehat{t}_i,1} Y_{t-1}  \right\|_2^2 + \eta_{\widehat{t}_i,1} \mathcal{P}(\psi_{\widehat{t}_i,1})  \Bigg\rbrace, \label{eq:lic:est1} \\
\widehat{\psi}_{\widehat{t}_i,2} &=& \mbox{argmin}_{\psi_{\widehat{t}_i,2}} \Bigg\lbrace \frac{1}{a_n}  \sum_{t= \widehat{t}_i}^{\widehat{t}_i+a_n-1} \left \| y_t - \psi_{\widehat{t}_i,2} Y_{t-1} \right \|_2^2 + \eta_{\widehat{t}_i,2}\mathcal{P}(\psi_{\widehat{t}_i,2})  \Bigg\rbrace; \label{eq:lic:est2}
\end{eqnarray}
if $\widehat{t}_i \in \widehat{\mathcal{A}}_n \backslash A $, then 
\begin{eqnarray}\label{eq:lic:est:2}
\widehat{\psi}_{\widehat{t}_i} &=& \mbox{argmin}_{\psi_{\widehat{t}_i}} \Bigg\lbrace \frac{1}{2a_n} \sum_{t= \widehat{t}_i-a_n}^{\widehat{t}_i+a_n-1}\left \| y_t - \psi_{\widehat{t}_i} Y_{t-1}  \right\|_2^2 + \eta_{\widehat{t}_i} \mathcal{P}(\psi_{\widehat{t}_i})   \Bigg\rbrace,
\end{eqnarray}
where the auxiliary tuning parameters $\eta_{\widehat{t}_i,1}$ and $\eta_{\widehat{t}_i,2}$ are for the left and right side of $\widehat{t}_i$, respectively
when $ \widehat{t}_i \in A $, we set it as  $\eta_{\widehat{t}_i} = \left(\eta_{\widehat{t}_i,1}, \eta_{\widehat{t}_i,2}\right)$. If $ \widehat{t}_i \in \widehat{\mathcal{A}}_n \backslash A $, then there is only one tuning parameter which is denoted by $ \eta_{\widehat{t}_i} $.
The selection of these tuning parameters are primarily based on theoretical results \citep{basu2019low}. The regularization function $\mathcal{P}$ is specified by user, in this work, we provide three options: \emph{sparse}, \emph{group sparse}, and \emph{low rank plus sparse} as introduced in Section \ref{sec:2.2}. Also, $a_n$ is the size of neighborhood in which the VAR model parameters are estimated.

Then, the localized information criterion is defined as:
\begin{eqnarray}\label{eq:lic}
\mbox{LIC}(A;\eta_n) &=& \left\lbrace \sum_{\widehat{t}_i \in A}  \left(  \sum_{t=\widehat{t}_i -a_n}^{\widehat{t}_i - 1} \left\| y_t - \widehat{\psi}_{\widehat{t}_i,1} Y_{t-1}  \right \|_2^2  +   \sum_{t=\widehat{t}_i}^{\widehat{t}_i+a_n-1} \left\| y_t - \widehat{\psi}_{\widehat{t}_i,2} Y_{t-1}  \right\|_2^2   \right) \right. \nonumber \\
 &+& \left. \sum_{\widehat{t}_i \in \widehat{\mathcal{A}}_n \backslash A}  \sum_{t=\widehat{t}_i -a_n}^{\widehat{t}_i+a_n-1} \left \| y_t - \widehat{\psi}_{\widehat{t}_i} Y_{t-1}  \right\|_2^2  \right\rbrace+ |A| \, \omega_n \nonumber \\
&\overset{\text{def}}{=}& L_n(A;\eta_n) + |A| \, \omega_n,
\end{eqnarray}
where $\eta_n = \left(\eta_{\widehat{t}_1}, \dots, \eta_{\widehat{t}_{\widehat{m}}} \right)$ and $\omega_n$ are auxiliary tuning parameters. Therefore, the screened selected change points are the minimizers satisfying
\begin{equation}
    \label{eq:lic-objective}
    (\widetilde{m}, \widetilde{t}_j; j=1,2,\dots, \widetilde{m}) = \argmin_{0 \leq m \leq \widehat{m}, s=(s_1, \dots, s_m)\subseteq \widehat{\mathcal{A}}_n}\text{LIC}(s; \eta_n).
\end{equation}
Then, we denote the set of screened change points from \eqref{eq:lic-objective} by $\widetilde{\mathcal{A}}_n = \{\widetilde{t}_1, \dots, \widetilde{t}_{\widetilde{m}}\}$.

\noindent 
{\bf Step 3. Exhaustive search:} The previous local screening step manages to remove redundant candidate change points which are located far away from any true change points. However, in $a_n$-neighborhoods of each true change point, there might be more than one estimated change points remaining in the set $\widetilde{\mathcal{A}}_n$. In order to avoid this issue, we employ an exhaustive search for each cluster to select the final estimated change point. 

For any set $A \subset \{1,2,\dots, T\}$, we denote $C_A(r)$ as the minimal partition of $A$, where the diameter for each subset is at most $r$ (i.e., for any $B \subseteq A$, $\text{diam}(B) \overset{\text{def}}{=} \max_{a,b\in b}|a-b| \leq r$). Now, denote the subsets in $C_{\widetilde{A}_n}(2a_n)$ by $\{C_1, C_2, \dots, C_{\widetilde{m}}\}$. 

Next, for each selected cluster of break points $C_i, i=1,2,\cdots,\widetilde{m}$, we define the \textit{search interval} $(l_i, u_i)$ whose lower and upper bounds are given by:
\begin{equation*}
    l_i = \begin{cases}
                  c_i - a_n,\quad &\text{if } |C_i| = 1, \\
                  \min\{C_i\},\quad &\text{otherwise},
              \end{cases}\quad
    u_i = \begin{cases}
                  c_i + a_n,\quad &\text{if } |C_i| = 1, \\
                  \max\{C_i\},\quad &\text{otherwise},
              \end{cases}          
\end{equation*}
where $c_i$ is the unique element in $C_i$, whenever $|C_i|=1$. Denote the subset of corresponding block indices in the interval $(l_i,u_i)$ by $J_i$ with $J_0 = \{1\}$ and $J_{\widetilde{m}+1} = \{k_T\}$. Further denote the closest block end to $\left( \max J_{i-1} + \min J_i \right)/2$ as $w_i$. Now, the \textit{local} parameter estimators are given by \eqref{eq:psi}. Finally, for each $i=1,2,\cdots, \widetilde{m}$, the final estimated break points is defined as:
\begin{equation}
    \label{eq:final-estimate}
    \widetilde{t}_i^f = \argmin_{s\in (l_i, u_i)}\left\{ \sum_{t=l_i}^{s-1}\|y_{t+1} - \widetilde{\Phi}^{(.,i)}Y_t \|_2^2 + \sum_{t=s}^{u_i-1}\|y_{t+1} - \widetilde{\Phi}^{(.,i+1)}Y_t \|_2^2 \right\},
\end{equation}

\noindent
{\bf Step 4. Model parameter estimation:} Once the final set of break points have been identified from exhaustive search step, we can estimated the transition matrices (and thus the Granger causal networks) by using the algorithms developed in \citep{lin2017regularized} for stationary data. To ensure that the data in the time segments between break points are strictly stationary, we remove all time points in a $R_T$-radius neighborhood of the break points obtained in Step 4. The length of $R_T$ needs to be at least $b_T$. Specifically, denote by $s_{j1} = \widetilde{t}_j - R_T - 1 $, $ s_{j2} = \widetilde{t}_j + R_T + 1 $ for $ j = 1, \cdots, \widetilde{m} $, and set $ s_{02} = q $ and $ s_{(\widetilde{m}+1)1} = T $. Next, define the intervals $ I_{j+1} = [ s_{j2}, s_{(j+1)1}  ] $ for $ j = 0, \cdots, \widetilde{m}$. The idea is to form a linear regression on $ \cup_{j=0}^{\widetilde{m}} I_{j+1} $ and estimate the auto-regressive parameters by minimizing an $\ell_1$-regularized least squares criterion. Specifically, we form the following linear regression similar to \eqref{eq:vectorform_block}: $\mathcal{Y}_s = \mathcal{X}_s B + E_s$, where $\mathcal{Y}_s = (y_q, \cdots, y_{s_{11}}, \cdots, y_{s_{\widetilde{m}2}}, \cdots, y_T)^\prime$, $B = (\beta_1, \beta_2, \cdots, \beta_{\widetilde{m}+1})$, the corresponding error term $E_s = (\zeta_q, \cdots, \zeta_{s_{11}}, \dots, \zeta_{s_{\widetilde{m}2}}, \cdots, \zeta_{T})^\prime$, and the design matrix is given by:
\begin{equation*}
    \mathcal{X}_s = 
 \begin{pmatrix} 
     \widetilde{\mathbf{Y}}_1 & \mathbf{0} & \cdots & \mathbf{0} \\
     \mathbf{0} & \widetilde{\mathbf{Y}}_2 & \cdots & \mathbf{0} \\
     \vdots & \vdots & \ddots & \vdots \\
     \mathbf{0} & \mathbf{0} & \cdots & \widetilde{\mathbf{Y}}_{\widetilde{m}},
 \end{pmatrix}.
\end{equation*}
where the diagonal elements are given by: $\widetilde{\mathbf{Y}}_j^\prime = (Y_{s_{j2}-1}, \cdots, Y_{s_{(j+1)1}-1})$, and $j=0,1,\dots, \widetilde{m}-1$. Then, we can rewrite it in compact form as: 
\begin{equation*}
 {\textbf{Y}}_{\textbf{s}} =  \textbf{Z}_{\textbf{s}} \textbf{B}  + \textbf{E}_{\textbf{s}},
\end{equation*}
where $\textbf{Y}_{\textbf{s}} = \mbox{vec}(\mathcal{Y}_{\textbf{s}})$, ${\textbf{Z}_{\textbf{s}}} = I_p \otimes \mathcal{X}_{\textbf{s}}$, $\textbf{B} = \mbox{vec}(B) $, ${\textbf{E}}_{\textbf{s}} = \mbox{vec}(E_{\textbf{s}}) $, and \textbf{s} is the collection of all $ s_{j1}$ and $s_{j2}$ for $j = 0, \ldots, m_0+1$. Let $\widetilde{\pi} = (\widetilde{m} +1) p^2 q$, $N_j = s_{(j+1)1}-s_{j2} $ be the length of the interval $I_{j+1}$ for $ j = 0, \cdots, \widetilde{m}$ and $N = \sum_{j=1}^{\tilde{m}} N_j $. Then, $\textbf{Y}_{\textbf{s}} \in \mathbb{R}^{N p \times 1} $, $ {\textbf{Z}_{\textbf{s}}} \in \mathbb{R}^{N p \times \widetilde{\pi}} $, $ \textbf{B} \in \mathbb{R}^{\tilde{\pi} \times 1} $, and $ {\textbf{E}}_{\textbf{s}} \in \mathbb{R}^{N p \times 1} $. Therefore, we estimate the VAR parameters by solving the following $\ell_1$ regularized optimization problem:
\begin{equation}
 \label{eq:estimation_third}
 \widehat{\textbf{B}} = \argmin_{\textbf{B}}\left\{ \frac{1}{N}\left\| \textbf{Y}_{\textbf{s}} - {\textbf{Z}_{\textbf{s}}} \textbf{B} \right\|_2^2 + \rho_T  \left\| \textbf{B} \right\|_1\right\}.
\end{equation}

{
\begin{remark}\label{remark:tbss}
\cite{safikhani2020joint} proposed a fused lasso based detection algorithm, that examines all time points in the data to select candidate change points. Hence, its corresponding time complexity is $\mathcal{O}(Tp^2q)$, with $T$ being the length of the time series, $p$ the number of time series in the VAR model and $q$ the lag. Further, the algorithm only considers sparse transition matrices. On the other hand, the developed TBSS algorithms improves time complexity to $\mathcal{O}(\sqrt{T}p^2q)$, and hence it represents the first algorithm exhibiting \textit{sublinear} complexity in the number of time points $T$. Further, the underlying transition matrices can be sparse, group sparse, or have a \emph{fixed} low rank component plus a time-varying sparse component. 
\end{remark}
}

\subsubsection{Low-rank and Sparse Two Step Procedure (LSTSP)}\label{sec:lstsp-model}
For the general reduced rank model proposed in \eqref{eq:reducedrank}, we investigate a VAR(1) model with low rank plus sparse structure. We start with a single change point detection method, let $\{X_t\}$ be a sequence of observations generated the VAR(1) model defined as follows:
\begin{equation}
    \label{eq:single-cp}
    \begin{aligned}
    X_t &= A_1^\star X_{t-1} + \epsilon_t^1, \quad t=1,2,\dots, \tau^\star, \\
    X_t &= A_2^\star X_{t-1} + \epsilon_t^2, \quad t=\tau^\star+1, \dots, n,
    \end{aligned}
\end{equation}
and it is further assumed that the transition matrices comprise of two time-varying components,
a low-rank and a sparse one:  
\begin{equation}
    \label{eq:15}
    A_1^\star = L_1^\star + S_1^\star \quad \text{and} \quad A_2^\star = L_2^\star + S_2^\star.
\end{equation}
Then, for any time point $\tau \in \{1,2,\dots, n\}$ we define the objective function with respect to the intervals $[1,\tau)$, and $[\tau, n)$:
\begin{equation}
    \label{eq:lps-left}
    \ell(L_1, S_1; \mathbf{X}^{[1:\tau)}) \overset{\text{def}}{=} \frac{1}{\tau-1}\sum_{t=1}^{\tau-1} \|X_t - (L_1+S_1)X_{t-1}\|_2^2 + \lambda_1\|S_1\|_1 + \mu_1\|L_1\|_*,
\end{equation}
\begin{equation}
    \label{eq:lps-right}
    \ell(L_2, S_2; \mathbf{X}^{[\tau:n)}) \overset{\text{def}}{=} \frac{1}{n-\tau}\sum_{t=\tau}^{n-1} \|X_t - (L_2+S_2)X_{t-1}\|_2^2 + \lambda_2\|S_2\|_1 + \mu_2\|L_2\|_*,
\end{equation}
where $\mathbf{X}^{[b:e)}$ denotes the data $\{X_t\}$ from time points $b$ to $e$, and the non-negative tuning parameters $\lambda_1$, $\lambda_2$, $\mu_1$, and $\mu_2$ control the regularization of the sparse and the low-rank components in the corresponding transition matrices.

Next, we introduce the objective function with respect to the time point $\tau$: for any time point $\tau \in \lbrace 1,2, \dots, n-1 \rbrace$, we obtain
\begin{equation*}
    \ell(\tau; L_1, L_2, S_1, S_2) \overset{\text{def}}{=} \frac{1}{n-1}\left( \sum_{t=1}^{\tau-1}\|X_t - (L_1+S_1)X_{t-1}\|_2^2 + \sum_{t=\tau}^{n-1}\|X_t - (L_2+S_2)X_{t-1}\|_2^2\right).
\end{equation*}
Then, the estimator $\widehat{\tau}$ of the change point $\tau^\star$ is given by:
\begin{equation}
    \label{eq:16}
    \widehat{\tau} \overset{\text{def}}{=} \argmin_{\tau \in \mathcal{T}}\ell(\tau; \widehat{L}_{1,\tau}, \widehat{L}_{2,\tau}, \widehat{S}_{1,\tau}, \widehat{S}_{2,\tau}),
\end{equation}
for the search domain $\mathcal{T} \subset \{1,2,\dots, n\}$, where, for each $\tau \in \mathcal{T}$, the estimators $\widehat{L}_{1,\tau}$, $\widehat{L}_{2,\tau}$, $\widehat{S}_{1,\tau}$, $\widehat{S}_{2,\tau}$ are derived from the optimization program \eqref{eq:16} with tuning parameters $\mu_{1,\tau}$, $\mu_{2,\tau}$, $\lambda_{1,\tau}$, and $\lambda_{2,\tau}$, respectively. 

Next, with the help of single change point detection method in \eqref{eq:16}, we are in the position to provide the rolling window method for detecting multiple change points. 
\begin{itemize}
    \item {\bf Step 1. Candidate change points selection:} It's based on the single change point detection method introduced in \eqref{eq:16}, additionally equipped with a \textit{rolling window} mechanism to select \textit{candidate} change points. We start by selecting an interval $[b_1,e_1) \subset \{1,2,\dots, n\}, b_1=1$, of length $h$ and employ on it the exhaustive search Algorithm 1 to obtain a candidate change point $\widetilde{\tau}_1$. Next, we shift the interval to the right by $l$ time points and obtain a new interval $[b_2, e_2)$, wherein $b_2=b_1+l$ and $e_2=e_1+l$. The application of Algorithm 1 to $[b_2,e_2)$ yields another candidate change point $\widetilde{\tau}_2$. This procedure continues until the last interval that can be formed, namely $[b_{\widetilde{m}},e_{\widetilde{m}})$, where $e_{\widetilde{m}}=n$ and $\widetilde{m}$ denotes the number of windows of size $h$ that can be formed. The following Figure \ref{fig:rolling-window} depicts this rolling-window mechanism. 
The blue lines represent the boundaries of each window, awhile the green dashed lines represent the candidate change point in each window. Note that the basic assumption for Algorithm 1 is that there exists a single change point in the given time series. However, it can easily be seen in Figure \ref{fig:rolling-window} that \textbf{not} every window includes a single change point.
\begin{figure}[ht!]
    \centering
    \resizebox{\textwidth}{!}{
    \begin{tikzpicture}
    \centering
        \draw[->](-8.2,0)--(8,0) node[pos=1,below]{$n$};
        \foreach \x/ \xtext in {-8/$b_1$, -7/$b_2$, -5/$e_1$, -4/$e_2$, 0/$b_j$, 3/$e_j$, 4/$b_{\widetilde{m}}$, 7/$e_{\widetilde{m}}$}{
            \draw (\x cm,-2pt) -- (\x cm,2pt) node[label={[label distance =0.1em]below:{\xtext}}]{};
            \draw[blue,thick] (\x cm,-0.2) -- (\x cm, 4);
        }
        \foreach \truept in {-4.5, 1, 5}{
            \draw(\truept,0)circle[radius=1.5pt];
        }
        \node[at={(-4.5,0)},label={[label distance=0.1em]above:{$\tau^\star_1$}}]{};
        \node[at={(1,0)},label={[label distance=0.1em]above:{$\tau^\star_2$}}]{};
        \node[at={(5,0)},label={[label distance=0.1em]above:{$\tau^\star_3$}}]{};
        
        \fill[red](-4.5,0)circle[radius=1.5pt];
        \fill[red](1,0)circle[radius=1.5pt];
        \fill[red](5,0)circle[radius=1.5pt];
        
        \foreach \estpt in {-6.44, -4.3, 0.8, 4.8}{
            \draw(\estpt,0)circle[radius=1.5pt];
            \draw[dashed,green,thick] (\estpt, -0.2) -- (\estpt, 4);
        }
        \node[at={(-6.44,0)},label={[label distance=0.1em]below:{$\widetilde{\tau}_1$}}]{};
        \node[at={(-4.3,0)},label={[label distance=0.1em]below:{$\widetilde{\tau}_2$}}]{};
        \node[at={(0.8,0)},label={[label distance=0.1em]below:{$\widetilde{\tau}_j$}}]{};
        \node[at={(4.8,0)},label={[label distance=0.1em]below:{$\widetilde{\tau}_{\widetilde{m}}$}}]{};
        
        \fill[green](-6.44,0)circle[radius=1.5pt];
        \fill[green](-4.3,0)circle[radius=1.5pt];
        \fill[green](0.8,0)circle[radius=1.5pt];
        \fill[green](4.8,0)circle[radius=1.5pt];
        
        \draw[->](-3.5,3) -- node[above]{rolling windows} (-0.5,3);
        
        \draw[<->,red,thick](-8,1) -- node[above]{$h$}(-5,1);
        \draw[->,red,thick](-8,3) -- node[above]{$l$}(-7,3);
        
    \end{tikzpicture}
    }
    \caption{Depiction of the rolling windows strategy. There are three true change points: $\tau_1^\star$, $\tau_2^\star$, and $\tau_3^\star$ (red dots); the boundaries of the rolling-window are represented in blue lines; the estimated change points in each window are plotted in green dashed lines, where the subscript indicates the index of the window used to obtain it.}
    \label{fig:rolling-window}
\end{figure}
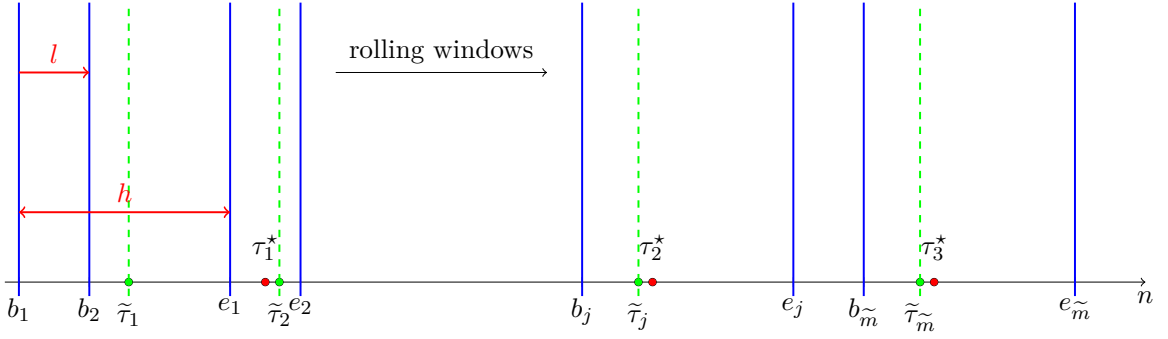

\item {\bf Step 2. Screening:} Let the candidate change points from Step 1 be denoted by $\{s_j\}$, $j=1,2,\cdots, \widetilde{m}$. Then, we define $0=s_0 < s_1 < s_2 < \dots < s_{\widetilde{m}} < s_{\widetilde{m}+1} = n$ and for ease of presentation use $L_i$ and $S_i$ instead of $L_{(s_{i-1}, s_i)}$ and $S_{(s_{i-1}, s_i)}$ for $i=1,2,\dots, m+1$. We also define matrices $\mathbf{L} \overset{\text{def}}{=} [L_1^\prime, L_2^\prime, \dots, L_{\widetilde{m}+1}^\prime]^\prime$ and $\mathbf{S} \overset{\text{def}}{=} [S_1^\prime, S_2^\prime, \dots, S_{\widetilde{m}+1}^\prime]^\prime$. Estimates for $\mathbf{L}$ and $\mathbf{S}$ are obtained as the solution to the following regularized regression problem:
\begin{equation*}
    (\widehat{\mathbf{L}}, \widehat{\mathbf{S}}) = \argmin_{L_i,S_i, 1\leq i \leq \widetilde{m}+1} \sum_{i=1}^{\widetilde{m}+1} \left\{
    \frac{1}{s_i - s_{i-1}}\sum_{t=s_{i-1}}^{s_i-1}\|X_t - (L_i+S_i)X_{t-1}\|_2^2 + \lambda_i\|S_i\|_1 + \mu_i \|L_i\|_*\right\},
\end{equation*}
with tuning parameters $(\bm{\lambda}, \bm{\mu}) = \{(\lambda_i, \mu_i)\}_{i=1}^{\widetilde{m}+1}$. Next, we define the objective function with respect to $(s_1,s_2, \dots, s_m)$:
\begin{equation*}
    \mathcal{L}_n(s_1,s_2,\dots,s_m; \bm{\lambda}, \bm{\mu}) \overset{\text{def}}{=} \sum_{i=1}^{\widetilde{m}+1} \left\{\sum_{t=s_{i-1}}^{s_i-1}\|X_t - (\widehat{L}_i+\widehat{S}_i)X_{t-1}\|_2^2 + \lambda_i\|\widehat{S}_i\|_1 + \mu_i \|\widehat{L}_i\|_*\right\}.
\end{equation*}
Then, for a suitably selected penalty sequence $\omega_n$, specified in the upcoming Assumption H5, we consider the following \textit{information criterion} defined as:
\begin{equation}
    \label{eq:ic}
    \text{IC}(s_1, s_2,\dots, s_m; \bm{\lambda}, \bm{\mu}, \omega_n) \overset{\text{def}}{=} \mathcal{L}_n(s_1,\dots,s_m; \bm{\lambda}, \bm{\mu}) + m\omega_n.
\end{equation}
The final selected change points are obtained by solving:
\begin{equation}
    \label{eq:lps-ic}
    (\widehat{m}, \widehat{\tau}_i, i=1,2,\dots, \widehat{m}) = \argmin_{0\leq m \leq \widetilde{m}, (s_1,\dots, s_m)} \text{IC}(s_1,\dots,s_m; \bm{\lambda}, \bm{\mu}, \omega_n).
\end{equation}

\item { {\bf Step 3. Model parameter estimation:} The following step provides the model parameter estimation procedure across the stationary segments and it is the analogue of Step 4 in the TBSS algorithm. Specifically, suppose the final estimated change points derived from Step 2 are denoted by $\widehat{\tau}_1, \dots, \widehat{\tau}_{\widehat{m}}$; then, for each estimated change point $\widehat{\tau}_j$, consider a radius $R_n>0$ and  denote the neighborhood of $\widehat{\tau}_j$ by $[r_{j2}, r_{(j+1)1})$ for $j=1,2,\dots, \widehat{m}$, where $r_{j1} = \widehat{\tau}_j - R_n-1$, $r_{j2} = \widehat{\tau}_j + R_n+1$, and also let $r_{02} = 1$ and $r_{(m_0+1)1} = T$. Then, for each neighborhood interval, the given process is stationary and we define the following optimization problem:
\begin{equation*}
    (\widehat{L}_j, \widehat{S}_j) = \argmin_{L_j, S_j}\left\{\frac{1}{r_{(j+1)1} - r_{j2}}\sum_{t=r_{j2}}^{r_{(j+1)1}}\|X_t - (L_j+S_j)X_{t-1}\|_2^2 + \lambda_j\|S_j\|_1 + \mu_j\|L_j\|_*\right\},
\end{equation*}
where the minimizers $(\widehat{L}_j, \widehat{S}_j)$ correspond to the estimated model parameters for the $j$-th estimated segment.
}
\end{itemize}

{
\begin{remark}
\label{remark:lstsp}
\cite{bai2020multiple} consider a VAR model, wherein the transition matrices can be decomposed to a \emph{fixed} low rank one, plus a time-varying sparse component. The paper developed a block fused lasso (BFL) based detection procedure. In the current paper, (i) the algorithm in \cite{bai2020multiple} is incorporated in the TBSS algorithm and (ii) we consider the much more challenging setting where \textit{both} the low rank and the sparse components are allowed to change. For this new setting, the TBSS algorithm will not work anymore, since it is unclear what the result of employing a fused lasso penalty on low rank matrices would lead to. Hence, the novel two-step LSTSP algorithm is introduced to solve the problem. 
\end{remark}
}

{
\begin{remark}\label{rem:estimation-VAR-parameters}
Estimation of VAR model parameters: It can be seen that Step 4 in Algorithm TBSS and Step 3 in Algorithm LSTSP focus on estimating the model parameters \textit{over the identified stationary segments} of the posited (e.g., sparse, group sparse, low rank plus sparse)
VAR models. Note that in change point analysis, detection of change points and estimation of model parameters are inerwined, since the search for change points depends on the nature of the VAR model one assumes (see Step 1 in both the TBSS and LSTSP algorithms).
\end{remark}
}

\section{Implementation}\label{sec:3}
A flowchart of the main modules of the \pkg{VARDetect} package is presented in Figure \ref{fig:main-structure}. The core of \pkg{VARDetect} is implemented in \proglang{C++} and the \pkg{Armadillo} library. \pkg{OpenMP} is also used for parallel programming in \proglang{C++}.

\tikzstyle{input} = [scale=.85, trapezium, trapezium left angle=70, trapezium right angle=110, minimum width=2cm, minimum height=1cm, inner sep=2.5pt, text centered, text width=2.5cm, draw=black, fill=blue!30]
\tikzstyle{output} = [scale=.85, trapezium, trapezium left angle=70, trapezium right angle=110, minimum width=2cm, minimum height=1cm, inner sep=2.5pt, text centered, text width=4cm, draw=black, fill=red!30]
\tikzstyle{decision} = [scale=.85, diamond, aspect=3, minimum width=2cm, minimum height=0.5cm, text centered, text width=3.5cm, inner sep=1pt, draw=black, fill=green!30]
\tikzstyle{process} = [scale=.85, rectangle, minimum width=1.5cm, minimum height=1cm, text centered, text width=3cm, draw=black, fill=orange!30]
\tikzstyle{arrow} = [thick,->,>=stealth]
\begin{figure}[!ht]
    \centering
    \begin{tikzpicture}[node distance=2.25cm]
        \node (dec0) [decision] {Simulation?};
        \node (pro0) [process, right of=dec0, xshift=3.75cm] {Generate synthetic VAR process $\{X_t\}$};
        \node (in1) [input, below of=dec0] {Input: Time series $\{X_t\}$};
        \node (dec1) [decision, below of=in1, yshift=-.5cm] {Low-rank structure?};
        \node (pro1) [process, below of=dec1, yshift=-.5cm] {Apply TBSS + penalty (A)/(B)};
        \node (dec2) [decision, right of=dec1, xshift=3.5cm] {Fixed low-rank?};
        \node (pro2) [process, below of=dec2, yshift=-.5cm] {Apply TBSS + penalty (C)};
        \node (pro3) [process, right of=pro2, xshift=2cm, yshift=0cm] {Apply LSTSP};
        \node (output) [output, below of=pro2] {Output: Estimated change points $\widehat{t}_j$ and model parameters $\widehat{\Phi}_j$.};
        
        \draw [arrow] (dec0) -- node[anchor=east] {no} (in1);
        \draw [arrow] (dec0) -- node[anchor=south] {yes} (pro0);
        \draw [arrow] (in1) -- (dec1);
        \draw [arrow] (pro0) |- (in1);
        \draw [arrow] (dec1) -- node[anchor=south] {yes} (dec2);
        \draw [arrow] (dec1) -- node[anchor=east] {no} (pro1);
        \draw [arrow] (dec2) -- node[anchor=east] {yes} (pro2);
        \draw [arrow] (dec2) -| node[anchor=west] {no} (pro3);
        \draw [arrow] (pro1) |- (output);
        \draw [arrow] (pro2) -- (output);
        \draw [arrow] (pro3) |- (output);
        \draw [dashed] (-3.5,-3) -- (-3.5,1) -- (7.5,1) -- (7.5,-3) -- node[anchor=south west, at end]{\textbf{Data generation}} (-3.5, -3);
        \draw [dashed] (-3.5,-3.25) -- node[anchor=south west, at end]{\textbf{Detection and estimation}} (-3.5, -9.75) -- (10.5, -9.75) -- (10.5, -3.25) -- (-3.5, -3.25);
    \end{tikzpicture}
    \caption{Main structure of \pkg{VARDetect} package.}
    \label{fig:main-structure}
\end{figure}
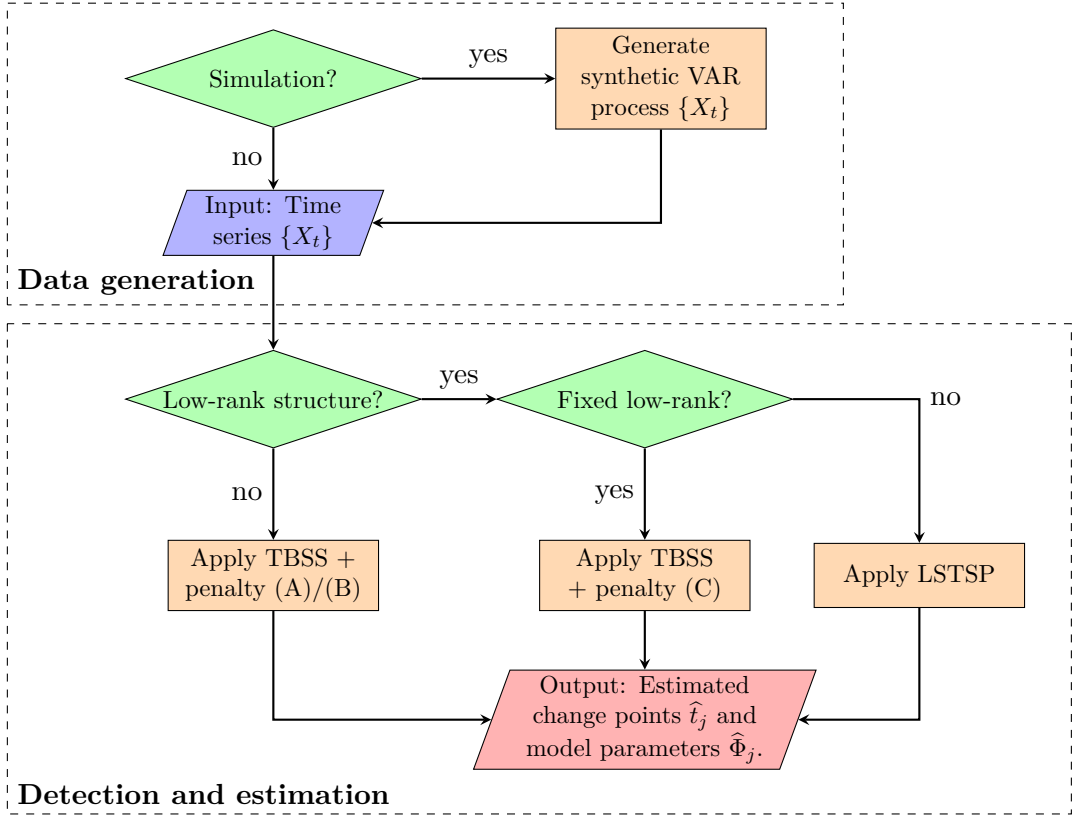

The package comprises of the following two modules: (i) the \textit{data generation} one that is useful for users interested in designing simulation studies and generates synthetic VAR data with specific structure for the corresponding transition matrices; (ii) the \textit{detection and estimation} module that is the main one.

Next, we describe in detail these modules and provide code snippets that illustrate their functionality. {{Version 0.1.5 is used to describe the package's functions and in all illustrative examples with both synthetic and real data.}}

\subsection{Data Generation}\label{sec:data-generation}
The structure of this module is shown in Figure \ref{fig:data-generation}. 
\tikzstyle{input} = [scale=.85, trapezium, trapezium left angle=70, trapezium right angle=110, minimum width=2cm, minimum height=1cm, inner sep=2.5pt, text centered, text width=2.5cm, draw=black, fill=blue!30]
\tikzstyle{output} = [scale=.85, trapezium, trapezium left angle=70, trapezium right angle=110, minimum width=2cm, minimum height=1cm, inner sep=2.5pt, text centered, text width=4cm, draw=black, fill=red!30]
\tikzstyle{decision} = [scale=.85, diamond, aspect=3, minimum width=2cm, minimum height=0.5cm, text centered, text width=3.5cm, inner sep=1pt, draw=black, fill=green!30]
\tikzstyle{process} = [scale=.85, rectangle, minimum width=1.5cm, minimum height=1cm, text centered, text width=3cm, draw=black, fill=orange!30]
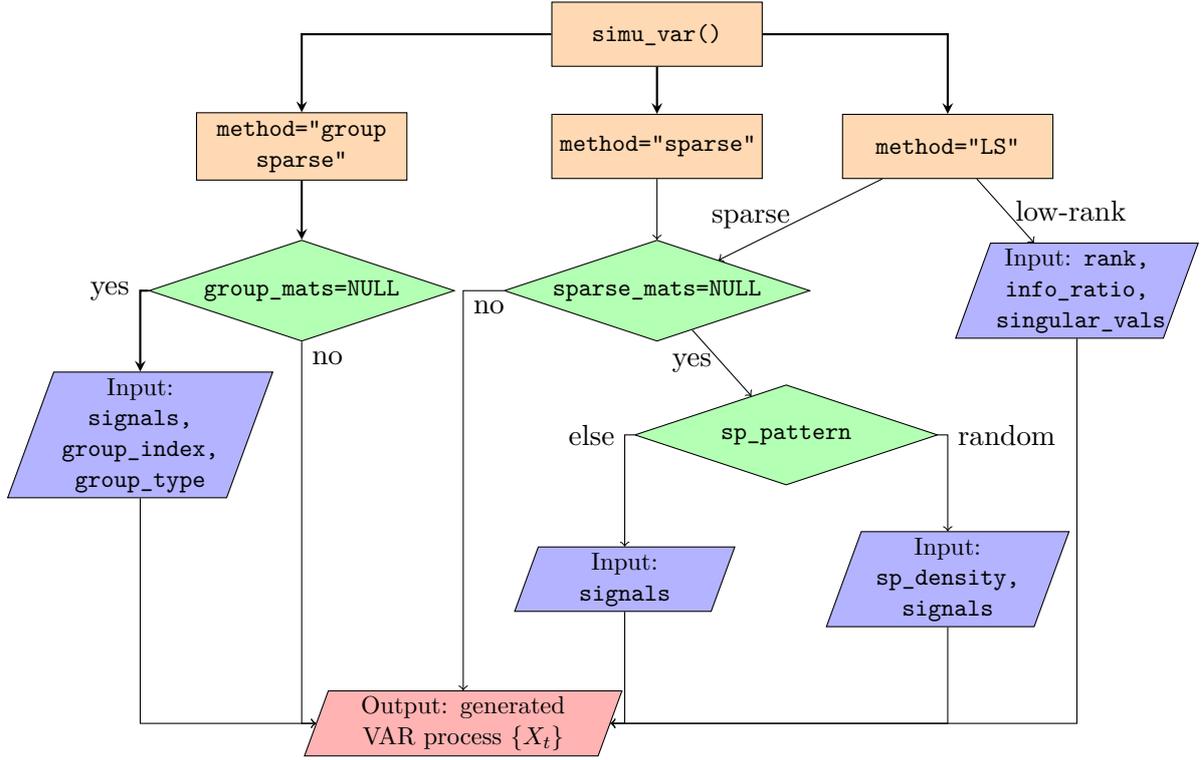
\begin{figure}[!ht]
    \centering
    \begin{tikzpicture}[node distance=2.25cm]
        \node (method) [process] {\code{simu_var()}};
        \node (group) [process, below of=method, xshift=-5.5cm, yshift=.5cm] {\code{method="group sparse"}};
        \node (sp) [process, below of=method, yshift=.5cm] {\code{method="sparse"}};
        \node (ls) [process, below of=method, xshift=4.5cm, yshift=.5cm] {\code{method="LS"}};
        
        \node (ls1) [input, below of=ls, xshift=2cm] {Input: \code{rank, info_ratio, singular_vals}};
        
        \node (group1) [decision, below of=group] {\code{group_mats=NULL}};
        \node (group2) [input, below of=group1, xshift=-2.5cm] {Input: \code{signals, group_index, group_type}};
        
        \node (sp1) [decision, below of=sp] {\code{sparse_mats=NULL}};
        \node (sp2) [decision, below of=sp1, xshift=2cm] {\code{sp_pattern}};
        \node (sp3) [input, below of=sp2, xshift=-2.5cm] {Input: \code{signals}};
        \node (sp4) [input, below of=sp2, xshift=2.5cm] {Input: \code{sp_density, signals}};
        
        \node (output) [output, below of=sp3, xshift=-2.5cm] {Output: generated VAR process $\{X_t\}$};
        
        \draw [arrow] (method) -| (group);
        \draw [arrow] (method) -- (sp);
        \draw [arrow] (method) -| (ls);
        \draw [arrow] (group) -- (group1);
        \draw [arrow] (group1) -| node[anchor=east]{yes} (group2);
        \draw [->] (group2) |- (output);
        \draw [->] (group1) |- node[anchor=north west, at start]{no}(output);

        \draw [->] (sp) -- (sp1);
        \draw [->] (sp1) -- node[anchor=east]{yes} (sp2);
        \draw [->] (sp2) -| node[anchor=east]{else} (sp3);
        \draw [->] (sp2) -| node[anchor=west]{random} (sp4);
        \draw [->] (sp3) |- (output);
        \draw [->] (sp4) |- (output);
        \draw [->] (sp1) -| node[anchor=north west]{no}(output);
        
        \draw [->] (ls) -- node[anchor=west]{low-rank} (ls1);
        \draw [->] (ls) -- node[anchor=east]{sparse} (sp1);
        \draw [->] (ls1) |- (output);
    \end{tikzpicture}
    \caption{Main structure and essential arguments of data generation function.}
    \label{fig:data-generation}
\end{figure}

Specifically, the data generation module is implemented around the function \code{simu_var}, and its signature is shown next.
\begin{Code}
simu_var(method = c("sparse", "group sparse", "fLS", "LS"), nob, k, lags = 1, 
 lags_vector = NULL, brk, sigma, skip = 50, signals = NULL, spectral_radius = 0.9, 
 group_mats = NULL, group_type = c("columnwise", "rowwise"), group_index = NULL, 
 sparse_mats = NULL, sp_density = NULL, rank = NULL, info_ratio = NULL, seed = 1,
 sp_pattern = c("off-diagonal", "diagonal", "random"), singular_vals = NULL)
\end{Code}
The arguments of function \code{simu_var} are listed as follows:
\begin{itemize}
\item \code{method}: An indicator of the specific structure of the transition matrices of the VAR process. There are four options: sparse (\code{"sparse"}), group sparse (\code{"group"}), fixed low rank plus sparse (\code{"fLS"}), and low rank plus sparse (\code{"LS"}).
\item \code{nob}: A numeric value that represents the number of observations $T$.
\item \code{k}: A numeric value, for the dimension of the VAR process.
\item \code{lags}: A numeric value, indicates the number of time lags of the VAR process. For example, we set \code{lags = q} for a VAR($q$) model. Default is \code{lags = 1} for a VAR(1) process.
\item \code{lags_vector}: A numeric vector, indicates the number of time lags in each stationary segment of the VAR model. 
\item \code{brk}: A numeric vector, includes all change points plus (\code{nob + 1}) as the last element. For example, suppose we have two change points at locations 1/3 and 2/3 of the total number of time points; we set \code{brk <- c(floor(nob / 3), floor(2 * nob / 3), nob + 1)}.
\item \code{sigma}: A numeric matrix. It represents the variance matrix for error term $\epsilon_t$.
\item \code{skip}: A numeric value. It is used to control the number of skipped leading data points in order to obtain a stationary time series. Specifically, when \code{simu_var} generates the process, it synthesizes a process with length of \code{nob + skip}, and returns the final process by removing the leading \code{skip} samples. Default is \code{skip = 50}.

\item \code{group_mats}: A list of numeric matrices. Includes all transition matrices for group sparse structure. It is only available for \code{method = "group sparse"}.
\item \code{group_type}: A character string used for indicating the types for group lasso. Set \code{group_type = "columnwise"}, \code{group_type = "rowwise"}, or \code{group_type = "index"} for column-wise, row-wise, and group index structures, respectively. Default setting is \code{"columnwise"}.
\item \code{group_index}: A list of numeric vectors, which indicates the group indices for a group sparse structure. Note that it is only available if both arguments \code{method = "group sparse"} and \code{group_type = "index"} are satisfied. For example, for a VAR($q$) model, we set the \code{group_index} as a list of length $q+1$, where the $l$-th element of the list stands for the non-zero row/column in the $l$-th lag, $l = 1, \dots, q$. The $(q+1)$-th element of the list stands for the zero row/column in the transition matrices.

\item \code{sparse_mats}: A list of numeric matrices. Includes all transition matrices across segments for sparse (components). It is available for \code{method = "sparse"} or \code{method = "LS"}.
\item \code{sp_pattern}: A character string used to choose the pattern of sparse component. Setting \code{sp_pattern = "diagonal"}, \code{sp_pattern = "off-diagonal"}, or \code{sp_pattern = "random"}. The random structure is generated by using Erd\"{o}s-R\'{e}nyi random graph provided in \pkg{igraph} \proglang{R} package. Default is \code{sp_pattern = "off-diagonal"}.
\item \code{sp_density}: A numeric vector, each element is in $(0,1)$. If \code{sp_pattern = "random"}, it dictates the density (proportion of non zero elements) in the transition matrices in each segment; it uses the \code{igraph} \proglang{R} package to generate a random sparse matrix. 
\item \code{signals}: A numeric vector. It assigns the magnitudes of sparse (components) for each segment. For example, if \code{signals = c(0.6, -0.5, 0.75)}, it indicates that all non-zero entries in the transition matrices of the three segments are 0.6, -0.5, and 0.75, respectively.

\item \code{rank}: A numeric vector, which indicates the rank of the transition matrix in each segment. Note that it is only available for \code{method = "LS"}.
\item \code{info_ratio}: A numeric vector, each element in the vector represents the information ratio for the corresponding segment, wherein the definition of information ratio is provided in Section \ref{sec:reduced-rank}. 
\item \code{singular_vals}: A numeric vector. It represents the singular values for the low rank components. Note that the length of \code{singular_vals} must be equal to the maximum rank in the corresponding \code{rank} argument.

\item \code{spectral_radius}: A numeric value in $(0,1)$ that ensures the generated time series is piece-wise stationary. Default value is 0.9, while it can be customized. Note that for each given transition matrix, the function will check stability for each segment and modify the amount of signals based on the provided argument \code{spectral_radius}, if the current segment is not stable. 
\item \code{seed}: A numeric value to control the random seed, with the default value set to 1.
\end{itemize}
The function \code{simu_var} returns a list comprising of a $n\times p$ numeric matrix containing all synthetic observations of the specified VAR model, as well as a numeric matrix of the same dimension containing the noise term, and a list with length equal to the number of segments consisting of model parameters. Usage examples are provided in Section \ref{sec:4}.

\subsection{TBSS for change points detection and model estimation}\label{sec:tbss}
As shown in Figure \ref{fig:main-structure}, the detection and estimation module comprises of the TBSS and LSTSP algorithms. Next, we summarize the key steps of TBSS in Algorithm \ref{algo:tbss} and provide the details of its implementation. 
\begin{algorithm}[!ht]
    \DontPrintSemicolon
    \KwInput{Time series data $\{X_t\}$, $t=1,2,\dots, n$; choice of regularization $\mathcal{R}$; regularization parameters $\lambda_{1,n}$ and $\lambda_{2,n}$ (optional); block size $b_n$ (optional).}
    
    \ {\bf Block fused lasso:} Partition time series into blocks of size $b_n$ and fix the coefficient parameters within each block. Then, estimate the model parameters for all blocks $\widehat{\Theta}$ by 
    solving \eqref{eq:general-penalty} with the relevant regularization penalty.
    
    \ {\bf Local screening:} For each candidate change point $\widehat{t}_j \in \widehat{\mathcal{A}}_n$, estimate the VAR model parameters \emph{locally} on the left and right hand sides of $\widehat{t}_j$ and compare them to those from a single VAR model that combines the data from the left and right sides of the candidate change points in a single stationary segment. This step uses equation \eqref{eq:lic}, and obtains a subset of the candidate change points defined in \eqref{eq:lic-objective}, and denoted by $\widetilde{\mathcal{A}}_n$.
    
    \ {\bf Exhaustive search:} applied to each selected cluster of break points $C_i$, for $i=1,2,\cdots, \widetilde{m}$ obtained from the local screening step. Define the search interval $(l_i, u_i)$, with these two quantities defined in Section \ref{sec:2.3}. Denote the subset of corresponding block indices in the interval $(l_i, u_i)$ by $J_i$ with $J_0 = \{1\}$ and $J_{\widetilde{m}+1} = \{k_T\}$. Then, denote the closest block end to $(\max J_{i-1} + \min J_{i})/2$ by $w_i$. 
    
    \For{$i=1,2,\dots, \widetilde{m}+1$}{
    \ Define \textit{local} parameter estimators by using \eqref{eq:psi}.
    
    \ The final selected break points $\widetilde{t}^f_i$ are derived by \eqref{eq:final-estimate}.
    }
    
    \ {\bf Model parameters estimation:} Based on the final estimated break points $\widetilde{\mathcal{A}}_n$, partition the input time series into $\widetilde{m}$ \textit{strictly stationary} segments, and estimate model parameters on all estimated segments by solving \eqref{eq:estimation_third}.
    
    \KwOutput{The final estimated break points $\widetilde{\mathcal{A}}_n = \{\widetilde{t}_1, \dots, \widetilde{t}_{\widetilde{m}}\}$, and corresponding model parameters $\widehat{\textbf{B}}$.}
    \caption{Threshold Block Segmentation Scheme (TBSS) Algorithm}
    \label{algo:tbss}
\end{algorithm}

The TBSS algorithm from Section \ref{sec:structure-sparse} for multiple change points detection under structural sparse VAR models is implemented in the function \code{tbss}:
\begin{Code}
tbss(data, method = c("sparse", "group sparse", "fLS"), q = 1, tol = 1e-2, 
  lambda.1.cv = NULL, lambda.2.cv = NULL, mu = NULL, group.index = NULL, 
  group.case = c("columnwise", "row-wise"), max.iteration = 50, refit = FALSE, 
  block.size = NULL, blocks = NULL, use.BIC = TRUE, an.grid = NULL)
\end{Code}
The accepted arguments of function \code{tbss} are described next:
\begin{itemize}
    \item \code{data}: A $n \times p$ numeric matrix, or dataframe, including all observations from the VAR process.
    \item \code{method}: A character string indicating the structure of the VAR model, with following three options: \code{"sparse"}, \code{"group sparse"}, and \code{"fLS"}. 
    \item \code{q}: A numeric value, it specifies the time lags for the given data. Default is 1.
    \item \code{tol}: A numeric value, indicates the tolerance of convergence in the first step of TBSS. Default is \code{1e-2}.
    \item \code{lambda.1.cv}: A numeric vector, indicates the tuning parameters for $\lambda_{1,n}$ introduced in the generic penalty function $\mathcal{R}$ in \eqref{eq:general-penalty}, \eqref{eq:fused-lasso}, \eqref{eq:group-lasso}, and \eqref{eq:lps-fixed}. 
    \item \code{lambda.2.cv}: A numeric vector, the values of tuning parameter for $\lambda_{2,n}$.
    \item \code{mu}: A numeric value, indicating the tuning parameter for the low rank component, only available when the method is set to \code{"fLS"}.
    \item \code{group.index}: A list of vectors of length $q+1$. For each element, it indicates the indices of non-zero rows or columns. Default is set to \code{NULL}.
    \item \code{group.case}: A character string, which is used to specify the settings for the group sparse case. It is only available when \code{method = "group sparse"}. 
    \item \code{max.iteration}: A numeric value, the maximum number of iterations for the first step of TBSS. The default value is 50.
    \item \code{refit}: A boolean argument, if \code{TRUE}, the function will refit the model parameters in each detected stationary segment. Default is \code{FALSE}. 
    \item \code{block.size}: A numeric value, that must be in the range $[2, n/2]$. If it is set to be \code{NULL}, then the function uses $\lfloor \sqrt{n} \rfloor$ as the block size. 
    \item \code{blocks}: A numeric vector, indicates the end-points of the blocks. The user can customize the size of each block by using this argument. 
    \item \code{use.BIC}: A boolean argument, if \code{TRUE}, the function will use BIC to choose the tuning parameter $\omega_n$ in \eqref{eq:lic-objective} of Step 2 of TBSS. 
    \item \code{an.grid}: A numeric vector, contains the values of $a_n$ which proposed in \eqref{eq:lic}.
\end{itemize}
The function \code{tbss} is flexible in that it allows the user to specify a custom block size or blocks by using the arguments \code{block.size} and \code{blocks}, respectively. In the next Section \ref{sec:guideline}, we also provide guidelines for selecting those arguments in applications.

When \code{tbss} is called, it returns an \proglang{S3} object of class \code{"VARDetect.result"}, including the following entries:
\begin{itemize}
    \item \code{data}: The original time series data of size $n\times p$.
    \item \code{q.t}: The time lags based on the user specified.
    \item \code{cp}: Final estimated change points, set as a numeric vector with length of $\widehat{m}$.
    \item \code{sparse_mats}: Estimated sparse components across all detected segments, as a list of numeric matrices of size $p \times p$ and length of $\widehat{m}+1$.
    \item \code{lowrank_mats}: Estimated low rank components across all detected segments, set as a list of numeric matrices.
    \item \code{est_phi}: The final estimated model parameters for all detected segments, set as a list of numeric $p\times p$ matrices of length $\widehat{m}+1$. 
    \item \code{time}: The computational time for applying the algorithm.
\end{itemize}
Note that \proglang{S3} objects of class \code{VARDetect.result} are supported by \code{plot}, \code{print}, and \code{summary} methods. We provide detailed descriptions of plot function \code{plot.VARDetect.result}, and summarize function \code{summary.VARDetect.result} in Section \ref{sec:visualization} and Section \ref{sec:evaluation}, respectively. Also, a series of usage examples are provided in Section \ref{sec:4}. 

\subsection{LSTSP for change points detection and model estimation}\label{sec:lstsp}
The key steps of the LSTSP algorithm are summarized in Algorithm \ref{algo:lstsp}.  
\begin{algorithm}[!ht]
    \DontPrintSemicolon
    \KwInput{Time series data $\{X_t\}$, $t=1,2,\dots, T$; Regularization parameters $\lambda_{j,n}$ and $\mu_{j,T}$ for $j=1,2,3$.}
    
    \ {\bf Rolling window scheme:} Suppose we have a rolling window $[b,e)$ of length $h=e-b$, and rolling step $l$. 
    
    \ \While{$e \leq T$}{
        \ Consider the process $\{X_t\}$ in the interval $[b,e)$ and detect a \emph{candidate} change point by minimizing the sum of squared errors as described in equation \eqref{eq:16} in Section \ref{sec:lstsp-model}.
        
        \ Append detected candidate change point into set $\widehat{\mathcal{A}}$.
        
        \ $b \gets b+l$ and $e \gets e+l$
    }
    
    \ {\bf Screening step:} Based on the candidate change points set $\widehat{\mathcal{A}}$, use the backward elimination algorithm to remove redundant candidate change points to obtain the final set of detected change points. 
    
    \ Partition the time axis based on candidate change points in $\widehat{\mathcal{A}}$. Set the initial value of the information criterion is $W_0=0$. 
    
    \ \While{$W_{m-1} \leq W_m$ and $m \neq 1$}{
        \ Let $\widetilde{\mathbf{t}} \overset{\text{def}}{=} \{\widetilde{t}_1, \dots, \widetilde{t}_{m}\}$ be the screened change points and define $W_m^* = \text{IC}(\widetilde{\mathbf{t}}; \bm{\lambda}, \bm{\mu}, \omega_n)$;
        
        \ For each $j=1,2,\dots, m$, we calculate $W_{m,-j} = \text{IC}(\widetilde{t}/\{\widetilde{t}_j\}; \bm{\lambda}, \bm{\mu}, \omega_n)$, and define $W_{m-1} = \min_j W_{m,-j}$;
    }
    
    \ Estimate model parameters $\Phi_j$ based on the partitioned segments determined by the final set of detected change points $\widetilde{t}_j$'s. 
    
    \KwOutput{Detected change points $\widehat{t}_j$; Estimated model parameters $\widehat{\Phi}_j$, for $j=1,2,\dots, \widehat{m}$.}
    \caption{Low rank plus Sparse Two Step Procedure (LSTSP) Algorithm}
    \label{algo:lstsp}
\end{algorithm}

The function \code{lstsp} implements the LSTSP algorithm. The signature of \code{lstsp} is described next.
\begin{Code}
lstsp(data, lambda.1 = NULL, mu.1 = NULL, lambda1.seq = NULL, mu1.seq = NULL,
 lambda.2, mu.2, lambda.3, mu.3, omega = NULL, h = NULL, step.size = NULL, 
 tol = 1e-4, niter = 100, backtracking = TRUE, skip = 5, cv = FALSE, 
 nfold = NULL, verbose = FALSE)
\end{Code}
\begin{itemize}
    \item \code{data}: A $n\times p$ numeric matrix, or dataframe. 
    \item \code{lambda.1}: A numeric vector that contains the tuning parameters for sparse component in the rolling window step of the LSTSP algorithm. It includes two values for the left and right side models of any time point $t$. 
    \item \code{mu.1}: A numeric vector that contains the tuning parameters for the low rank components in the rolling window step of the LSTSP algorithm. Similar to \code{lambda.1}.
    \item \code{lambda.1.seq}: A numeric vector of a sequence of tuning parameters for the sparse component for the left segment for the first step of cross validation. If \code{lambda1.seq} is set to \code{NULL}, the function sets it according to the theoretical results in \citep{basu2019low}.
    \item \code{mu.1.seq}: A sequence of low rank tuning parameters for the left segment in cross validation. If it is \code{NULL}, then the function sets \code{mu1.seq} based on the theoretical results. 
    \item \code{lambda.2}: A numeric vector that contains the tuning parameters for the sparse components in the screening step of the LSTSP algorithm. Similar to \code{lambda.1}.
    \item \code{mu.2}: A numeric vector that contains the tuning parameters for the low rank components in the screening step of the LSTSP algorithm. Similar to \code{lambda.1}.
    \item \code{lambda.3}: A numeric number, tuning parameter for re-estimating the sparse components on detected stationary segments after change point detection. 
    \item \code{mu.3}: A numeric number, tuning parameter for re-estimating the low rank components on detected stationary segments after change point detection. 
    
    \item \code{omega}: A numeric value, for the tuning parameter in the screening step of LSTSP; the larger the \code{omega}, the smaller the number of final selected change points obtained. If \code{omega = NULL}, then the function will use the theoretical results according to \cite{safikhani2020joint}.
    \item \code{h}: A numeric value that indicates the length of the window size; must be in the range of $[1,n]$. If it is \code{NULL}, the function will set \code{h} as $\lfloor \sqrt{n} \rfloor$.
    \item \code{step.size}: A numeric value, the step size for the rolling window scheme; if \code{step.size = NULL}, then the function will use $\lfloor h/4 \rfloor$ as the step size.
    \item \code{tol}: A numeric value, controls the algorithm convergence. Default value is \code{1e-4}.
    \item \code{niter}: A numeric value, specifies the maximum number of iterations for algorithm. Default is 100. 
    \item \code{backtracking}: A boolean argument, if \code{TRUE}, the function will apply backtracking method to FISTA algorithm in VAR model parameters estimation. 
    \item \code{skip}: A numeric value, it indicates the number of observations near the boundaries the function should skip. Default is 5. 
    \item \code{cv}: A boolean argument, indicates if it uses cross validation to select tuning parameter in the first step.
    \item \code{nfold}: A positive integer, represents the number of folds applied in cross validation.
    \item \code{verbose}: A boolean argument, if \code{TRUE}, the function will print all information during the detection and estimation procedure. Default is \code{FALSE}.
\end{itemize}
When called, \code{lstsp} returns the same \proglang{S3} object of class \code{"VARDetect.result"} described in Section \ref{sec:tbss}. An illustration of the algorithm and the code is given in Section \ref{sec:4}.

\subsection{Visualization and summary}\label{sec:visualization}
Plots of the locations of the detected change points, as well as the estimated model parameters by \pkg{VARDetect} are discussed next. The \code{plot} method is supported for \proglang{S3} objects of class \code{"VARDetect.result"}; it is designed to plot the given time series along with the detected change points, the estimated model parameters as heatmaps of the transition matrices and also as network layouts of the Granger causal effects an model parameter through Granger network layouts, and finally the density level (\% of non-zero coefficients) of the estimated sparse components across all detected stationary segments.

\proglang{S3} objects of class \code{"VARDetect.result"} for the output are supported by \code{plot} method. The arguments of the function are described next:
\begin{Code}
plot.VARDetect.result(x, display = c("cp", "param", "granger", "density"), 
  cp.col = "red", threshold = 0.1, layout = c("circle", "star", "nicely"), ...)
\end{Code}
It accepts the following arguments:
\begin{itemize}
    \item \code{x}: An object of class \code{"VARDetect.result"}.
    \item \code{display}: A character string indicating the object to plot. \code{display = "cp"} plots the detected change points together with the input time series, \code{display = "param"} plots the estimated model parameters as a heatmap, \code{display = "granger"} plots the sparse components in a network layout format, while \code{display = "density"} provides the density level across all segments.
    \item \code{cp.col}: A character string determining the color to plot the location of detected change points with the default value being red.
    \item \code{threshold}: A positive number, representing the threshold to include an entry from the estimated transition matrix in the Granger causal network and density plot.
    \item \code{layout}: A character string indicating the layout used for the the Granger causal network. The following three options are available: \code{layout = "circle"}, \code{layout = "star"}, and \code{layout = "nicely"}.
\end{itemize}

We illustrate the function on the following synthetic data set generated by the \code{simu_var} and fitted by the \code{tbss} functions, respectively:
\begin{CodeChunk}
\begin{CodeInput}
R> nob <- 4000; p <- 15
R> brk <- c(floor(nob / 3), floor(2 * nob / 3), nob + 1)
R> m <- length(brk); q.t <- 1
R> sp_density <- rep(0.05, m * q.t)
R> signals <- c(-0.6, 0.6, -0.6)
R> try <- simu_var(method = "sparse", nob = nob, k = p, lags = q.t, 
+                  brk = brk, sigma = diag(p), signals = signals, 
+                  sp_density = sp_density, sp_pattern = "random", seed = 1)
R> data <- as.matrix(try$series)
R> fit <- tbss(data, method = "sparse", q = 1)
\end{CodeInput}
\end{CodeChunk}

\begin{figure}[!ht]
    \centering
    \includegraphics[width=.33\textwidth, trim={0 10 0 50}, clip]{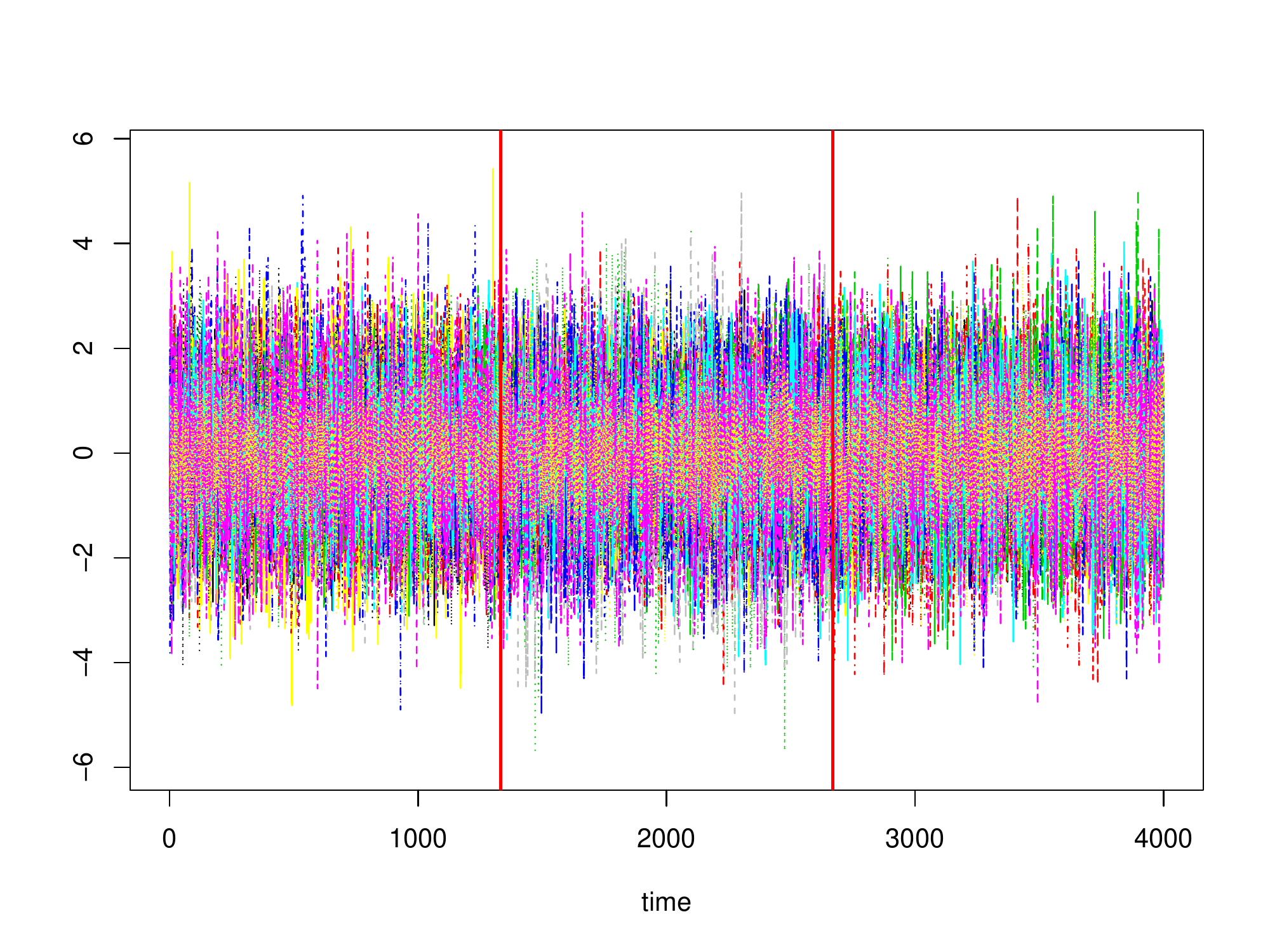}%
    \includegraphics[width=.33\textwidth, trim={0 10 0 50}, clip]{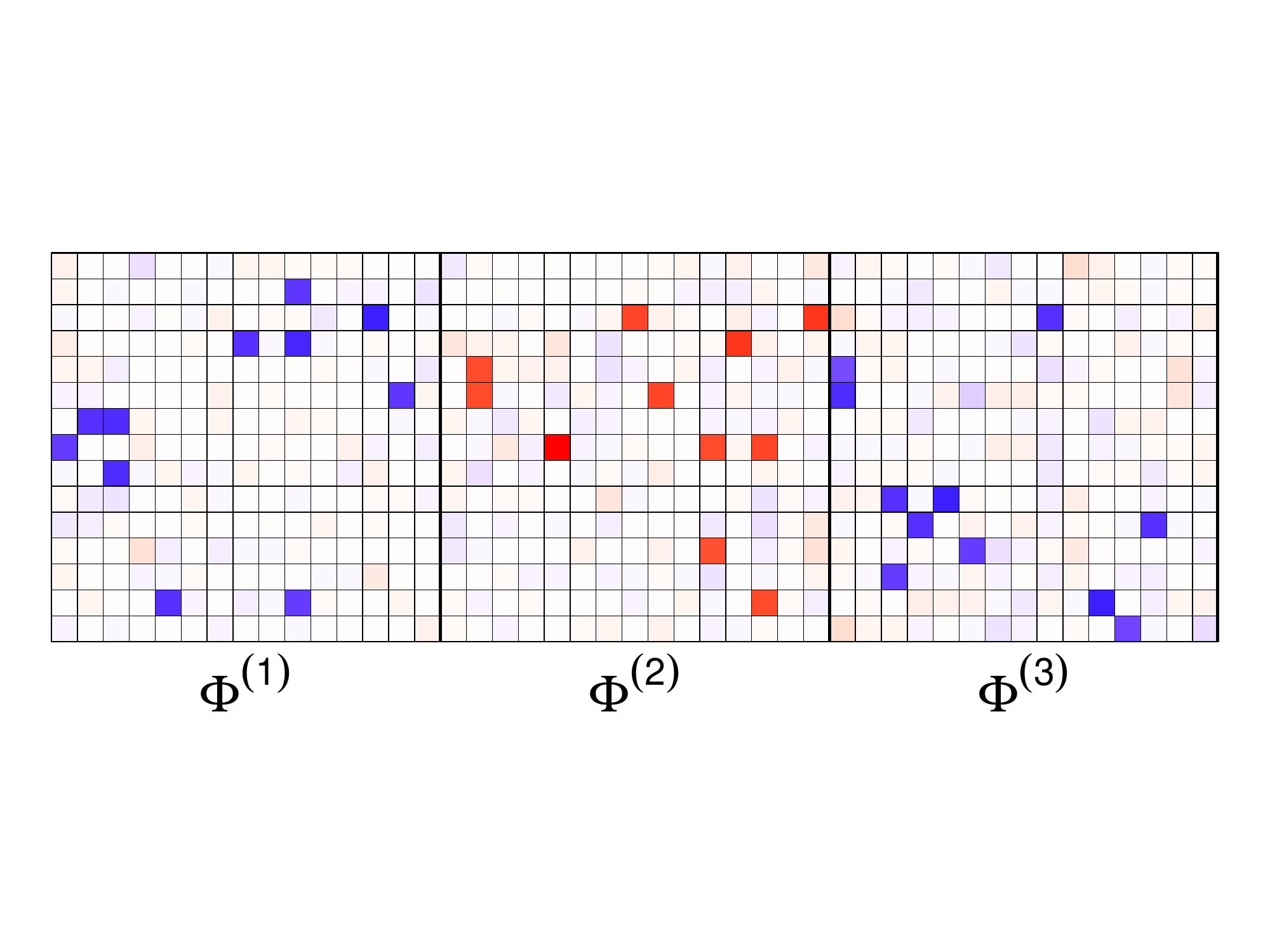}%
    \includegraphics[width=.33\textwidth, trim={0 10 0 50}, clip]{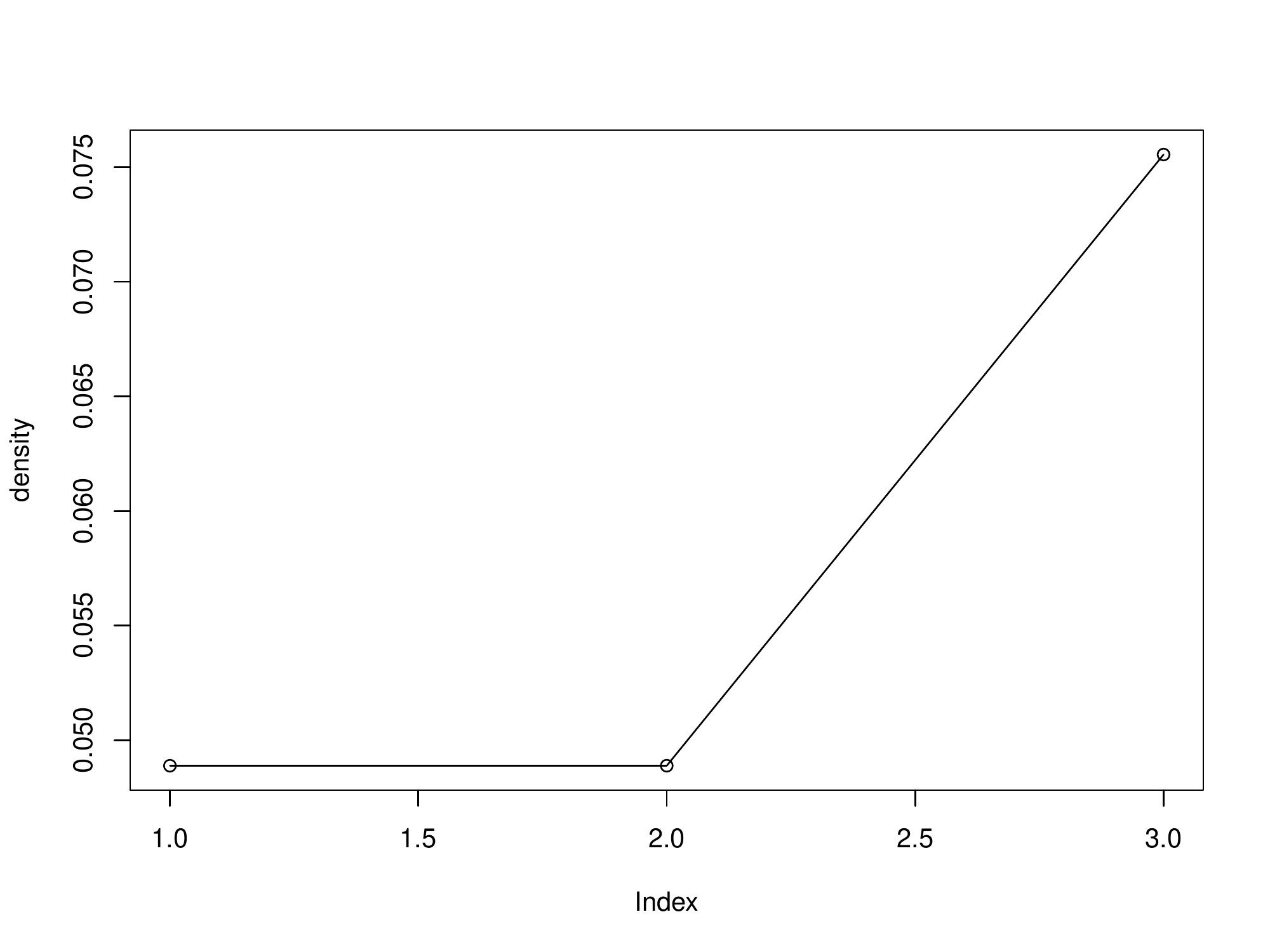}%
    \vfill
    \includegraphics[width=.33\textwidth, trim={0 10 0 50}, clip]{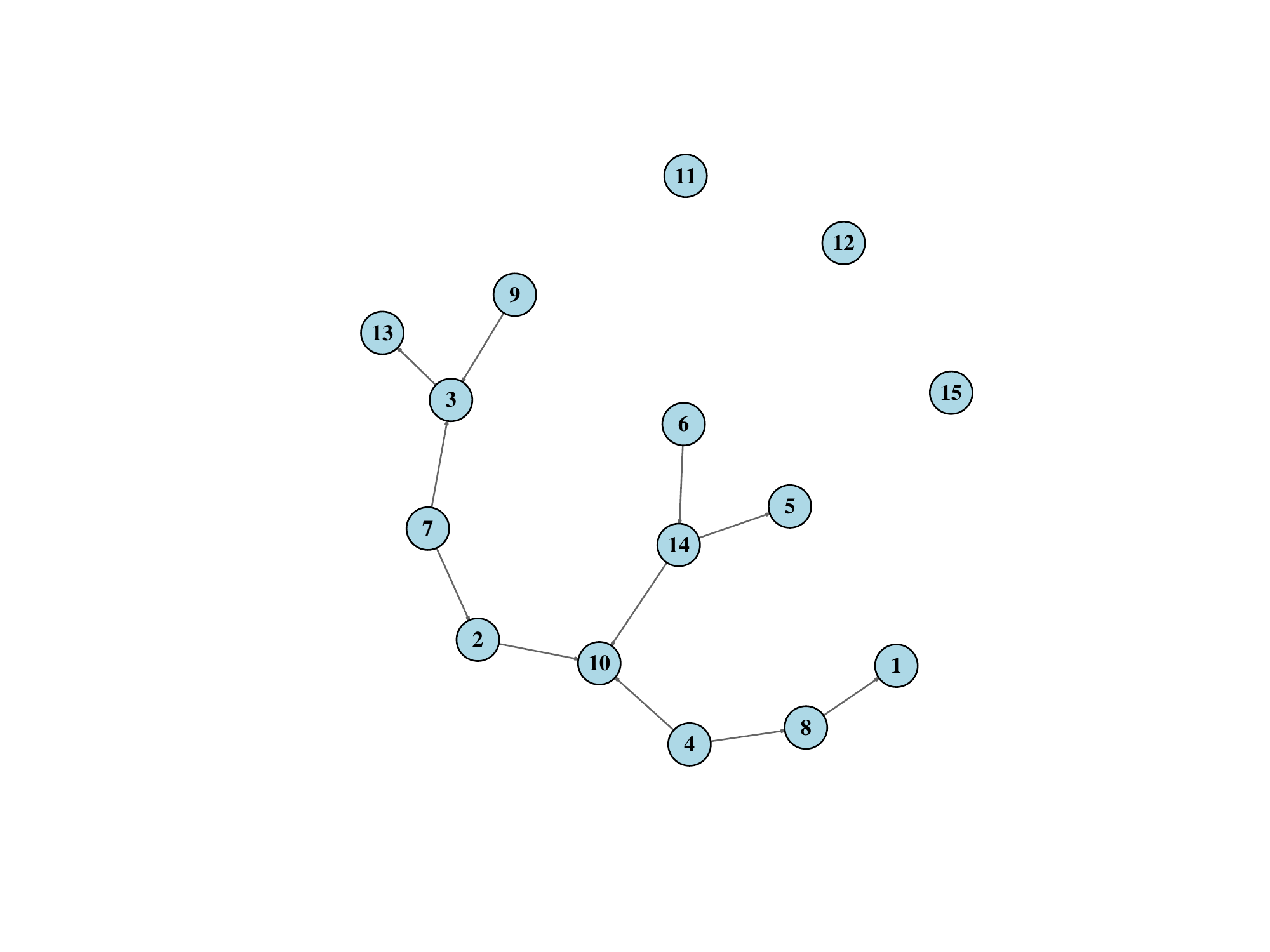}%
    \includegraphics[width=.33\textwidth, trim={0 10 0 50}, clip]{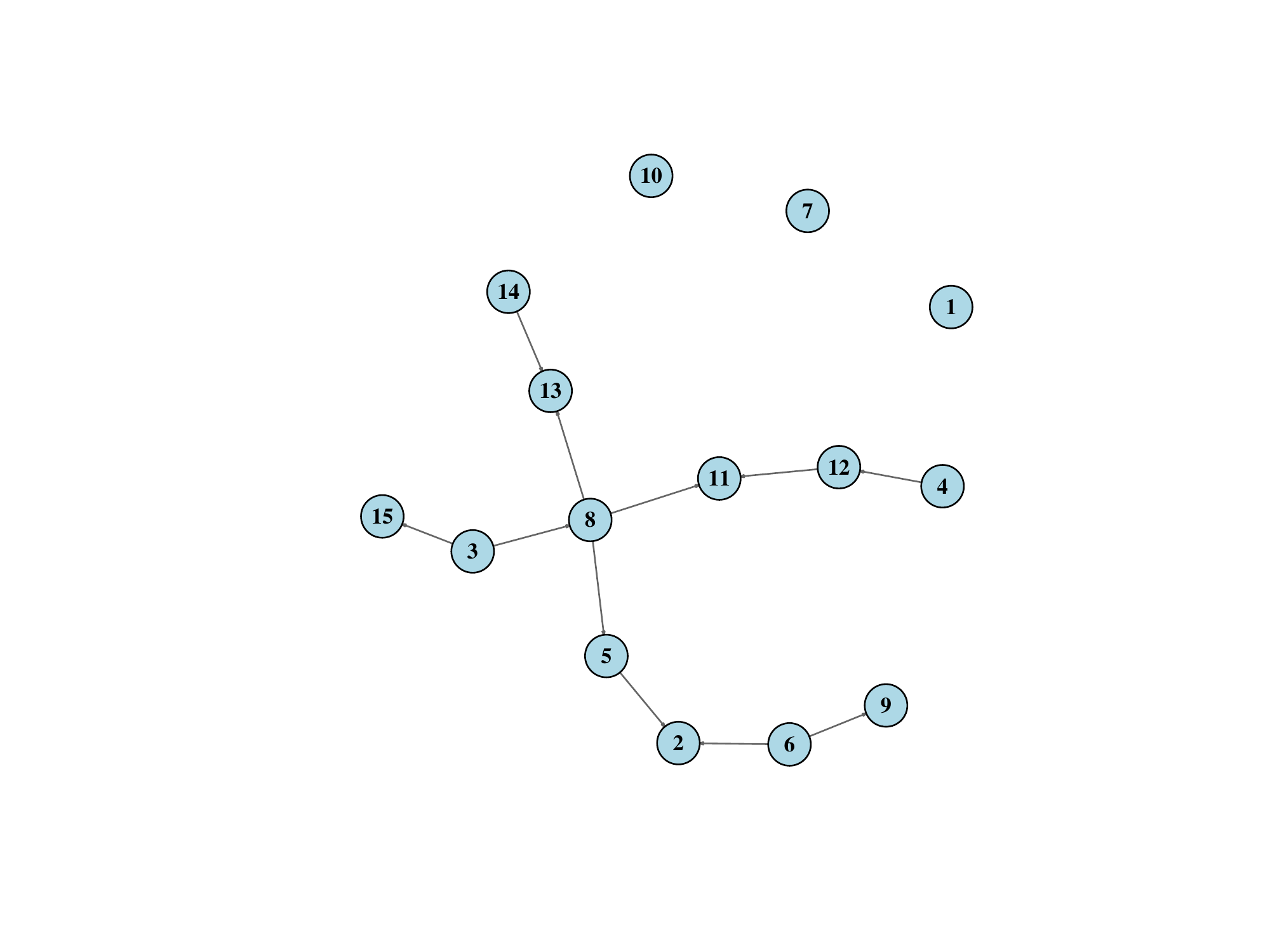}%
    \includegraphics[width=.33\textwidth, trim={0 10 0 50}, clip]{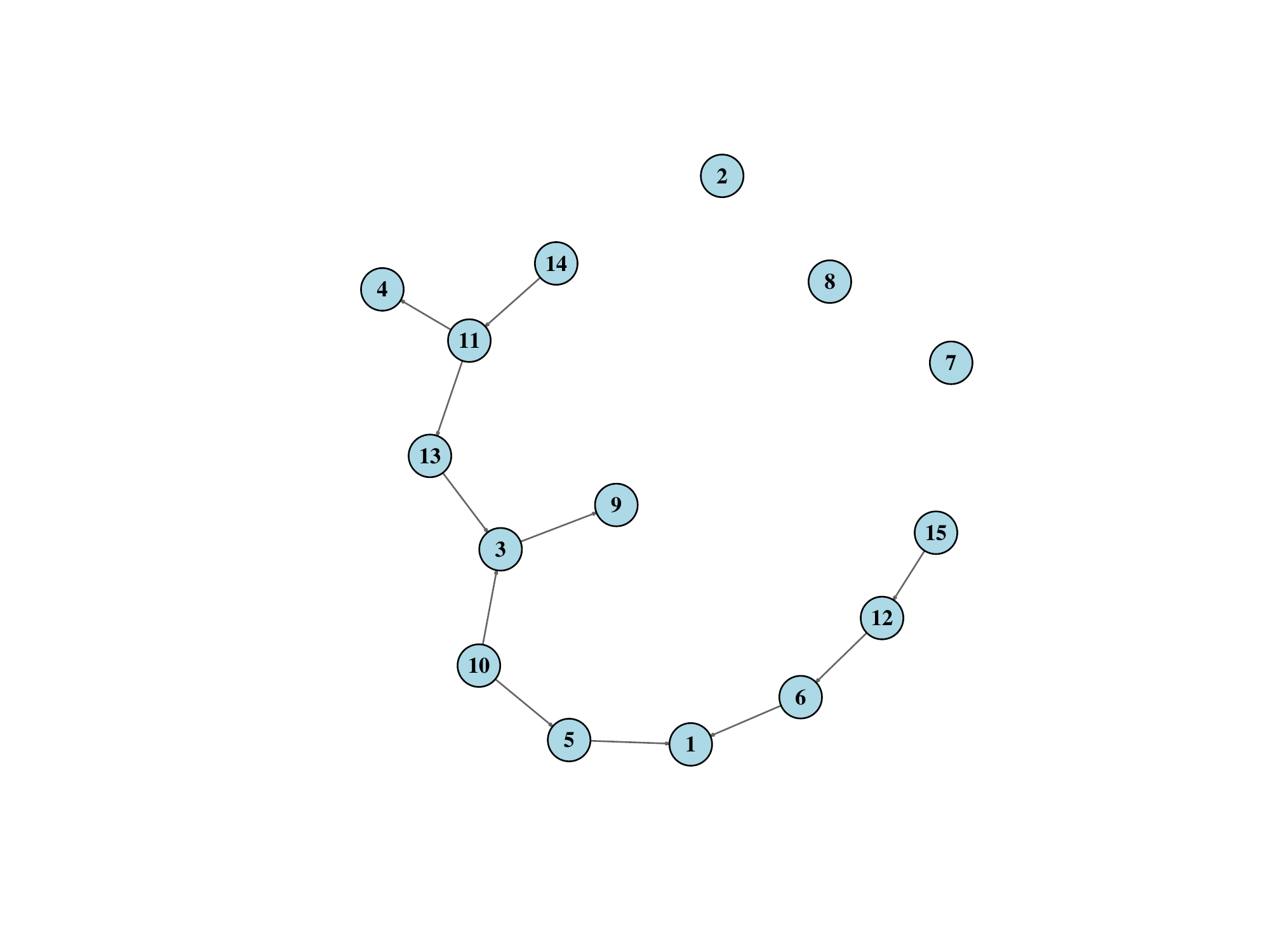}
    \caption{Top left: The input time series data along with the detected change points located at $\widehat{t}_1 = 1333$ and $\widehat{t}_2 = 2668$. Top middle: heatmap of the estimated model parameters for all three segments. Top right: density level for the estimated model parameter thresholded at 0.1. Bottom: Granger causal networks for the estimated sparse components across the three stationary segments}
    \label{fig:visualize-examples}
\end{figure}

We summarize the analysis pf the data set through the following plots: (i) the input time series together with the detected change points, (ii) a heatmap of the estimated model parameters, (iii) the density plot of the sparse components, and (iv) Granger causal networks obtained from the estimated sparse components across all partitioned stationary segments.
\begin{CodeChunk}
\begin{CodeInput}
R> plot(fit, display = "cp")
R> plot(fit, display = "param")
R> plot(fit, display = "density", threshold = 0.1)
R> plot(fit, display = "granger", threshold = 0.2, layout = "nicely")
\end{CodeInput}
\end{CodeChunk}
These plots are depicted in Figure \ref{fig:visualize-examples}.

In addition to the \code{plot} function, the \proglang{S3} object of class \code{"VARDetect.result"} also supports the \code{summary} and \code{print} functions, the signatures of functions are illustrated next:
\begin{Code}
print.VARDetect.result(object, ...)
summary.VARDetect.result(object, ...)
\end{Code}
where \code{object} argument indicates the \proglang{S3} object of class \code{"VARDetect.result"}. The usage example is demonstrated as follows:
\begin{CodeChunk}
\begin{CodeInput}
R> print(fit)
R> summary(fit)
\end{CodeInput}
The output is given by:
\begin{CodeOutput}
Estimated change points are: 1333 2668
============================= Summary ============================
Detected 2 change points, located at: 
1333 2668
==================================================================
Sparsity levels for estimated sparse components are: 
0.1955556 0.1866667 0.1688889
==================================================================
There is no low rank components in the current model! 
==================================================================
Running time is: 
2.55 seconds
==================================================================
\end{CodeOutput}
\end{CodeChunk}

\subsection{Simulation functions and performance evaluation}\label{sec:evaluation}
In \pkg{VARDetect}, the functions \code{simu_tbss} and \code{simu_lstsp} generate data for numerical experiments. The features of \code{simu_tbss} are:
\begin{Code}
simu_tbss(nreps, simu_method = c("sparse", "group sparse", "fLS"), 
  nob, k, lags = 1, lags_vector = NULL, brk, sigma, skip = 50, 
  group_mats = NULL, group_type = c("columnwise", "rowwise"), 
  roup_index = NULL, sparse_mats = NULL, sp_density = NULL, 
  signals = NULL, rank = NULL, info_ratio = NULL, 
  sp_pattern = c("off-diagonal", "diagoanl", "random"), 
  singular_vals = NULL, spectral_radius = 0.9, 
  est_method = c("sparse", "group sparse", "fLS"), 
  q = 1, tol = 1e-2, lambda.1.cv = NULL, lambda.2.cv = NULL, mu = NULL, 
  group.index = NULL, group.case = c("columnwise", "rowwise"), 
  max.iteration = 100, refit = FALSE, block.size = NULL, blocks = NULL, 
  use.BIC = TRUE, an.grid = NULL)
\end{Code}
then, the signature of \code{simu_lstsp} is given next:
\begin{Code}
simu_lstsp(nreps, simu_method = c("LS"), nob, k, lags = 1, 
  lags_vector = NULL, brk, sigma, skip = 50, group_mats = NULL, 
  group_type = c("columnwise", "rowwise"), group_index = NULL, 
  sparse_mats = NULL, sp_density = NULL, signals = NULL, rank = NULL, 
  info_ratio = NULL, sp_pattern = c("off-diagonal", "diagonal", "random"), 
  singular_vals = NULL, spectral_radius = 0.9, 
  lambda.1 = NULL, mu.1 = NULL, lambda.1.seq = NULL, mu.1.seq = NULL, 
  lambda.2, mu.2, lambda.3, mu.3, alpha_L = 0.25, omega = NULL, h = NULL, 
  step.size = NULL, tol = 1e-4, niter = 100, backtracking = TRUE, 
  rolling.skip = 5, cv = FALSE, nfold = NULL, verbose = FALSE)
\end{Code}
As one can see, both functions accept the extra argument \code{nreps}, in addition to the arguments accepted by functions \code{simu_var} and \code{tbss}:
\begin{itemize}
    \item \code{nreps}: A positive integer, indicating the number of simulation replications.
\end{itemize}
These two functions return the same \proglang{S3} object of class named \code{"VARDetect.simu.result"}, which supports the \code{summary} method to summarize the detected change points, as well as the estimated model parameters. To implement the \code{summary} method, we consider the following performance metrics: \emph{selection rate}, \emph{Hausdorff distance}, and some commonly used \emph{statistical metrics} including \emph{sensitivity}, \emph{specificity}, \emph{accuracy}, and  \emph{Matthew's correlation coefficient (MCC)}, respectively. 

(a) The metric \emph{selection rate} is defined as follows: for any true change point $t_j$, there exists an estimated change point $\widehat{t}_j$ satisfying
\begin{equation}
    \label{eq:success}
    t_j - \frac{t_j - t_{j-1}}{L} \leq \widehat{t}_j \leq t_j + \frac{t_{j+1} - t_j}{L},
\end{equation}
then a \emph{success} is denoted by, where $L$ corresponds to a critical value to control the range of the success. Then, the \emph{selection rate} for the $j$-th true change point is defined as
\begin{equation*}
    \text{selection rate} \overset{\text{def}}{=} \frac{\# \text{success}}{\# \text{replications}}.
\end{equation*}

(b) The Hausdorff distance between the set of estimated change points $\widetilde{\mathcal{A}}$ and that of true change points $\mathcal{A}$ is defined as follows: for any two sets $A$ and $B$ let
\begin{equation*}
    d_H(A, B) = \max\left\{ \max_{a\in A}\min_{b\in B}|a-b|,\ \max_{b\in B}\min_{a\in A}|a-b| \right\}.
\end{equation*}
In the \code{summary} method, it returns the mean, standard deviation, and the median of the Hausdorff distance between the set of detected change points and that of the true change points.

(c) The remaining evaluation  metrics correspond to the popular
\emph{sensitivity}, \emph{specificity}, \emph{accuracy}, and \emph{Matthew's correlation coefficient (MCC)}, respectively. 

In summary, the \proglang{S3} object of class \code{"VARDetect.simu.result"} provides the true change points and model parameters together with the fitted results for each replicate, and the \code{summary} methods return a detailed analysis of the change point selection rate, Hausdorff distance, together with standard performance metrics for the estimated model parameters. 
\begin{Code}
summary.VARDetect.simu.result(object, critical = 5, ...)
\end{Code}
The accepted arguments in this function are:
\begin{itemize}
    \item \code{object}: An object of \proglang{S3} class \code{"VARDetect.simu.result"}.
    \item \code{critical}: A positive numeric value, representing the critical value $L$ defined in \eqref{eq:success}.
\end{itemize}

As an illustration, we revisit the synthetic time series data used in Section \ref{sec:visualization} and apply \code{simu_tbss} to conduct a toy simulation study with 5 replicates:
\begin{CodeChunk}
\begin{CodeInput}
R> try_simu <- simu_tbss(nreps = 5, simu_method = "sparse", nob = nob, 
+                        k = p, lags = q.t, brk = brk, sigma = diag(p), 
+                        signals = signals, sp_pattern = "off-diagonal",
+                        est_method = "sparse", q = q.t, refit = TRUE)
R> summary(try_simu, critical = 5)
\end{CodeInput}
The output is illustrated in the following:
\begin{CodeOutput}
========================== Selection rate: ==========================
    Truth   Mean Std Selection rate
1 0.33325 0.3332   0              1
2 0.66650 0.6665   0              1
======================== Hausdorff distance: ========================
  Mean Std Median
1    0   0      0
====================== Statistical Measurment: ======================
     SEN    SPC    ACC    MCC
Mean   1 0.8306 0.8412 0.4876
Std    0 0.0362 0.0340 0.0484
Incorrect estimation replication: NULL
======================== Computational Time: ========================
Averaged running time: 4.272 seconds
=====================================================================
\end{CodeOutput}
\end{CodeChunk}

\subsection{Guidelines for tuning parameter selection}\label{sec:guideline}
There are a number of tuning parameters in the TBSS and LSTSP algorithms. The theoretical rates for those tuning parameters are provided in relevant papers \citep{safikhani2020joint, bai2020multiple}. Next, we provide guidelines for their selection.
\begin{itemize}
    \item[($\lambda_{1,n}$, $\lambda_{2,n}$):] Both $ \lambda_{1,n} $ and $ \lambda_{2,n} $ can be selected through cross-validation. In the package, we randomly select $ 20\% $ of the blocks equally spaced with a random initial point for splitting the training set and the validation set. Denote the last time point in these selected blocks by $ \mathcal{T}$. The data without observations in $ \mathcal{T}$ can then be used in the first step of our procedure to estimate $\Theta$ for a range of values for $ \lambda_{1,n} $ and $ \lambda_{2,n} $. The parameters estimated in the first step are then used to predict the series at time points in $ \mathcal{T}$. The values of $ \lambda_{1,n} $ and $\lambda_{2,n} $  which minimize the mean squared prediction error over $ \mathcal{T} $ correspond to the cross-validated choices. The sequence for $ \lambda_{1,n} $ is selected by constructing a sequence of $K_1$ values, decreasing from $\lambda_{1,\text{max}}$ to $\lambda_{1,\text{min}}$ on the log scale, where the maximum value  $\lambda_{1,\text{max}}$ is the smallest value for which the entire estimated parameter $\widehat{\theta_i} = 0$, for all $i = 1, \cdots, k_n$, while the minimum value is set to be $\lambda_{1,\text{min}} = \epsilon\lambda_{1,\text{max}}$. We choose $ \lambda_{2,n} = c\sqrt{\frac{\log p}{n}}$, where $c$ is a decreasing sequence of $K_2$ values. 
In the \pkg{VARDetect} package, we set  $K_1=10$, $K_2 = 3$, $\epsilon = 10^{-3}$ when the block size $b_n$ is smaller than $2p$ and $\epsilon = 10^{-4}$ otherwise.
    \item[$b_n$:] The TBSS method is robust to the choice of block size $b_n$; the default setting corresponds to $\lfloor \sqrt{n} \rfloor$.  
    \item[$a_n$:] This tuning parameter can be selected through a grid search. We apply an exhaustive search procedure on a grid of $a_n$'s ranked  from  the minimum to the maximum  and record the number of selected break points for each $a_n$, and  then stop  this  process  when  the  number  of  break  points  selected  does  not  change any more.  
Specifically, we select $ a_n^{(1)} , a_n^{(2)}, \cdots, a_n^{(\ell)} $ as an equally spaced sequence from the interval $[a_n^{(lb)}, a_n^{(ub)}]$ in a increasing order where $a_n^{(lb)}=\text{max}(\lfloor \overline{b_n}\rfloor, \lfloor\text{log}n \, \text{log}p  \rfloor )$, $a_n^{(ub)}=\text{min}(10a_n^{(lb)},(\widehat{t}_1-q - 1), (T-q -\widehat{t}_{\widehat{m}} -1) )$, and $\overline{b_n}$ is the mean of the block sizes. Denote the number of selected break points using BSS with $ a_n^{(i)} $ as the neighborhood size by $ n_i $, for $ i = 1, 2, \ldots, \ell$. The optimal neighborhood size can be defined as the first time $ n_i $ remains unchanged. In other words, $ a_n^{(\ell^\star)} $ is the optimal neighborhood size when $ \ell^\star = \min\{ {1 \leq i \leq \ell}: n_i = n_{i+1} = n_{i+2}\} $. In all simulation scenarios, we set $\ell = 5$.
\item[$ \eta_n $:] All $ \eta_{\widehat{t}_i } $, $ \eta_{\widehat{t}_i ,1} $ and $ \eta_{\widehat{t}_i ,2} $ are set to $ { \left( \log (2 a_n) \log p  \right) }/{\left( 2 a_n \right) } $, $ i = 1, \ldots, \widehat{m} $.

\item[$ \omega_n $:] Selecting $ \omega_n $ represents a challenge, since it depends on the magnitude of the changes in the VAR parameters so as to consider them as inducing change points. We consider a data-driven approach, wherein we cluster changes in the objective function $L_n$  into two subgroups, containing small and large ones, respectively. The proposed algorithm is summarized next:
\begin{itemize}
    \item Denote the candidate break points selected in the first step by $\widehat{t}_1, \cdots, \widehat{t}_{\widehat{m}}$. For each $k=1,2,\cdots, \widehat{m}$, compute $v_k  = L_n(\widehat{\mathcal{A}}_n \backslash \left \{\widehat{t}_{k} \right \}; \eta_n)- L_n(\widehat{\mathcal{A}}_n; \eta_n)  $.
    \item Consider two boundary points $t_1^r = a_n + q$ and    $t_2^r = T - a_n$ as reference points. Compute $v^{r}_i  = L_n(\widehat{\mathcal{A}}_n ; \eta_n)- L_n(\widehat{\mathcal{A}}_n \cup \{ t_i^r\}; \eta_n)  $ for $i = 1,2$, and set the reference value to $v^r = \max(v^{r}_1,v^{r}_2)$.
    \item Combine the jumps $v_k$ for each candidate break point and the reference value $v^r$ (with $2$ replicates) into one vector $V = (v_1,  v_2,\dots, v_{\hat{m}}, v^r, v^r )$. Apply the k-means clustering algorithm \citep{hartigan1979algorithm} to the vector $V$ with two centers. Denote the sub-vector with the smaller center as the small subgroup, $V_S$, and the other sub-vector as the large subgroup, $V_L$.
    \item 
If (between-group SS/total SS) in k-means clustering is high and the reference value $v^r$  is not in $V_L$, set $ \omega_n = \min V_L $; otherwise, set $ \omega_n = \max V$.  

\end{itemize}
One could also combine the $k$-means clustering method \citep{hartigan1979algorithm} with the BIC criterion \citep{schwarz1978estimating} to cluster the changes in the parameter matrix into two subgroups.
\end{itemize}

The proposed LSTSP algorithm also relies on multiple tuning parameters, whose selection is discussed next.
\begin{itemize}
    \item[$(\lambda_{j,n}, \mu_{j,n})$:] To select the tuning parameters related to the regularized linear regression models \eqref{eq:lps-left}, \eqref{eq:lps-right}, and \eqref{eq:lps-ic}, the theoretical values of $\lambda_{j,n}$ and $\mu_{j,n}$ introduced in \cite{basu2019low, safikhani2020joint} are applied. The function also allows the user to customize these tuning parameters. 
    \item[$h$:] The window size $h$ (or argument \code{h} in the \code{lstsp} function) should satisfy that the length of $h$ must be smaller than the minimum distance between two consecutive change points; in other words, the size of the window must ensure there is only one change point within it. Hence, in accordance to the theoretical results and assumption A4 in \cite{safikhani2020joint}, $h$ is recommended to  beset to $\lfloor \sqrt{n}\rfloor$.
    \item[$l$:] The rolling step size (or argument \code{step.size} in function \code{lstsp}) is determined by the user. It must be in the range of $[1,h]$, where $h$ is the window size, and our empirical recommendation it to set it to a value no larger than $\lfloor h/4 \rfloor$.
    \item[$\omega_n$:] The tuning parameter $\omega_n$ is introduced in the screening step of LSTSP and it is used to remove change points that lead to small changes in $\mathcal{L}_n$ given in \eqref{eq:ic}. In practice, we recommend the user to set it to $\omega_n = C(\log n\log p)$ for some constant $0 < C < 1$. An alternative method is to select this parameter similar to $\omega_n$ for the TBSS algorithm by replacing the function $L_n$ with $\mathcal{L}_n$.
\end{itemize}

\subsection{Guidelines for selecting the lag of the VAR model}\label{sec:bic}
{
We consider the following time lag selection procedure for the various posited VAR models. It uses the BIC criterion and the main steps are given next:
\begin{itemize}
    \item[1.] Employ VAR($d$) models with the appropriate structure for the transition matrices (e.g., sparse, group sparse, etc.) for different values of $d=1,2,\cdots,D$ to detect change points in the given time series data; the upper bound $D$ can be specified based on exploratory analysis of the time series (e.g., by plotting the partial autocorrelation function of the various series) or based on prior information.
    \item[2.] Based on the detected change points, fit a VAR($d$) model for the segment $[\widehat{t}_j, \widehat{t}_{j+1})$, for $j=1,2,\cdots, \widehat{m}$, and calculate the BIC value as follows:
    \begin{equation*}
        \text{BIC}_j = \log\det(\widehat{\Sigma}_j) + \frac{d_j \log (\widehat{t}_{j+1} - \widehat{t}_j)}{\widehat{t}_{j+1} - \widehat{t}_j},
    \end{equation*}
    where $\widehat{\Sigma}_j$ is the estimated covariance matrix for the $j$th segment $[\widehat{t}_j, \widehat{t}_{j+1})$, and $d_j$ is the number of non-zero elements in the transition matrices. Then, define the total BIC as:
    \begin{equation*}
        \text{BIC} = \sum_{j=1}^{\widehat{m}}\text{BIC}_j.
    \end{equation*}
\end{itemize}
Code for implementing this strategy is provided in Section \ref{sec:sparse-example} point 3.  
}

\section{Illustrative Examples}\label{sec:4}

\subsection{Change point detection on simulated data from structured sparse VAR models}\label{sec:sparse-example}
We generate a $p$-dimensional VAR($q$) process with $T=4000$ observations, lag $q=2$ and two change points:
\begin{CodeChunk}
\begin{CodeInput}
R> nob <- 4000; p <- 15                     
R> brk <- c(floor(nob / 3), floor(2 * nob / 3), nob + 1)
R> m <- length(brk); q.t <- 2     
R> signals <- c(-0.6, -0.4, 0.6, 0.4, -0.6, -0.4)
\end{CodeInput}
\end{CodeChunk}
Next, we present different options for the sparsity pattern.
\begin{itemize}
    \item {\bf Random sparse:} The first example shows how the function \code{simu_var} can be used to generate a VAR process with a random sparse transition matrix. Note that it requires sparsity levels (\code{sp_density}) for each segment.
    \begin{CodeChunk}
    \begin{CodeInput}
R> sp_density <- rep(0.05, m * q.t)
R> try <- simu_var(method = "sparse", nob = nob, k = p, lags=q.t, 
+                 brk = brk, sigma = diag(p), signals = signals,
+                 sp_density = sp_density, sp_pattern = "random")
R> print(plot_matrix(do.call("cbind", try\$model_param), m * q.t))
    \end{CodeInput}
    \end{CodeChunk}
    \begin{figure}[!ht]
        \centering
        \includegraphics[width = .7\textwidth, trim={0 150 0 120}, clip]{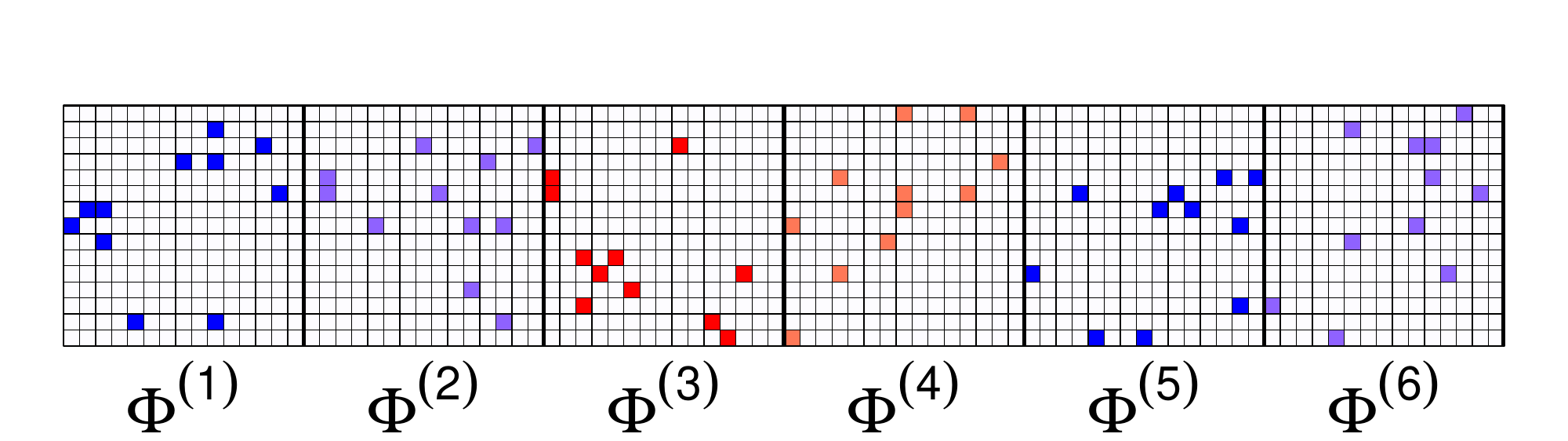}
        \caption{Example of a sparse transition matrix (random pattern with 5\% sparsity level).}
        \label{fig:sparse_random}
    \end{figure}
    
    \item {\bf One-off diagonal:} Under this setting, we assume the 1-off diagonal sparse structure for the transition matrices across all time lags. The input process and true model parameters are shown in Figure \ref{fig:sparse_offdiag}
    \begin{CodeChunk}
    \begin{CodeInput}
R> try <- simu_var(method = "sparse", nob = nob, k = p, lags = q.t, 
+                 brk = brk, sigma = diag(p), signals = signals,
+                 sp_pattern =  "off-diagonal", seed = 1)
R> MTS::MTSplot(data)
R> print(plot.matrix(do.call("cbind", try\$model_param), m * q.t))
    \end{CodeInput}
    \end{CodeChunk}
    \begin{figure}[!ht]
        \centering
        \includegraphics[width=.475\textwidth, trim={0 10 0 50}, clip]{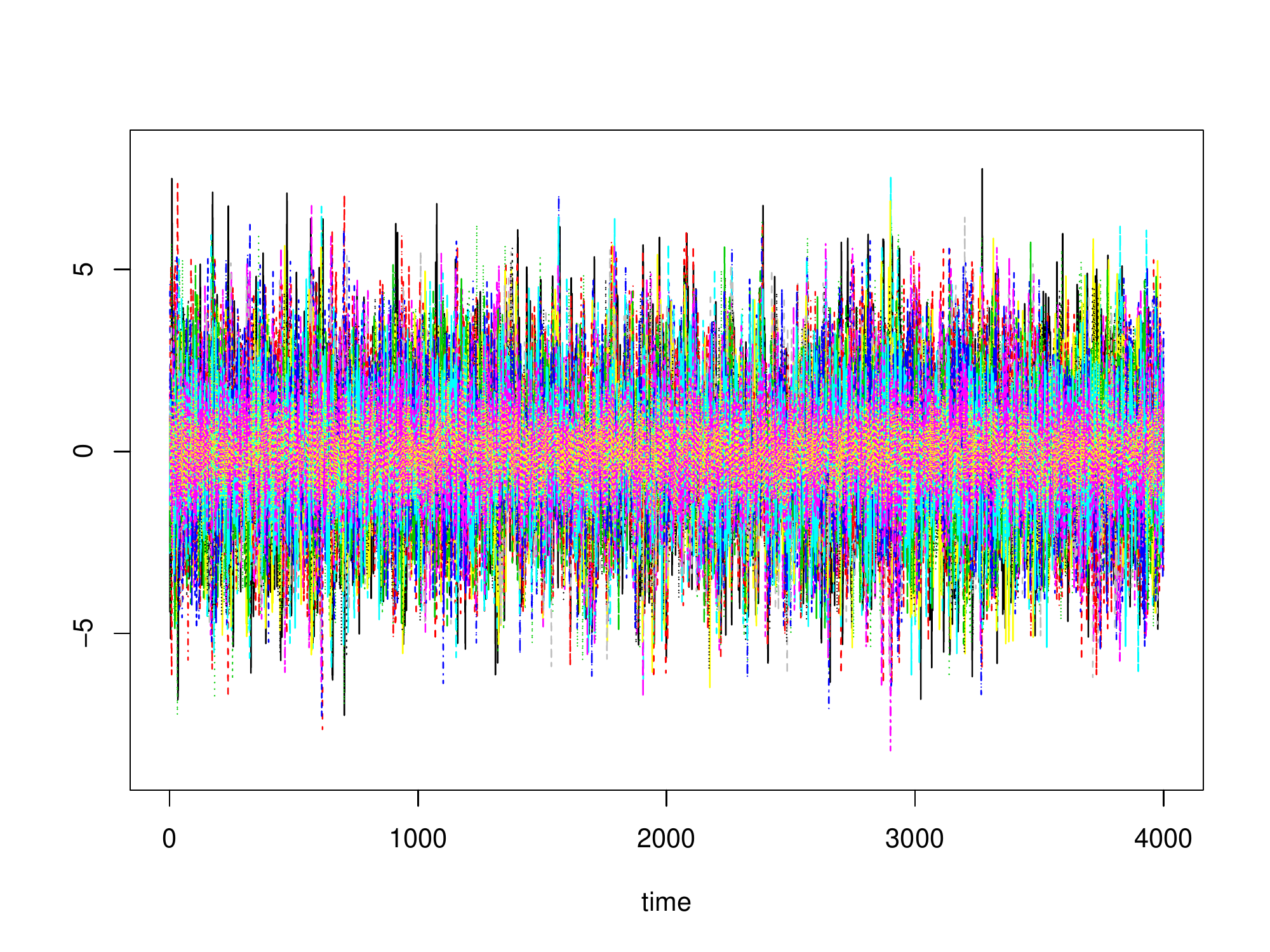}
        \includegraphics[width=.475\textwidth, trim={0 10 0 50}, clip]{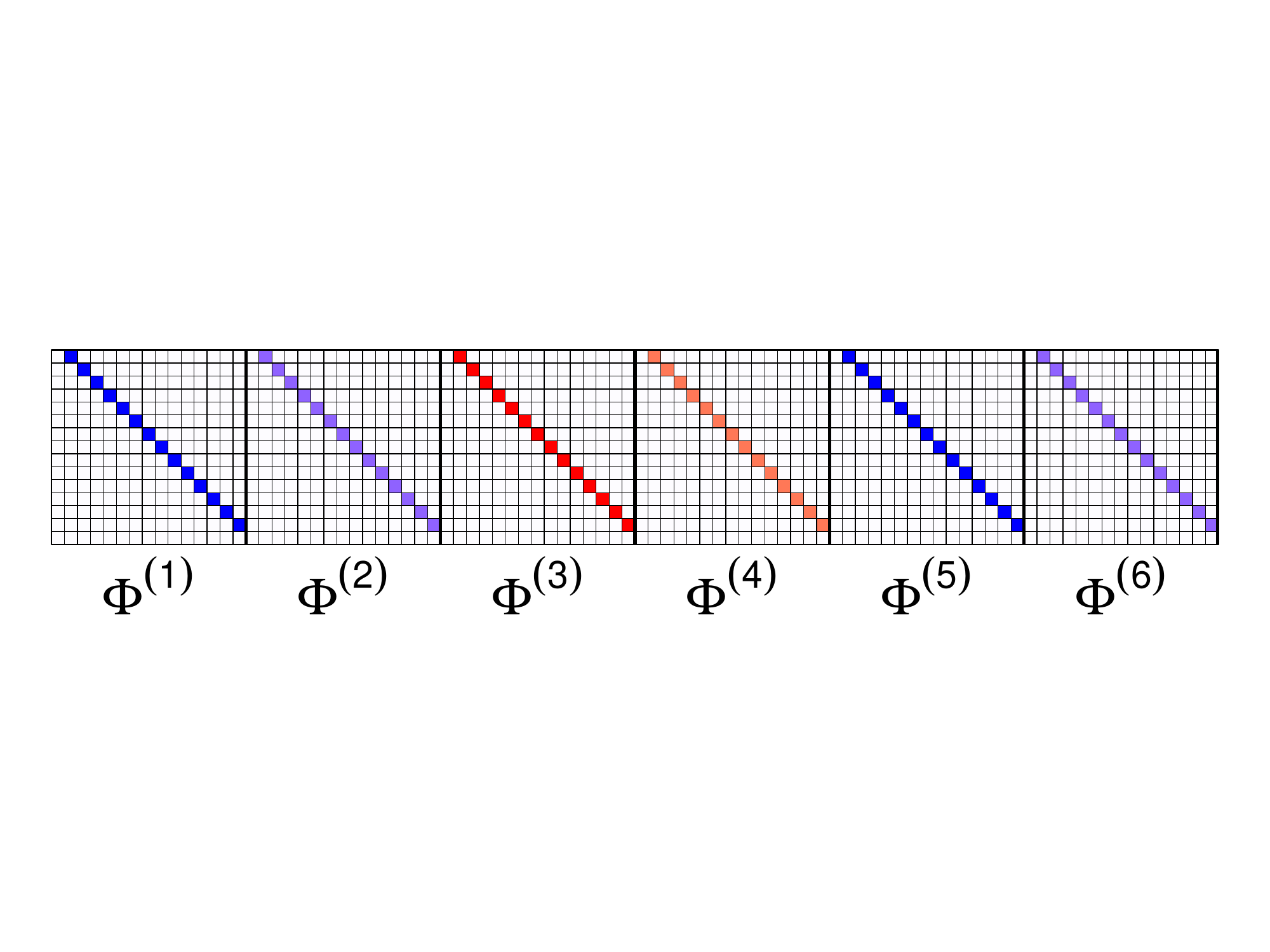}
        \caption{Example of a VAR(2) random sparse process. Left panel: generated process with two change points $t_1=1333$ and $t_2=2666$; Right panel: histogram of transition matrices (1-off diagonal pattern).}
        \label{fig:sparse_offdiag}
    \end{figure}
    The code for detecting the change points and estimating the model parameters is given next.
    \begin{CodeChunk}
\begin{CodeInput}
R> fit <- tbss(data, method = "sparse", q = q.t)
R> print(fit)
\end{CodeInput}
\begin{CodeOutput}
Estimated change points are: 1333 2666
\end{CodeOutput}
The following Figure \ref{fig:sparse-estimation} illustrates the detected change points in the generated VAR(2) process, as well as the estimated model parameters. Recall that in Section \ref{sec:visualization}, we discuss the \code{plot} function for \proglang{S3} object of class \code{"VARDetect.result"}. Hence, one can apply the following code to present the estimated change points and model parameters.
\begin{CodeInput}
R> plot(fit, display = "cp")
R> plot(fit, display = "param")
\end{CodeInput}
\end{CodeChunk}
The output is shown in Figure \ref{fig:sparse-estimation}.
\begin{figure}[!ht]
    \centering
    \includegraphics[width=.475\textwidth, trim={0 10 0 50}, clip]{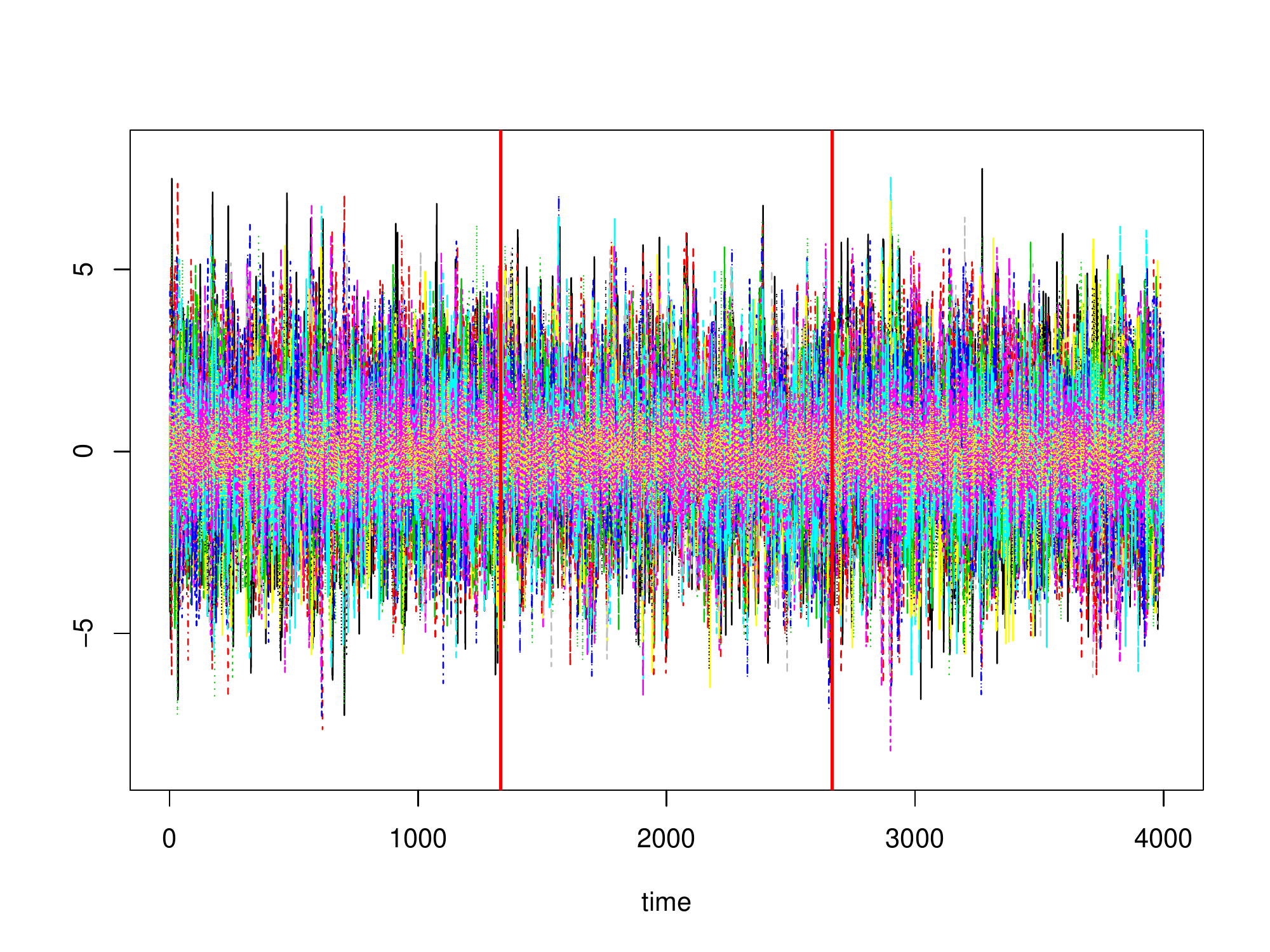}%
    \includegraphics[width=.475\textwidth, trim={0 10 0 50}, clip]{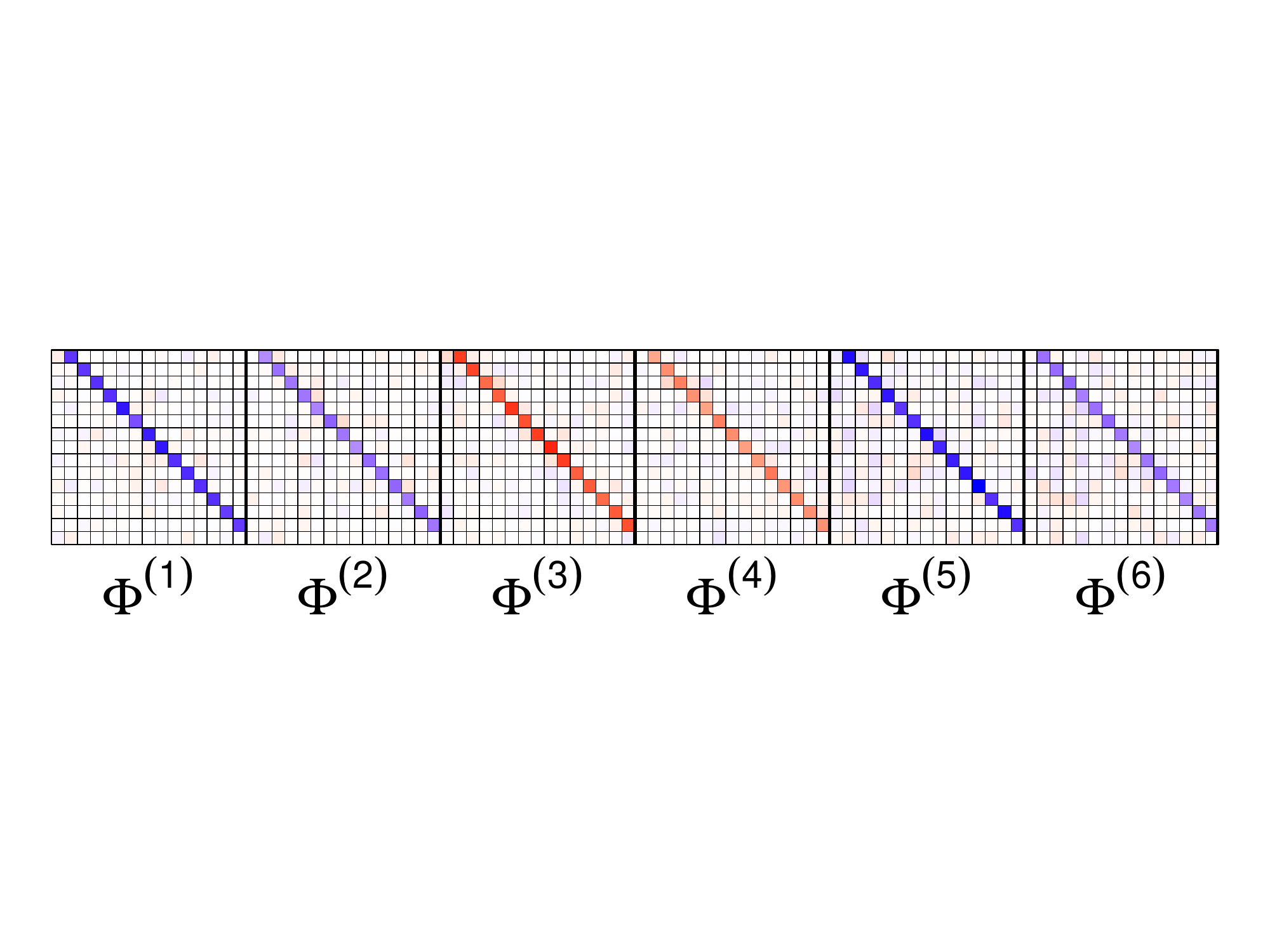}
    \caption{Left panel: results for detected change points $\hat{t}_1=1333$, and $\hat{t}_2=2666$ (red lines); Right panel: plot for estimated model parameters.}
    \label{fig:sparse-estimation}
\end{figure}

    \item {\bf Sparse VAR model with different time lags:} It is assumed that the time lags are different in each stationary segment. Under this setting, the data generation function \code{simu_var} requires the user to provide a vector of positive integers representing the time lags for each segment via argument \code{lags_vector}. As an example, the following code is used for generating a time series with $T=1000$ and $p=15$, and one change point at $t=500$. The time lags for the segments to the left and to the right of the change point are 1 and 2, respectively. The time series together with the true model parameters are depicted in Figure \ref{fig:sparse_var12_true}.
\begin{CodeChunk}
\begin{CodeInput}
R> nob <- 1000; p <- 15
R> brk <- c(floor(nob / 2), nob + 1)
R> m <- length(brk); q.t <- 2
R> signals <- c(-0.8, 0.6, 0.4)
R> try <- simu_var(method = "sparse", nob = nob, k = p, brk = brk, 
+                  signals = signals, lags_vector = c(1, 2), 
+                  sp_pattern = "off-diagonal")
R> data <- try$series; data <- as.matrix(data)
R> MTS::MTSplot(data)
R> print(plot_matrix(do.call("cbind", try$model_param), m * q.t))
\end{CodeInput}
\end{CodeChunk}
    \begin{figure}[!ht]
        \centering
        \includegraphics[width=.475\textwidth, trim={0 10 0 50}, clip]{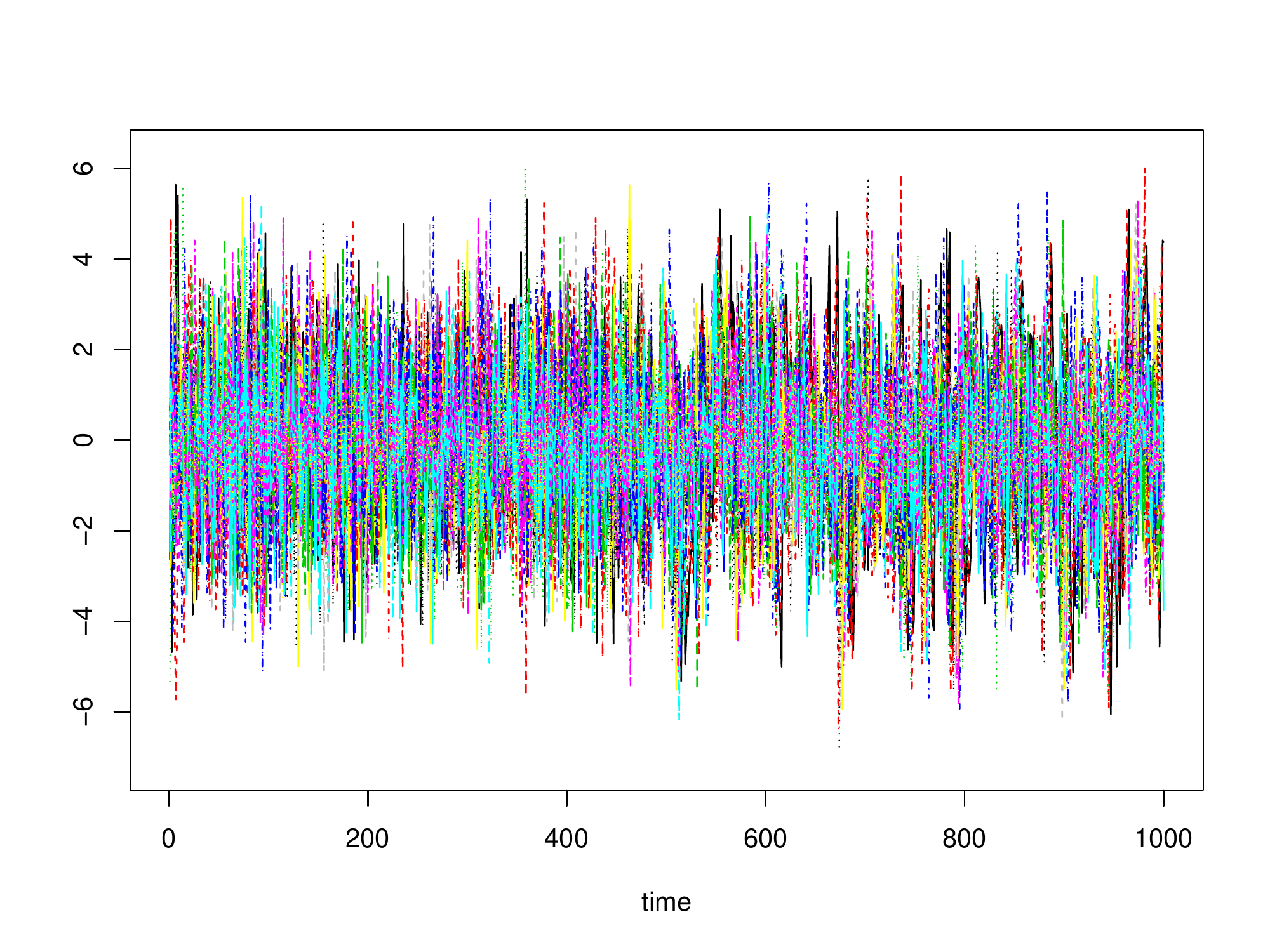}
        \includegraphics[width=.475\textwidth, trim={0 10 0 50}, clip]{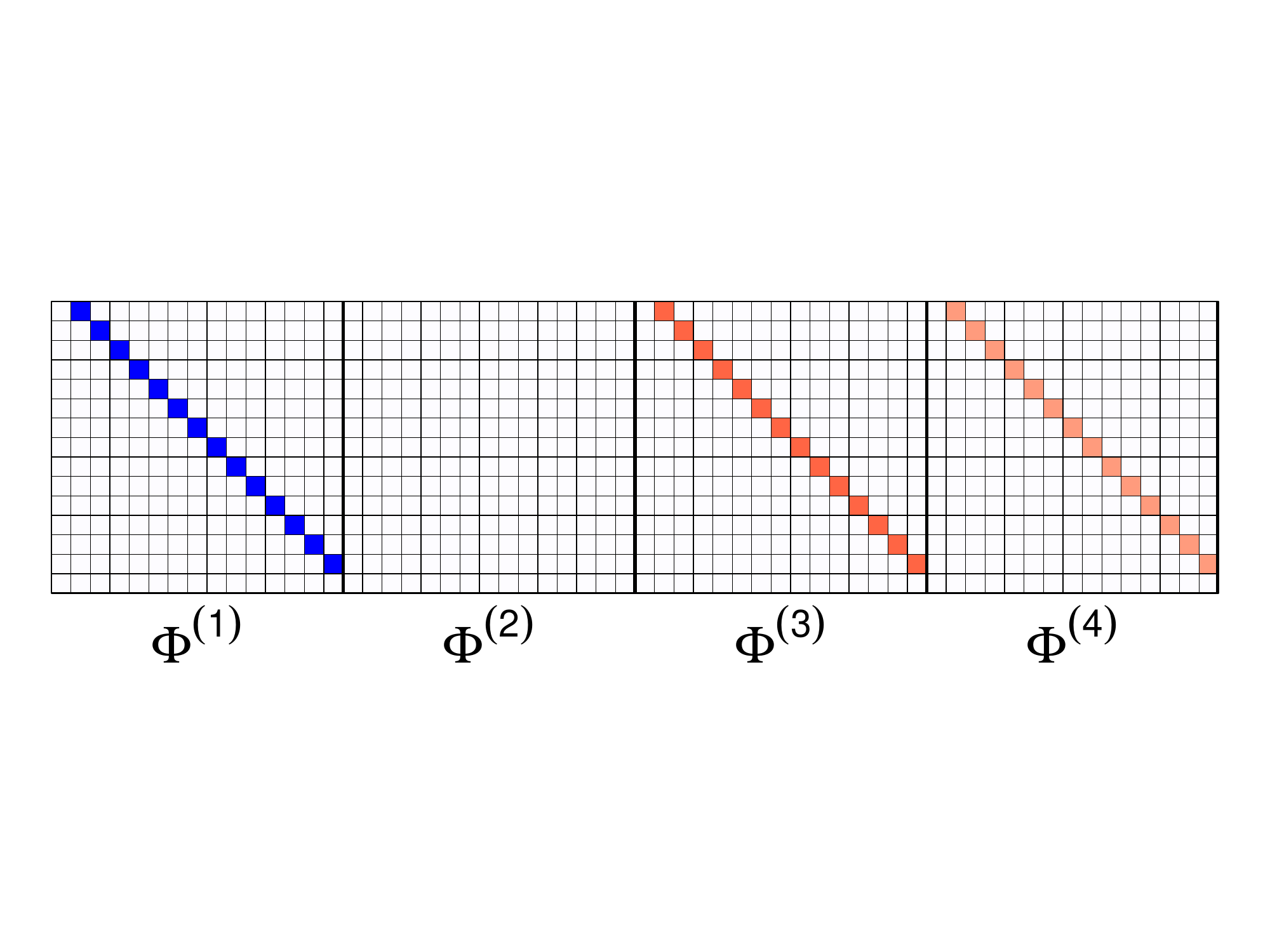}
        \caption{Example for a sparse VAR model with different lags to the left and to the right of the single change point.  Left panel: generated data with single change point at $t=500$; Right panel: model parameters (1-off diagonal pattern).}
        \label{fig:sparse_var12_true}
    \end{figure}
    The code for detection and estimation is provided next. Note that we require the refit argument to be set as \code{TRUE} in order to obtain an accurate estimate of the model parameters with different lags. The estimated change points and model parameters are illustrated in Figure \ref{fig:sparse_var12_est}.

{
Before we apply the TBSS algorithm to detect the change points, we first determine the time lag for the given time process. The following code is used to implement the time lag selection strategy introduced in Section \ref{sec:bic}. 
\begin{CodeChunk}
\begin{CodeInput}
R> # lags selection using BIC (unknown lags)
R> library(sparsevar)
R> d_full <- c(1, 2, 3, 4)
R> BIC_full <- rep(0, length(d_full))
R> for(i in 1:length(d_full)){
+   d <- d_full[i]
+   fit <- tbss(data, method = "sparse", q = d, refit = TRUE)
+   sparse_mats <- fit$sparse_mats
+   cp_est <- fit$cp   
+   cp_full <- c(1, cp_est, nob+1)
+   BIC <- 0
+   for(j in 1:(length(cp_est)+1)){
+     data_temp <- as.matrix(data[(cp_full[j]): (cp_full[j+1]-1), ])
+     n_temp <- dim(data_temp )[1]
+     sparse_mats_temp <- sparse_mats[[j]]
+     residual <- c()
+     for(t in ((d+1):n_temp)){
+       y_pred <- 0
+       for(dd in 1:d){
+         phi <- sparse_mats_temp[, ( (dd-1)*p +1) :( (dd)*p ) ]
+         y_pred <- y_pred + phi %*% (data_temp[t-dd,])
+       }
+       residual <- cbind(residual, data_temp[t,] - y_pred)
+     }
+     sigma.hat <- 0*diag(p);
+     for(t in 1:(n_temp-d)){
+         sigma.hat <- sigma.hat +  residual[, t]%*%t(residual[, t])
+     }
+     sigma.hat <- (1/(n_temp -d))*sigma.hat;
+     log.det <- log(det(sigma.hat));
+     count <- sum(sparse_mats_temp !=0)
+     BIC <- BIC + log.det + log((n_temp - d))*count/(n_temp - d)
+   }
+   BIC_full[i] <- BIC
+ }
R> BIC_full
[1] 3.003772 0.974745 1.203904 7.663452
R> #choose the one with the smallest BIC
R> d_full[which.min(BIC_full)]
[1] 2
\end{CodeInput}
\end{CodeChunk}
The result shows that the best lag $q=2$ and hence this is used in the TBSS algorithm, based on the following code.
}

\begin{CodeChunk}
\begin{CodeInput}
R> fit <- tbss(data, method = "sparse", q = q.t, refit = TRUE)
R> print(fit)
\end{CodeInput}
\begin{CodeOutput}
Estimated change points are: 500
\end{CodeOutput}
\begin{CodeInput}
R> plot(fit, display = "cp")
R> plot(fit, display = "param")
\end{CodeInput}
\end{CodeChunk}
    \begin{figure}[!ht]
        \centering
        \includegraphics[width=.475\textwidth, trim={0 10 0 50}, clip]{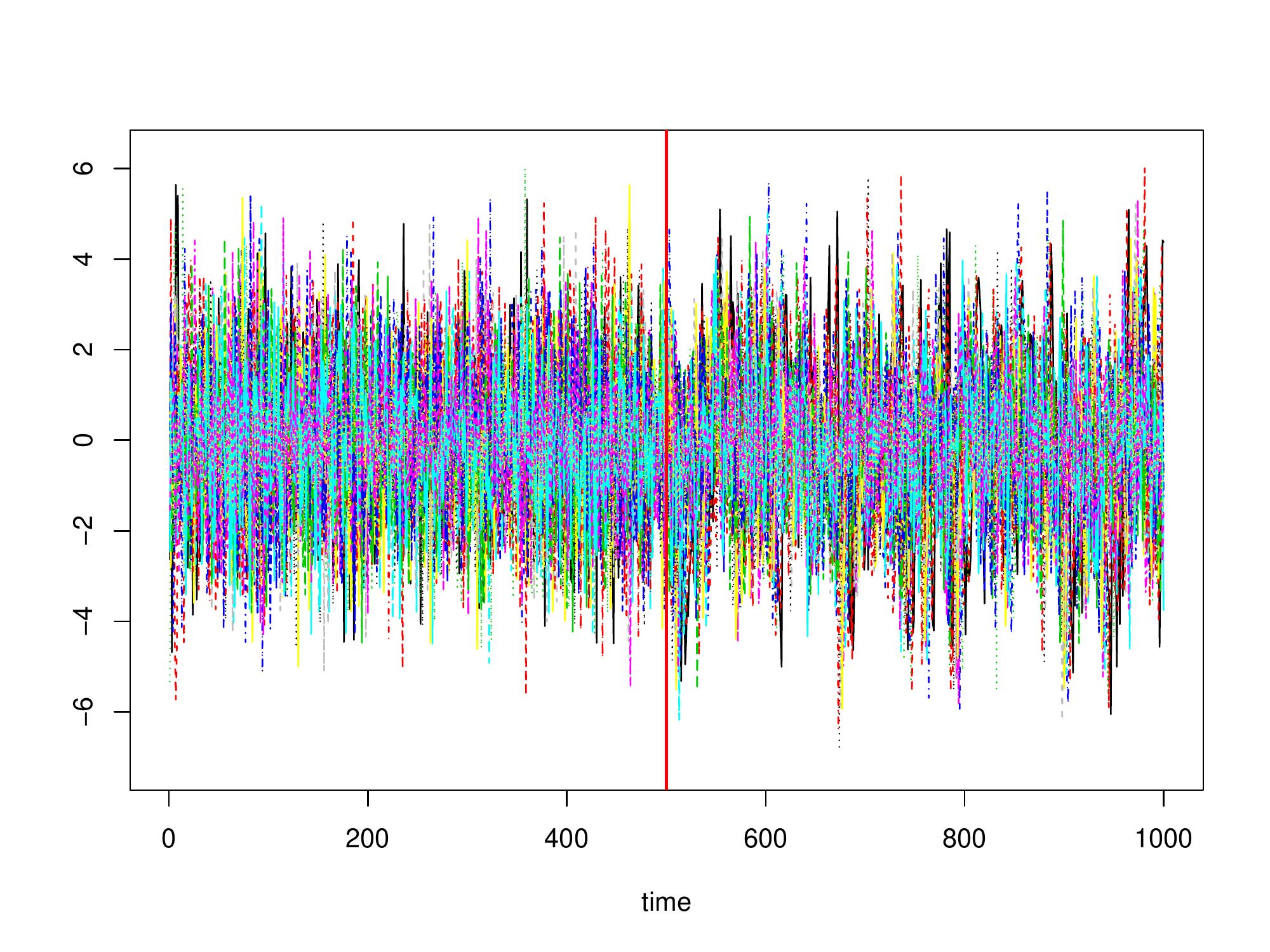}
        \includegraphics[width=.475\textwidth, trim={0 10 0 50}, clip]{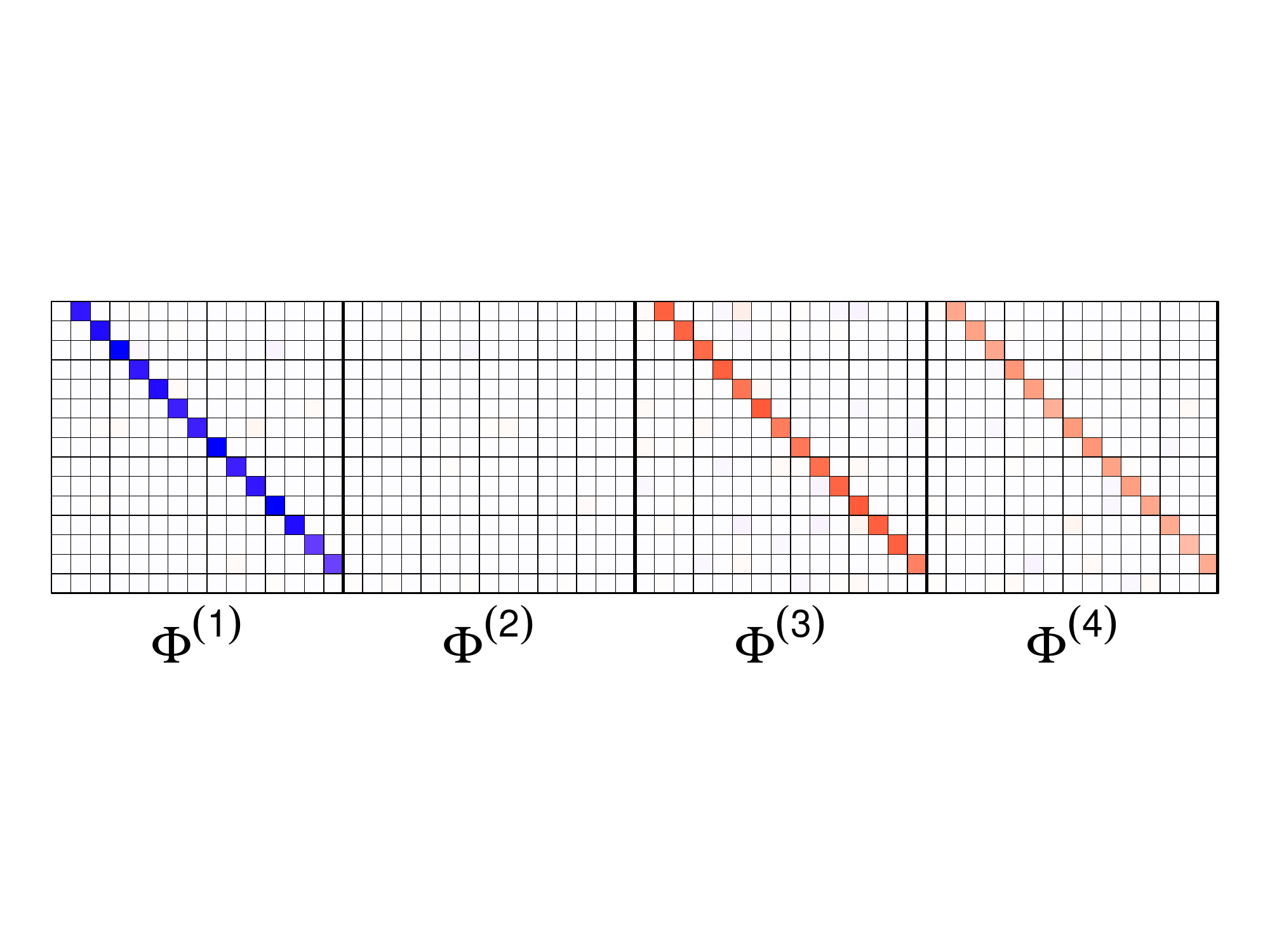}
        \caption{Left panel: generated data with estimated change points $\widehat{t}=500$; Right panel: estimated model parameters.}
        \label{fig:sparse_var12_est}
    \end{figure}
\end{itemize}

\subsection{Group sparse VAR model}\label{sec:grpsp-example}
An example with column-wise group sparse transition matrices is plotted in Figure~\ref{fig:group_sparse_col}. The data are generated from a $p=20$ dimensional VAR($q$) process with $T=4000$ observations including two change points:
\begin{CodeChunk}
\begin{CodeInput}
R> nob <- 4000; p <- 20                  
R> brk <- c(floor(nob / 3), floor(2 * nob / 3), nob + 1)
R> m <- length(brk); q.t <- 2     
\end{CodeInput}
\end{CodeChunk}
We consider the following settings:
\begin{itemize}
    \item {\bf Column-wise group sparse:} Under the first setting, we investigate a column-wise (separate across all lags) group sparse structure. A VAR(2) process is generated by using the following code.
    \begin{CodeChunk}
    \begin{CodeInput}
R> signals <- c(-0.8, -0.4, 0.6, -0.4, -0.8, -0.4)
R> num_group <- 3
R> group_index <- vector('list', num_group)
R> group_index[[1]] <- c(1, 5)
R> group_index[[2]] <- c(31)
R> try <- simu_var(method = "group sparse", nob = nob, k = p, 
+                  lags = q.t, brk = brk, sigma = diag(p), 
+                  signals = signals, group_index = group_index, 
+                  group_type = "columnwise")
R> MTS::MTSplot(data)
R> print(plot.matrix(do.call("cbind", try\$model_param), m * q.t))
    \end{CodeInput}
    \end{CodeChunk}
    \begin{figure}[!ht]
    \centering
    \includegraphics[width=.475\textwidth, trim={0 10 0 50}, clip]{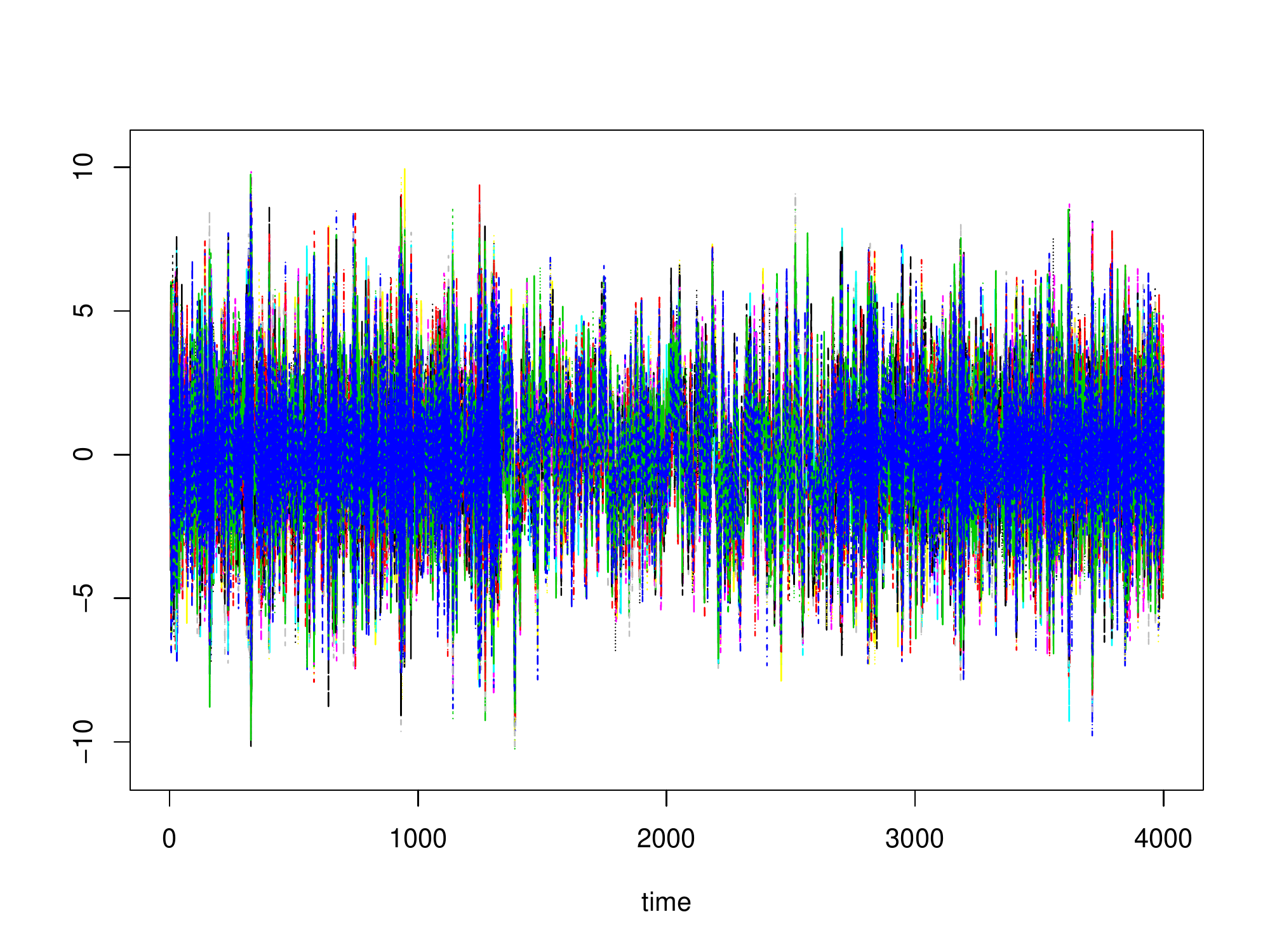}%
    \includegraphics[width=.475\textwidth, trim={0 10 0 50}, clip]{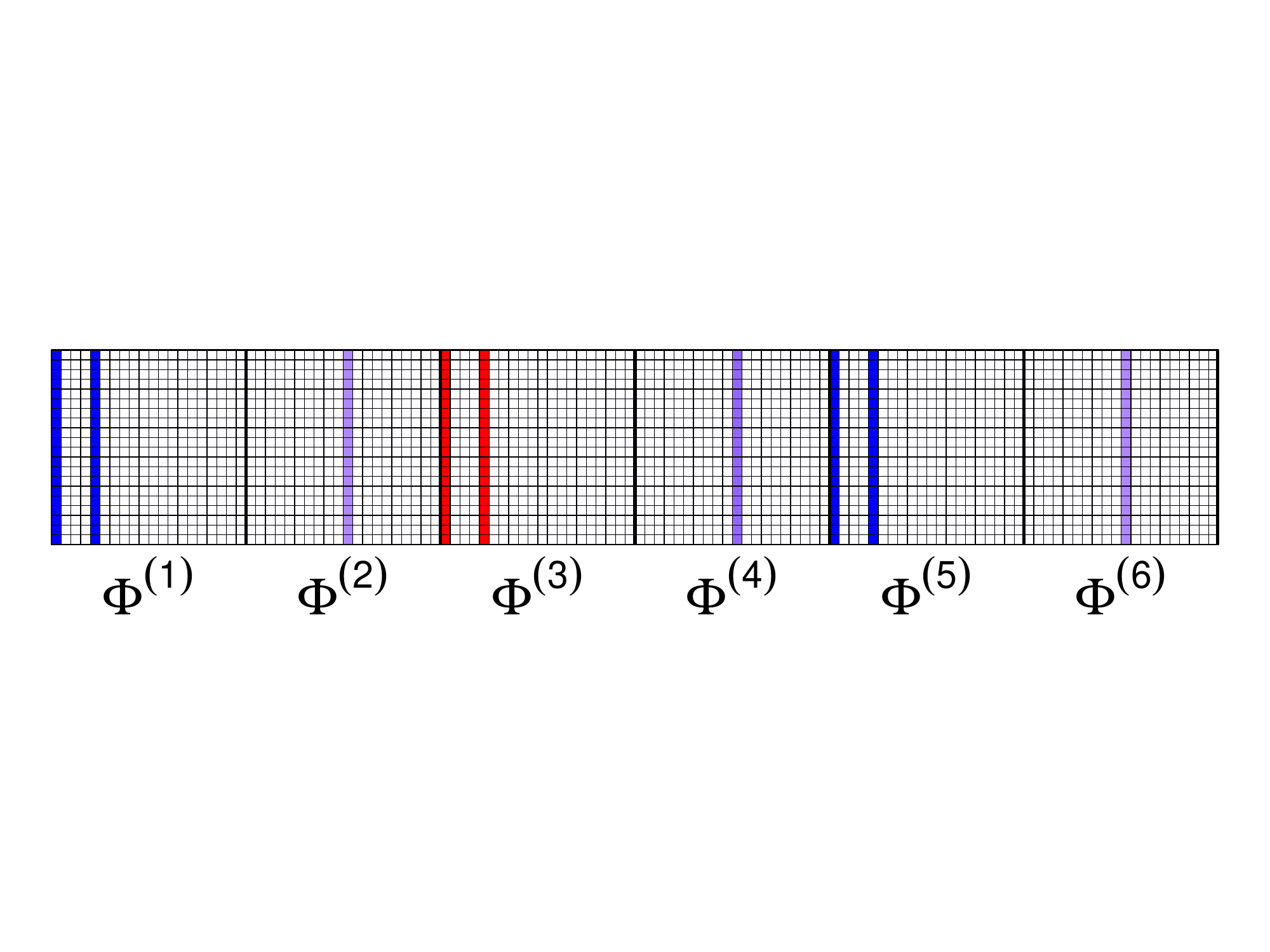}
    \caption{Example of group sparse transition matrices (group by column). Left panel: generated VAR process with change points $t_1=1333$ and $t_2=2666$; Right panel: true model parameters.}
    \label{fig:group_sparse_col}
\end{figure}
    The code for detection and estimation is shown next.
\begin{CodeChunk}
\begin{CodeInput}
R> fit <- tbss(data, method = "group sparse", q = q.t, 
+             group.case = "columnwise", 
+             group.index = as.list(c(0: (p * q.t - 1))))
R> print(fit)
\end{CodeInput}
\begin{CodeOutput}
Estimated change points are: 1332 2665
\end{CodeOutput}
Figure \ref{fig:col_group_sparse-estimation} illustrates the detected change points in the generated VAR(2) process, as well as the estimated model parameters. 
\begin{CodeInput}
R> plot(fit, display = "cp")
R> plot(fit, display = "param")
\end{CodeInput}
\end{CodeChunk}
\begin{figure}[!ht]
    \centering
    \includegraphics[width=.475\textwidth, trim={0 10 0 50}, clip]{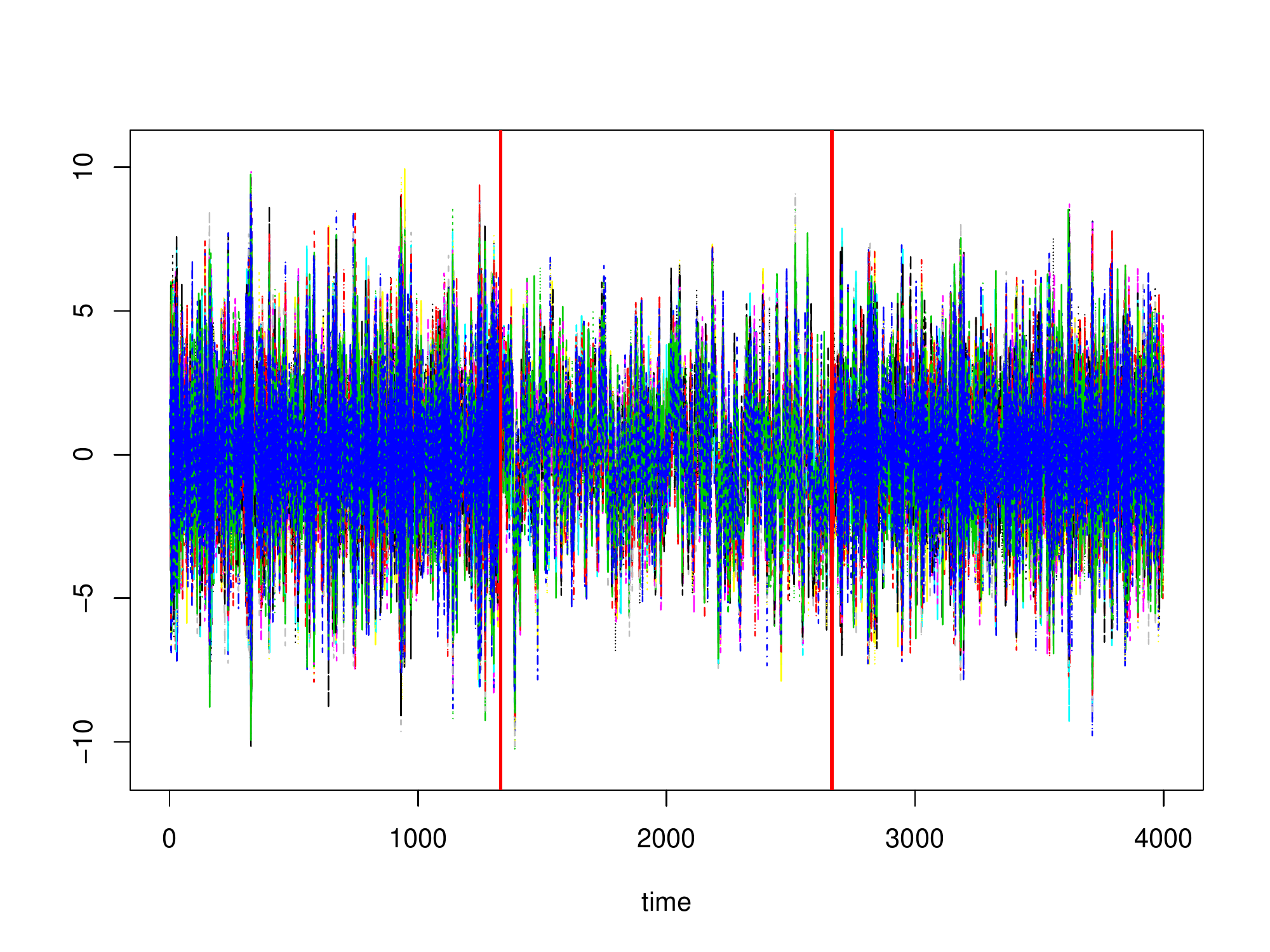}%
    \includegraphics[width=.475\textwidth, trim={0 10 0 50}, clip]{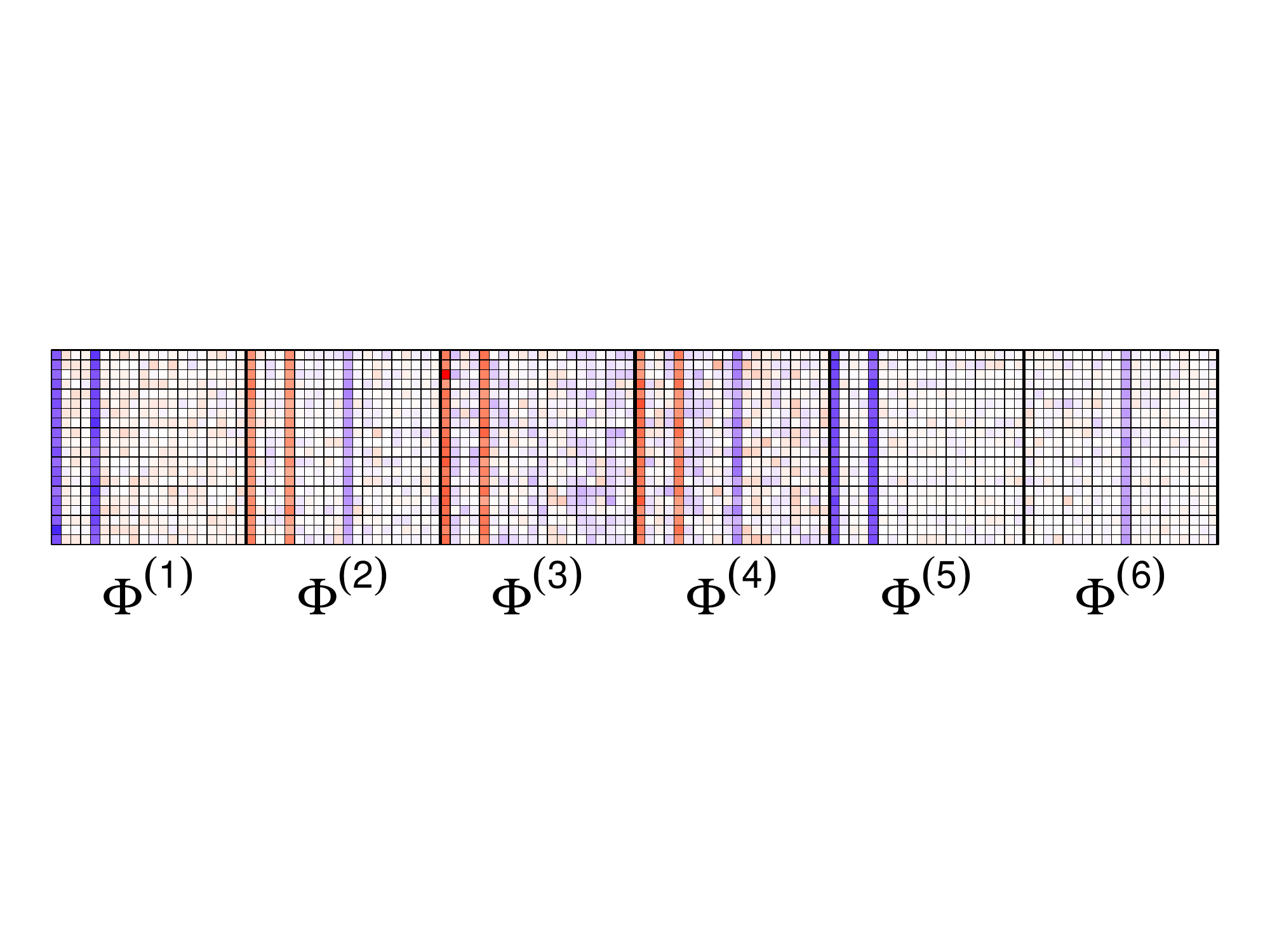}
    \caption{Left panel: results for detected change points $\hat{t}_1=1332$, and $\hat{t}_2=2665$ (red lines); Right panel: plot of estimated model parameters.}
    \label{fig:col_group_sparse-estimation}
\end{figure}

\item {\bf Row-wise group sparse:} Next, we consider a row-wise (simultaneous across all lags) group sparse structure, generated by
\begin{CodeChunk}
\begin{CodeInput}
R> signals <- c(-0.8, 0.4, 0.6, -0.3, -0.8, 0.4)
R> num_group <- q.t + 1
R> group_index <- vector('list', num_group)
R> group_index[[1]] <- c(1, 3)
R> group_index[[2]] <- c(1, 3) + p
R> try <- simu_var(method = 'group sparse', nob = nob, k = p, lags = q.t, 
+                  sigma = diag(p), brk = brk, signals = signals, 
+                  group_index = group_index, group_type = "rowwise")
R> data <- as.matrix(try$series)
R> MTS::MTSplot(data)
R> print(plot_matrix(do.call("cbind", try$model_param), m * q.t))
\end{CodeInput}
\end{CodeChunk}
\begin{figure}[!ht]
    \centering
    \includegraphics[width=.475\textwidth, trim={0 10 0 50}, clip]{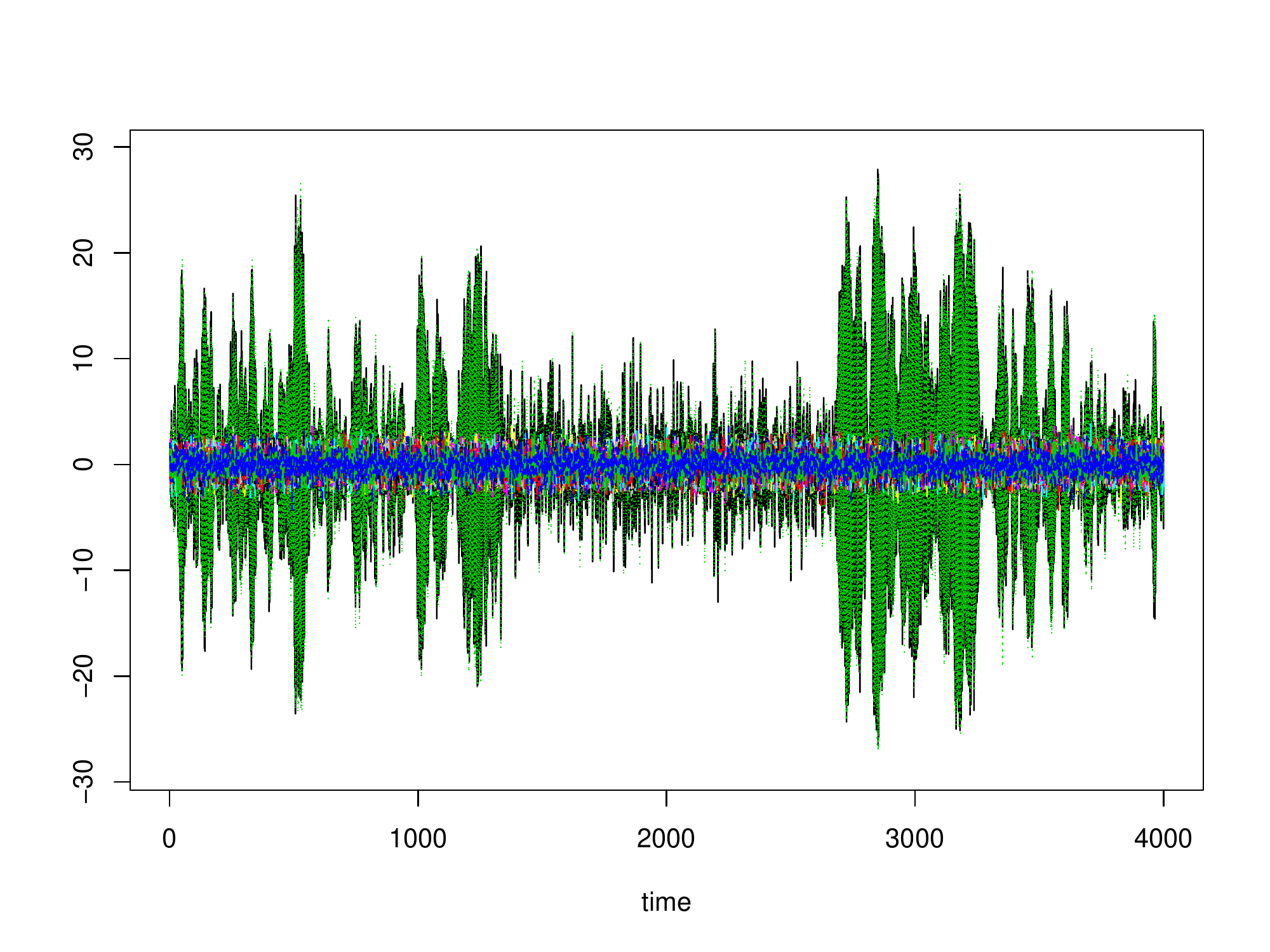}%
    \includegraphics[width=.475\textwidth, trim={0 10 0 50}, clip]{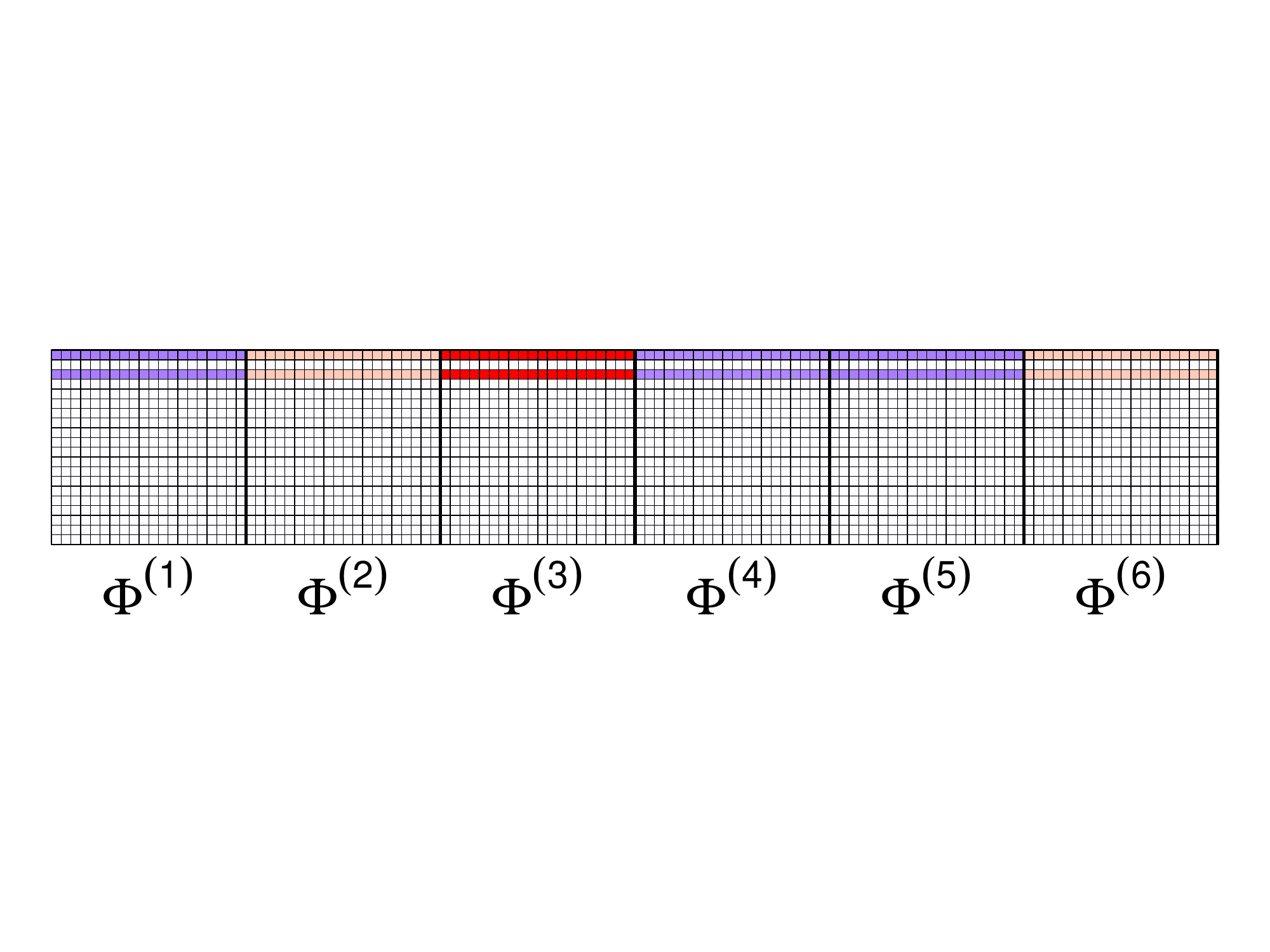}
    \caption{Example of group sparse transition matrices (group by row). Left panel: generated VAR process with change points at $t_1=1333$ and $t_2=2666$; Right panel: true model parameters.}
    \label{fig:group_sparse_row}
\end{figure}
The detection and estimation code is given next.
\begin{CodeChunk}
\begin{CodeInput}
R> group.index <- vector("list", p)
R> for(i in 1:p){
+      group.index[[i]] <- rep(i - 1, q.t) + seq(0, p * (q.t - 1), p)
+ }
R> fit <- tbss(data, method = "group sparse", q = q.t, 
+             group.case = "rowwise", group.index = group.index)
R> print(fit)              
\end{CodeInput}
\begin{CodeOutput}
Estimated change points are: 1333 2667
\end{CodeOutput}
Similar to the previous example, we also provide the detected change points together with the estimated model parameters in Figure \ref{fig:row_group_sparse-estimation}. The code is provided as follows.
\begin{CodeInput}
R> plot(fit, display = "cp")
R> plot(fit, display = "param")
\end{CodeInput}
\end{CodeChunk}
\begin{figure}[!ht]
    \centering
    \includegraphics[width=.475\textwidth, trim={0 10 0 50}, clip]{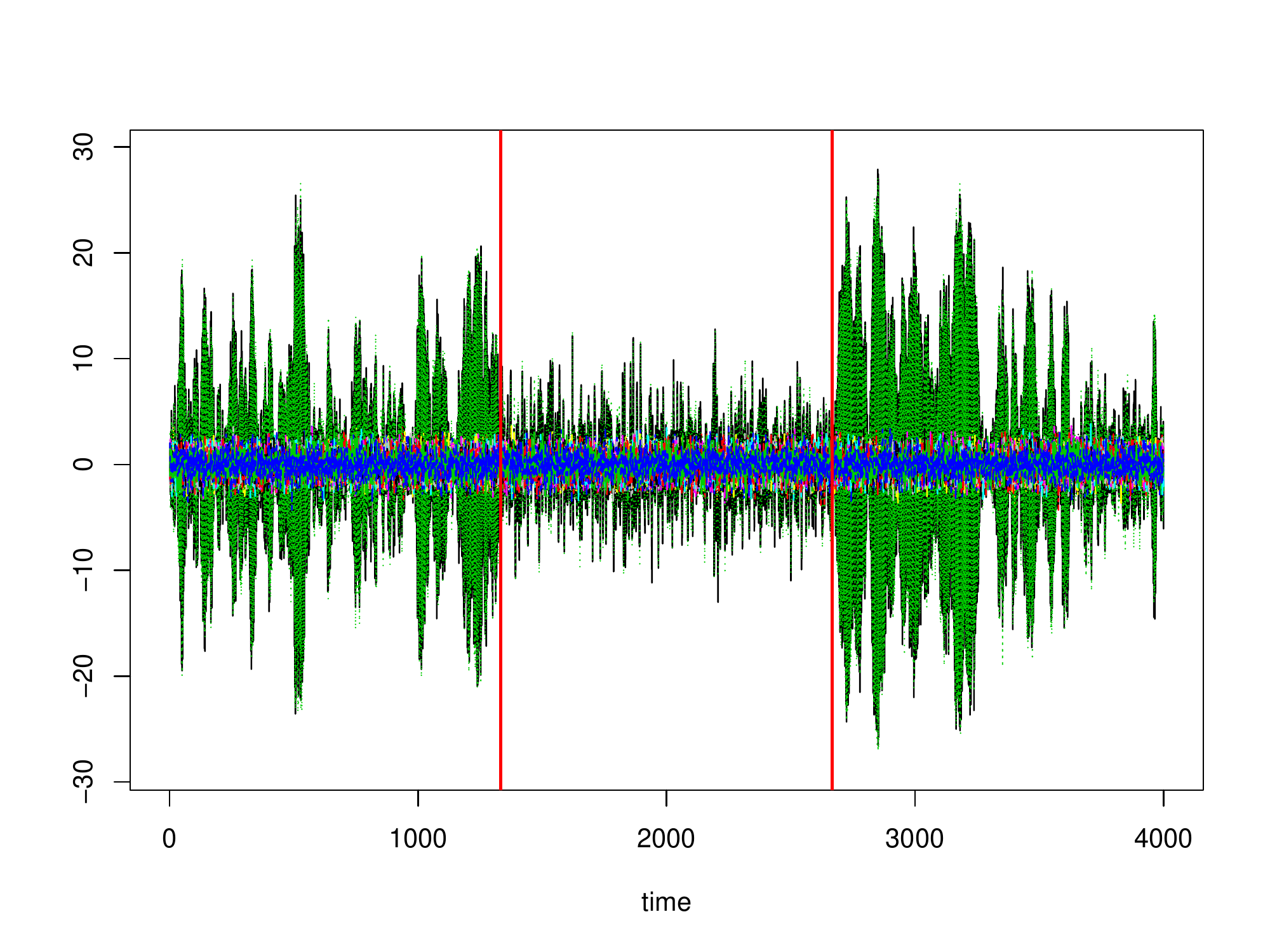}%
    \includegraphics[width=.475\textwidth, trim={0 10 0 50}, clip]{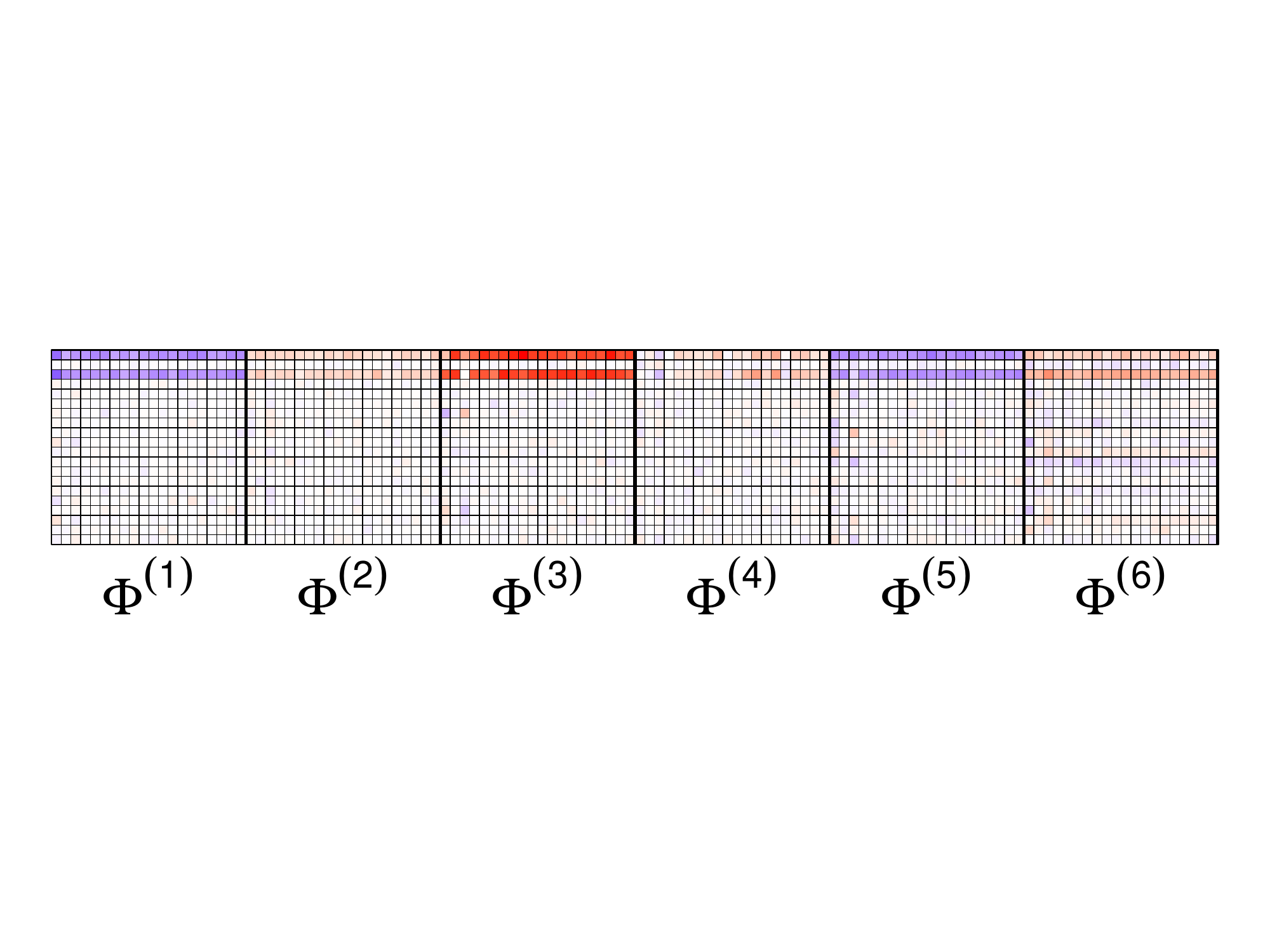}
    \caption{Left panel: results for detected change points at $\hat{t}_1=1333$, and $\hat{t}_2=2667$ (red lines); Right panel: plot of estimated model parameters.}
    \label{fig:row_group_sparse-estimation}
\end{figure}

\item {\bf Hierarchical lag group sparse:} The last setting corresponds to a hierarchical lag group sparse, generated by the following code.
\begin{CodeChunk}
\begin{CodeInput}
R> signals <- c(-0.4, -0.4, 0.4, -0.4, -0.4, -0.4)
R> num_group <- q.t + 1
R> group_index <- vector('list', num_group)
R> group_index[[1]] <- c(1, 3, 10)
R> group_index[[2]] <- c(3 + p)
R> try <- simu_var(method = 'group sparse', nob = nob, k = p, lags = q.t, 
+                  sigma = as.matrix(diag(p)), brk = brk, signals = signals,
+                  group_index = group_index, group_type = "rowwise")
R> data <- try$series
R> data <- as.matrix(data)
R> ts.plot(data)
R> print(plot_matrix(do.call("cbind", try$model_param), m * q.t))
\end{CodeInput}
\end{CodeChunk}
\begin{figure}[!ht]
    \centering
    \includegraphics[width=.475\textwidth, trim={0 10 0 50}, clip]{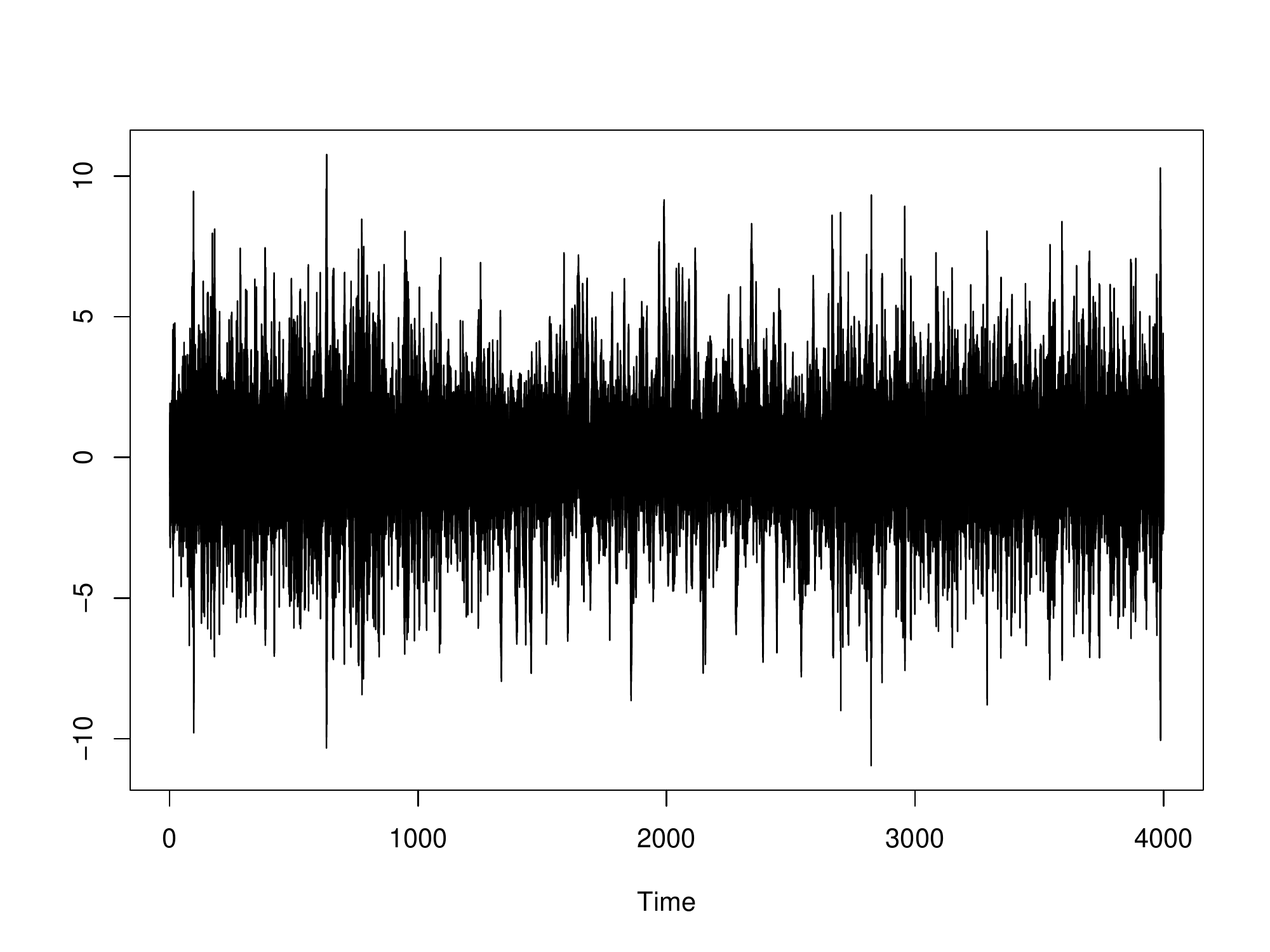}%
    \includegraphics[width=.475\textwidth, trim={0 10 0 50}, clip]{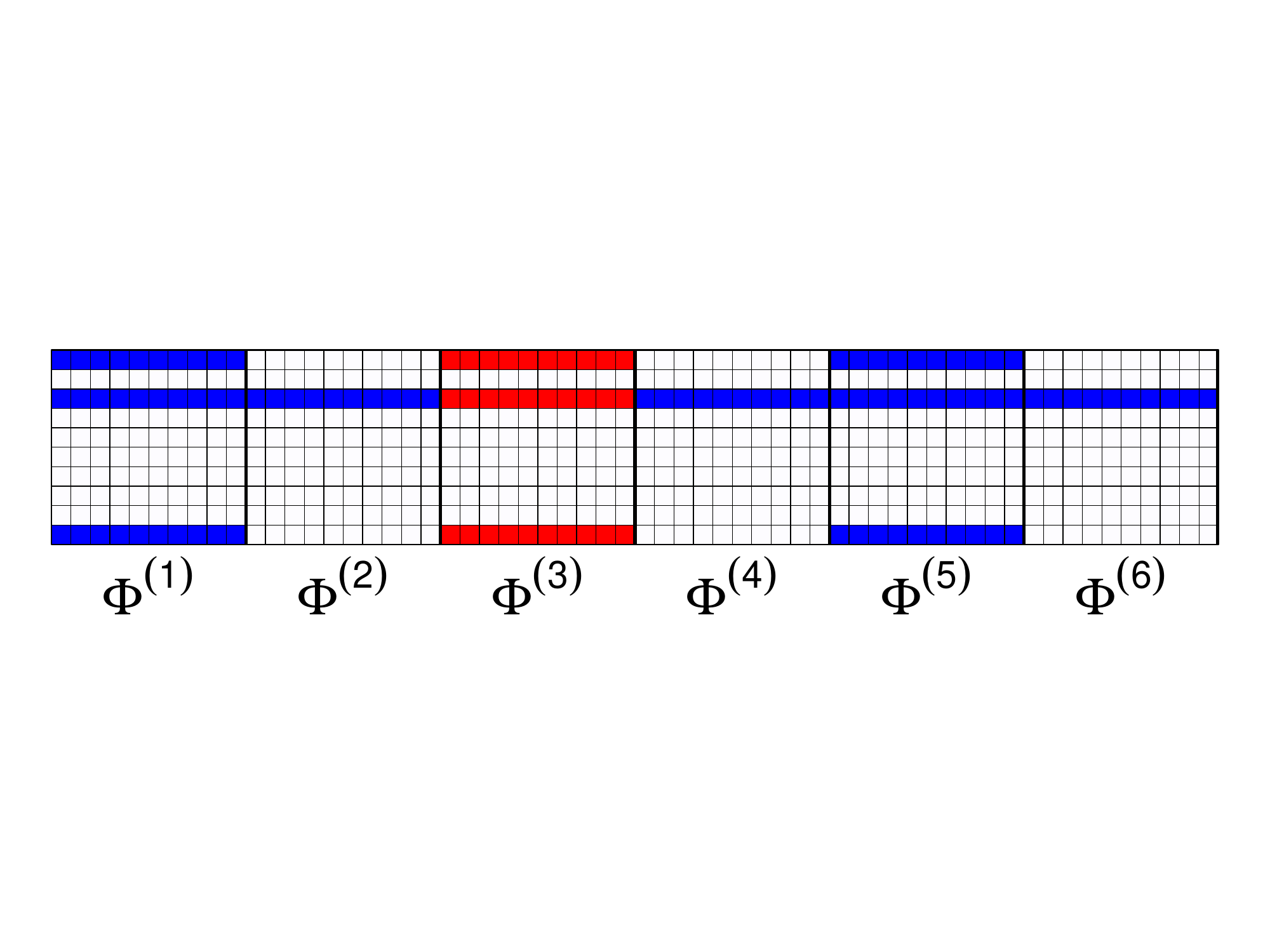}
    \caption{Example of group sparse transition matrix (group by hierarchical lag). Left panel: generated VAR process with change points at $t_1=1333$ and $t_2=2666$; Right panel: true model parameters.}
    \label{fig:group_sparse_hlag}
\end{figure}
Then, the detection and estimation results can be obtained by the following code.
\begin{CodeChunk}
\begin{CodeInput}
R> group.index <- vector("list", p * q.t)
R> for(i in 1:p){
+   for(j in 1:q.t){
+     if(j == 1){
+       group.index[[(j - 1) * p + i]] <- c((q.t - 1) * p + i) - 1
+     }else{
+       group.index[[(j - 1) * p + i]] <- rep(i - 1, q.t) 
+                                          + seq(0, p * (q.t - 1), p)
+     }
+   }
+ }
R> fit <- tbss(data, method = "group sparse", q = q.t,
+             group.case = "rowwise", group.index = group.index)
R> print(fit)
\end{CodeInput}
\begin{CodeOutput}
Estimated change points are: 1332 2663
\end{CodeOutput}
Next, we demonstrate the detected change points in the VAR process and present the estimated model parameters.
\begin{CodeInput}
R> plot(fit, display = "cp")
R> plot(fit, display = "param")
\end{CodeInput}
\end{CodeChunk}
\begin{figure}[!ht]
    \centering
    \includegraphics[width=.475\textwidth, trim={0 10 0 50}, clip]{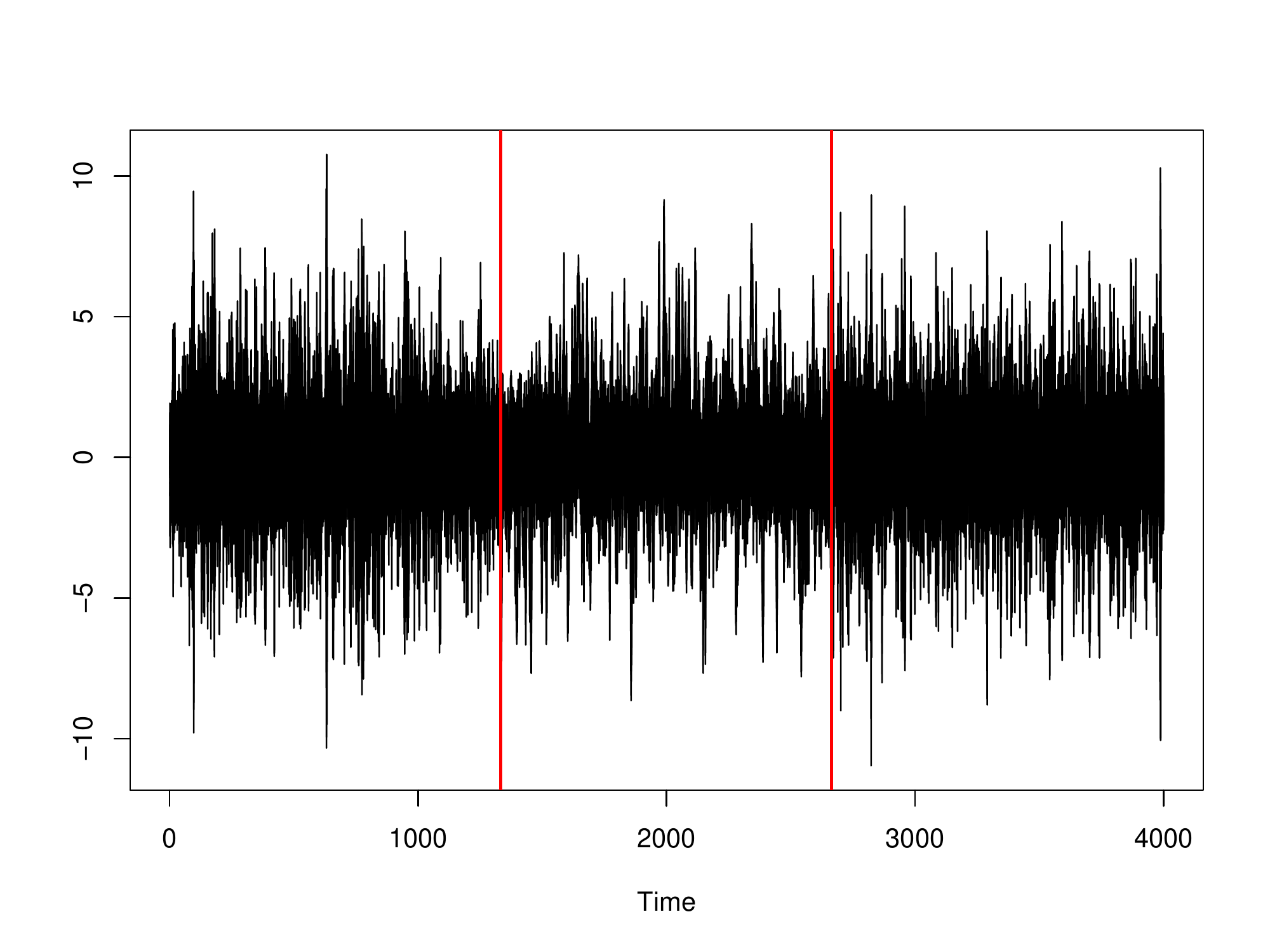}%
    \includegraphics[width=.475\textwidth, trim={0 10 0 50}, clip]{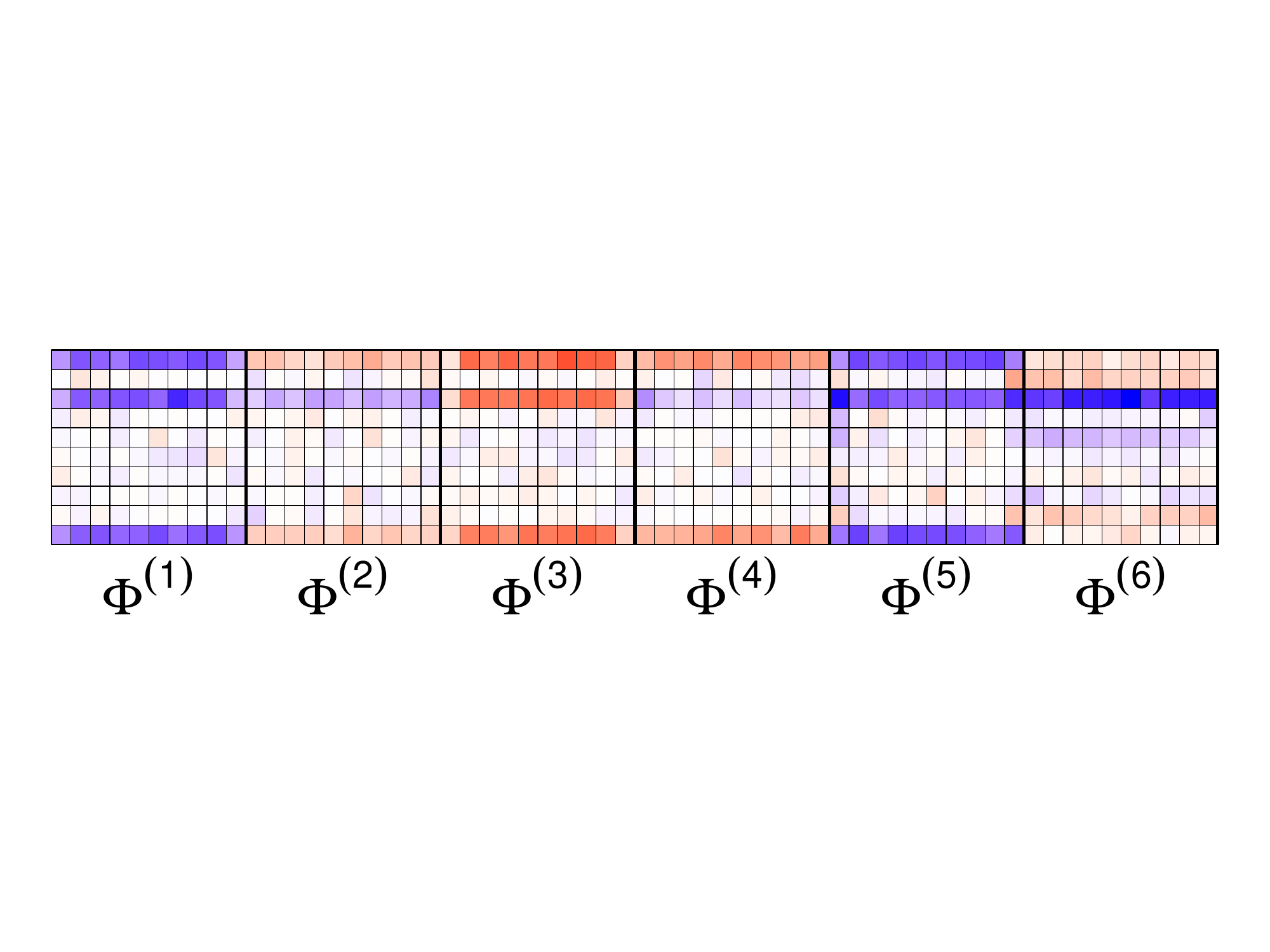}
    \caption{Left panel: results for detected change points at $\hat{t}_1=1332$, and $\hat{t}_2=2663$ (red lines); Right panel: plot of estimated model parameters.}
    \label{fig:hlag_group_sparse-estimation}
\end{figure}
\end{itemize}

\subsection{Change point detection on simulated data from a fixed low rank plus sparse VAR model}\label{sec:flps-example}

{
Next, we investigate change point detection for a VAR model with a \textit{fixed} low rank component plus a time-varying sparse component. First, we employ function \code{simu_var} to generate data according to the posited model and use function \code{tbss} to detect change points as well as estimate the model parameters. 

Specifically, data are generated from a $p=15$ dimensional $VAR(1)$ process with $T=300$ observations, with two change points located at $t_1=100$ and $t_2=200$, respectively. The rank of the low rank component is fixed as 2, and the information ratio is set to 0.35 for all intervals. 
\begin{CodeChunk}
\begin{CodeInput}
R> nob <- 300; p <- 15;
R> brk <- c(floor(nob/3), floor(2*nob/3), nob+1)
R> m <- length(brk); q.t <- 1
R> rank <- rep(2, m)
R> signals <- c(-0.7, 0.85, -0.7)
R> singular_vals <- c(1, 0.75); info_ratio = rep(0.35, 3)
\end{CodeInput}
\end{CodeChunk}
The simulated time series data is generated by \code{simu_var} function with method set as \code{fLS},
as shown next.
\begin{CodeChunk}
\begin{CodeInput}
R> try <- simu_var(method = "fLS", nob = nob, k = p, lags = q.t, brk = brk,
+                 sigma = as.matrix(diag(p)), signals = signals, seed = 1,
+                 rank = rank, singular_vals = singular_vals, 
+                 info_ratio = info_ratio, sp_pattern = "off-diagonal", 
+                 spectral_radius = 0.9)
R> data <- as.matrix(try$series)
R> MTS::MTSplot(data)
R> print(plot_matrix(do.call("cbind", try$model_param), m * q.t))
\end{CodeInput}
\end{CodeChunk}
\begin{figure}[!ht]
    \centering
    \includegraphics[width=.475\textwidth, trim={0 10 0 50}, clip]{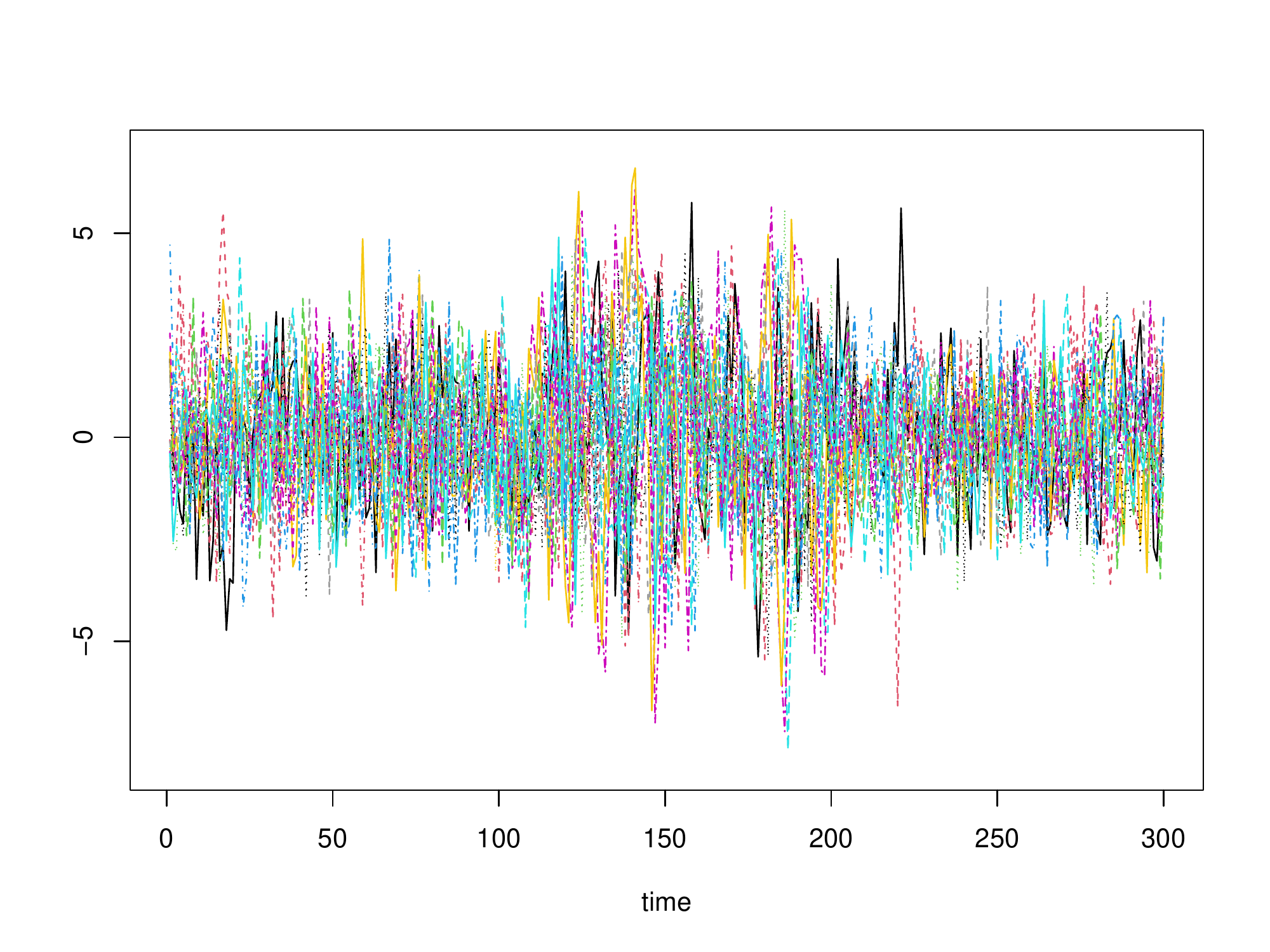}%
    \includegraphics[width=.475\textwidth, trim={0 10 0 50}, clip]{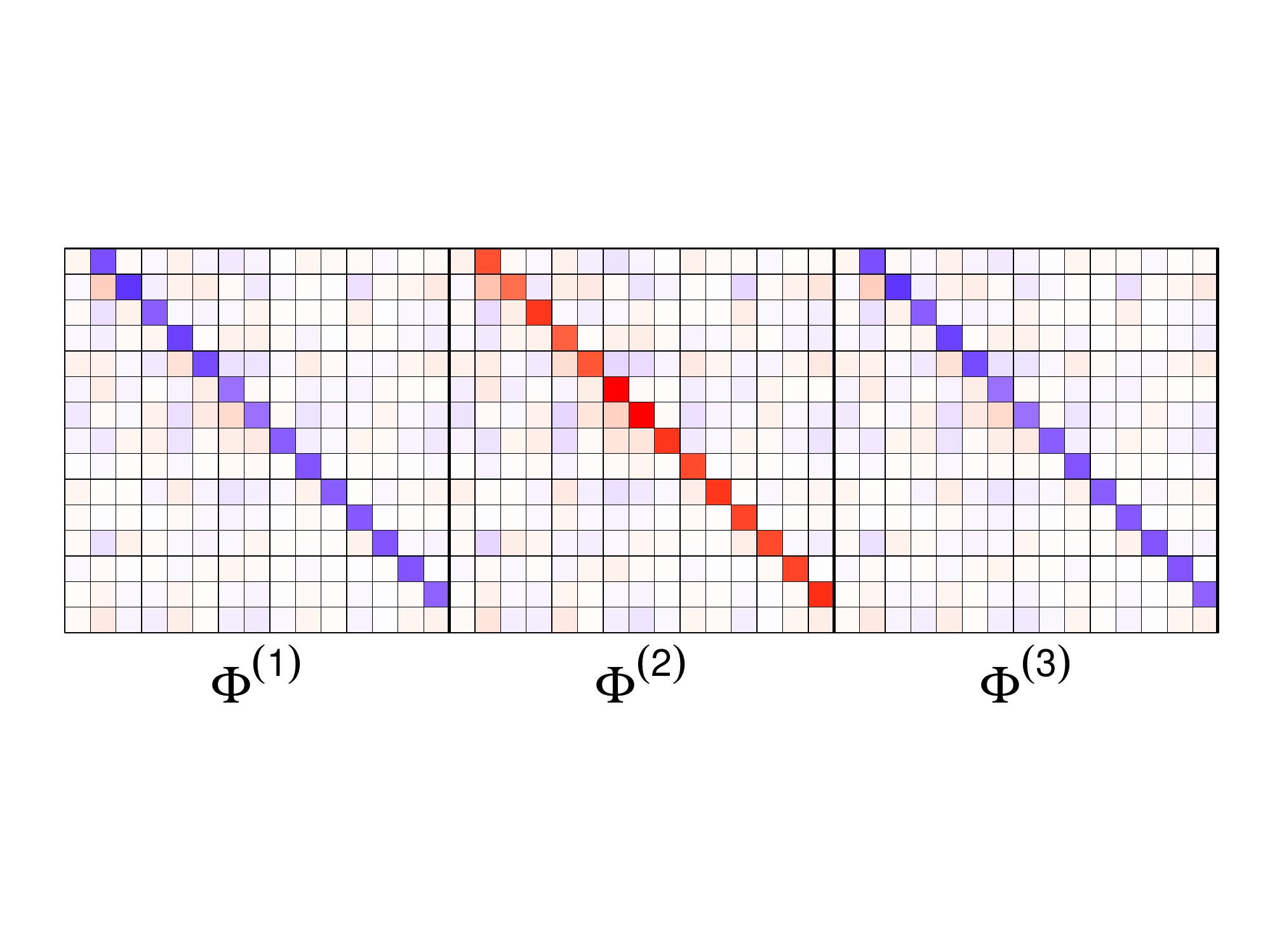}
    \caption{Example for generated fixed low rank plus sparse structured VAR(1) process. Left panel: generated process with two change points at $t_1=100$ and $t_2=200$; Right panel: true model parameters.}
    \label{fig:fLS-example}
\end{figure}

Next, change points are detected and the model parameters of the resulting stationary segments estimated using the following code.
\begin{CodeChunk}
\begin{CodeInput}
R> fit <- tbss(method = "fLS", mu = 150)
R> print(fit)
\end{CodeInput}
\begin{CodeOutput}
Estimated change points are: 100 200
\end{CodeOutput}
Figure \ref{fig:fLS-example-est} depicts the detected change points together with the time series data, as well as the estimated model parameters in the form of  heatmaps of the transition matrices.
\begin{CodeInput}
R> plot(fit, display = "cp")
R> plot(fit, display = "param")
\end{CodeInput}
\end{CodeChunk}
\begin{figure}[!ht]
    \centering
    \includegraphics[width=.475\textwidth, trim={0 10 0 50}, clip]{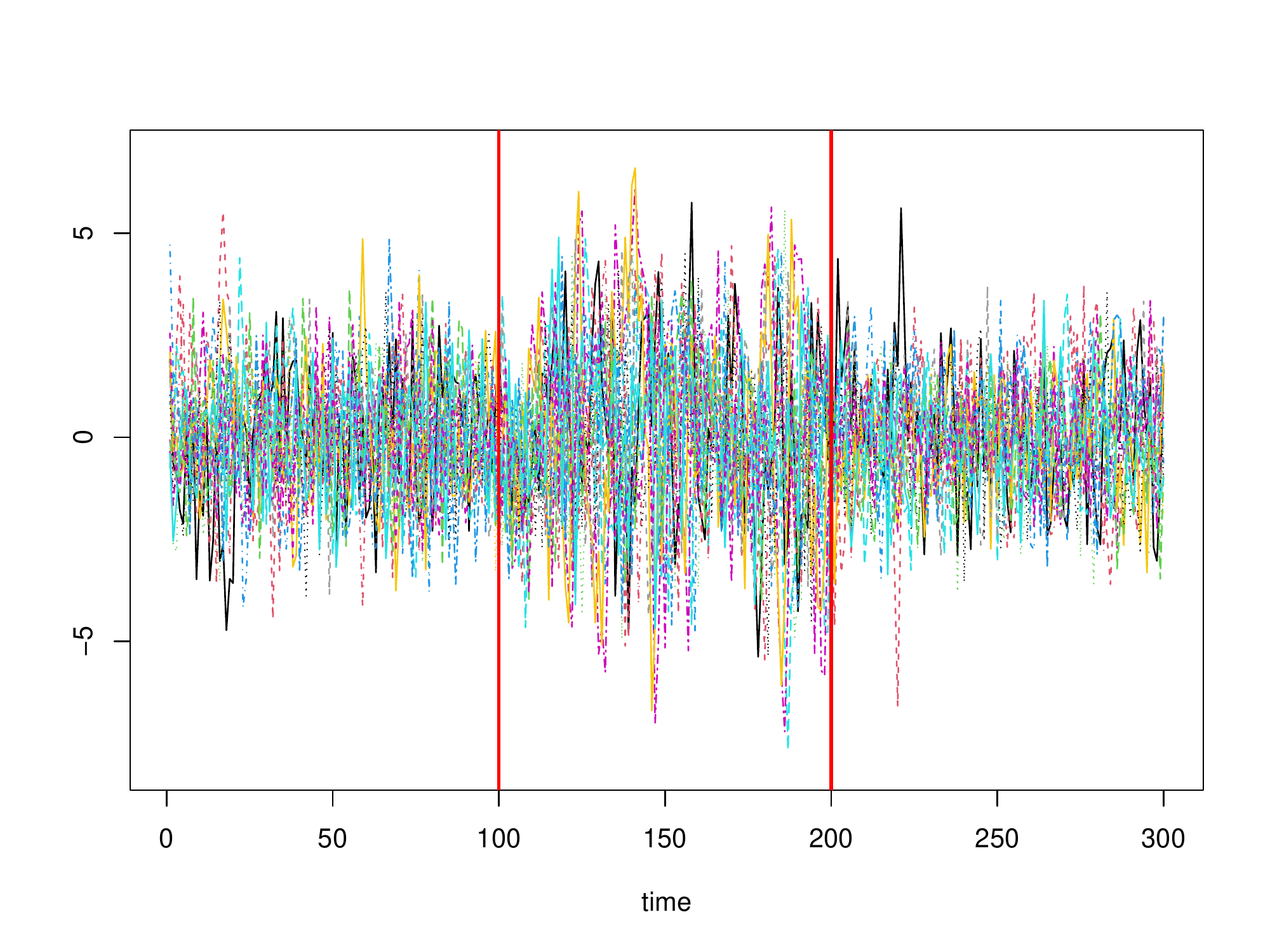}%
    \includegraphics[width=.475\textwidth, trim={0 10 0 50}, clip]{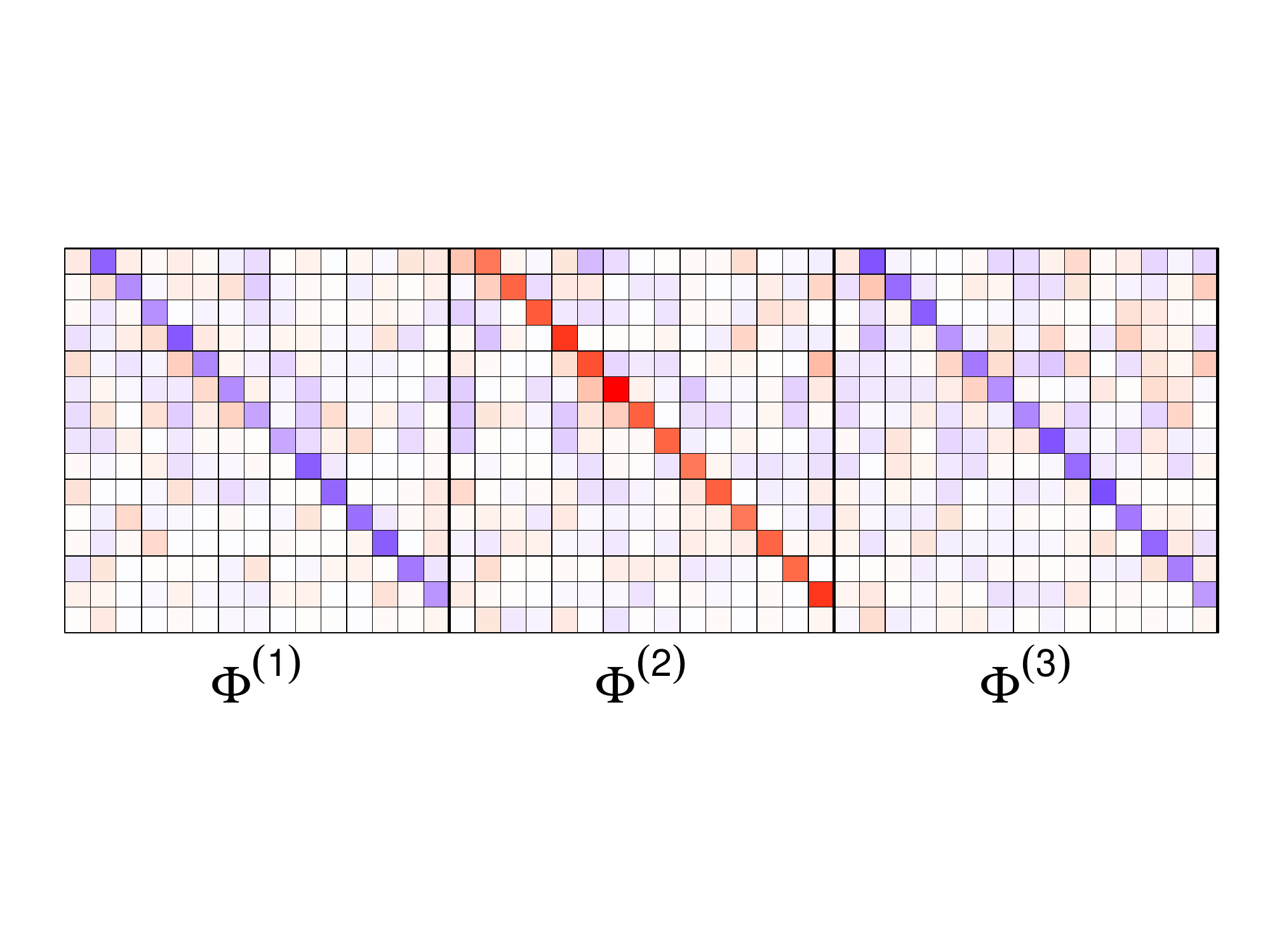}
    \caption{Left panel: results for detected change points at $\hat{t}_1=100$, and $\hat{t}_2=200$ (red lines); Right panel: plot of estimated model parameters.}
    \label{fig:fLS-example-est}
\end{figure}
}

\subsection{Change point detection on simulated data from a time-varying low rank plus sparse VAR model}\label{sec:lps-example}
We use function \code{simu_var} to generate data from a low rank plus sparse VAR process with multiple change points and use the function \code{lstsp} for change point detection and model parameter estimation. 

Suppose the dimension of the synthetic VAR process is $p=20$, and it includes $T=300$ observations with two change points located at $t_1=100$ and $t_2=200$. The sparsity pattern is set to 1-off diagonal, and the information ratios equals 0.35 across all three segments. The rank of the transition matrix in each of the three segments is set to 1, 3, and 1, respectively. 
\begin{CodeChunk}
\begin{CodeInput}
R> nob <- 300; p <- 20;
R> brk <- c(floor(nob / 3), floor(2 * nob / 3), nob + 1); 
R> signals <- c(-0.7, 0.8, -0.7)
R> rank <- c(1, 3, 1)
R> singular_vals <- c(1, 0.75, 0.5)
R> info_ratio <- rep(0.35, 3)
R> try <- simu_var(method = "LS", nob = nob, k = p, lags = 1, brk = brk, 
+                  sigma = as.matrix(diag(p)), signals = signals, 
+                  rank = rank, singular_vals = singular_vals, 
+                  info_ratio = info_ratio, sp_pattern = "off-diagonal", 
+                  spectral_radius = 0.9)
R> data <- try$series
R> MTS::MTSplot(data)
R> print(plot_matrix(do.call("cbind", try$model_param), m * q.t))
\end{CodeInput}
\end{CodeChunk}
\begin{figure}[!ht]
    \centering
    \includegraphics[width=.475\textwidth, trim={0 10 0 50}, clip]{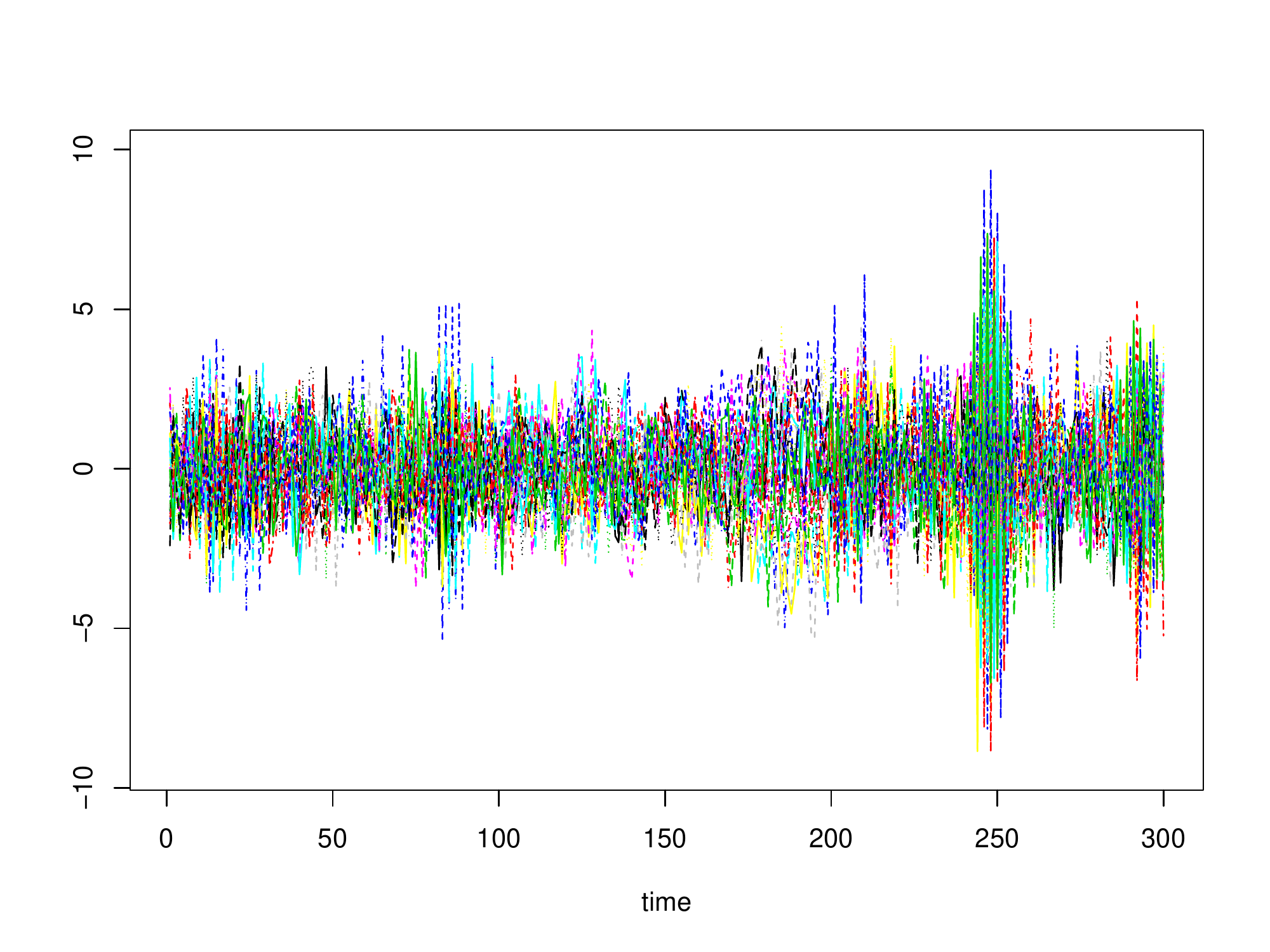}%
    \includegraphics[width=.475\textwidth, trim={0 10 0 50}, clip]{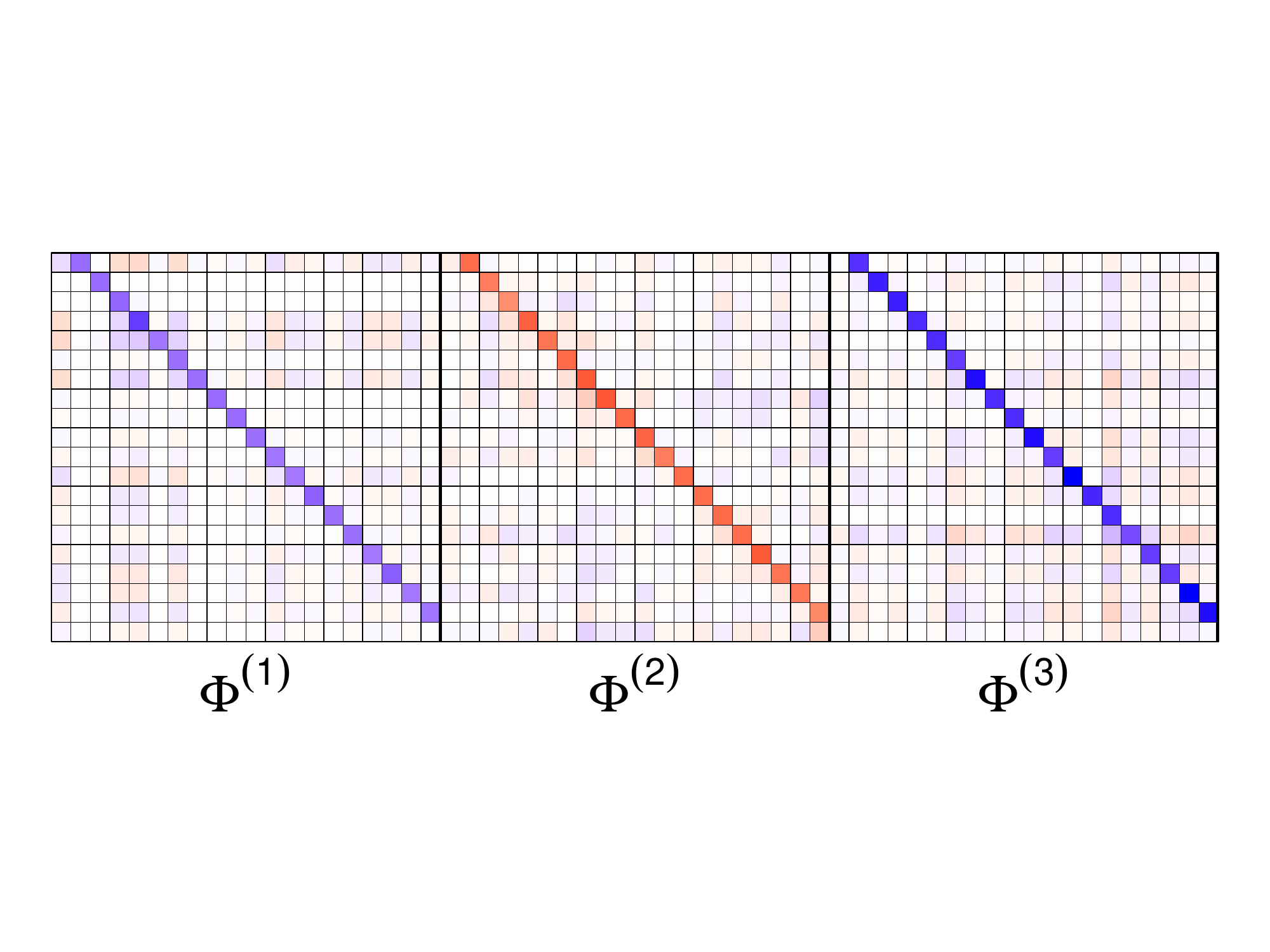}
    \caption{Example for generated low rank plus sparse structure VAR(1) process. Left panel: generated process with two change points at $t_1=100$ and $t_2=200$; Right panel: true model parameters.}
    \label{fig:lowrank-sparse-example}
\end{figure}
The code of detection and the final estimated change points are given by the following code; note the options used for selecting the tuning parameters. 
\begin{CodeChunk}
\begin{CodeInput}
R> lambda1 = lambda2 = lambda3 <- c(2.5, 2.5)
R> mu1 = mu2 = mu3 <- c(15, 15)
R> fit <- lstsp(data, lambda.1 = lambda1, mu.1 = mu1, 
+               lambda.2 = lambda2, mu.2 = mu2, 
+               lambda.3 = lambda3, mu.3 = mu3, 
+               step.size = 5, niter = 20, skip = 5, verbose = FALSE)
R> print(fit)
\end{CodeInput}
\begin{CodeOutput}
Estimated change points are: 101 200
\end{CodeOutput}
The following Figure \ref{fig:lowrank-sparse-estimation} illustrates the detected change points in the generated process, as well as the estimated model parameters. 
\begin{CodeInput}
R> plot(fit, display = "cp")
R> plot(fit, display = "param")
\end{CodeInput}
\end{CodeChunk}
\begin{figure}[!ht]
    \centering
    \includegraphics[width=.475\textwidth, trim={0 10 0 50}, clip]{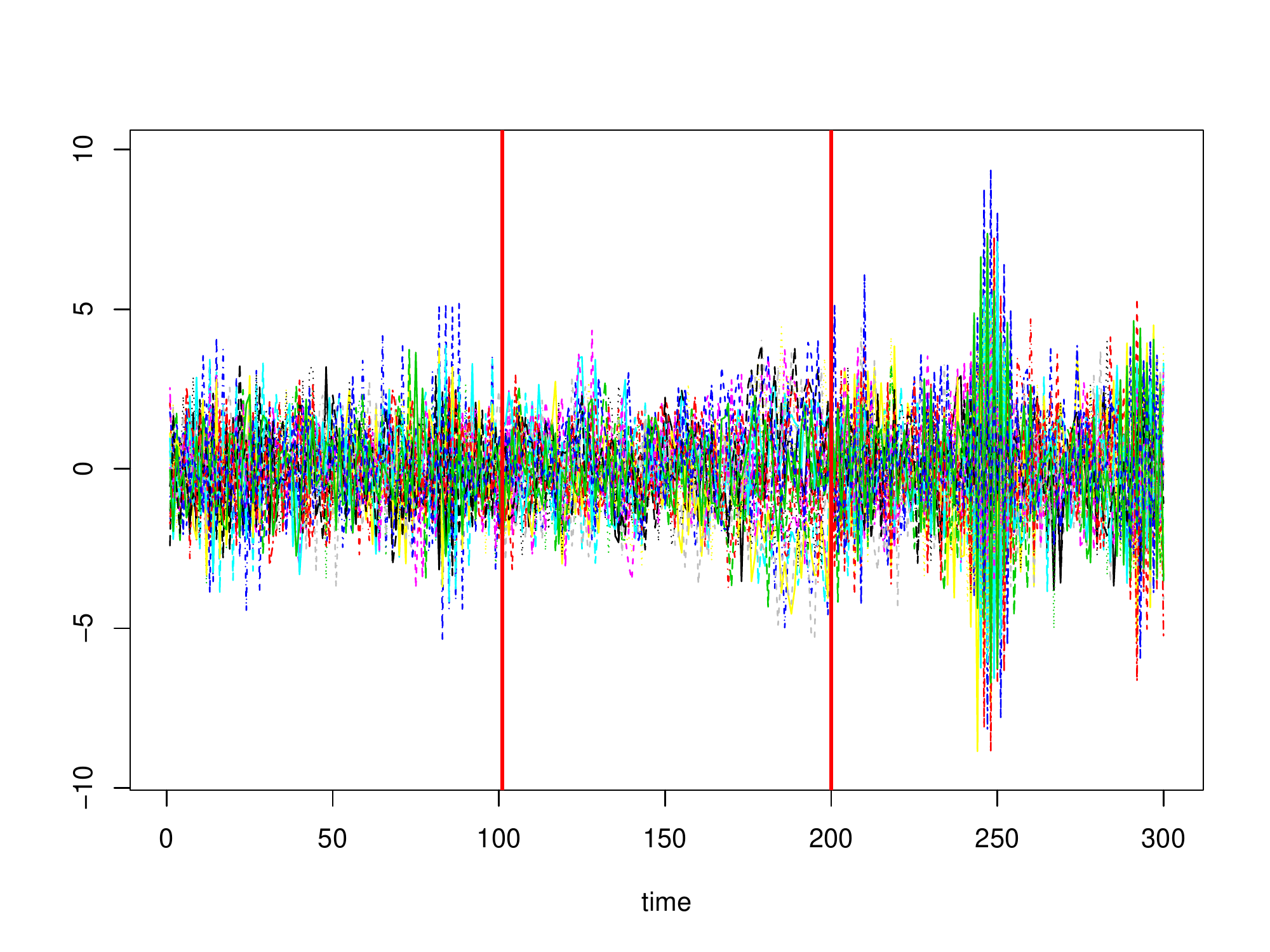}%
    \includegraphics[width=.475\textwidth, trim={0 10 0 50}, clip]{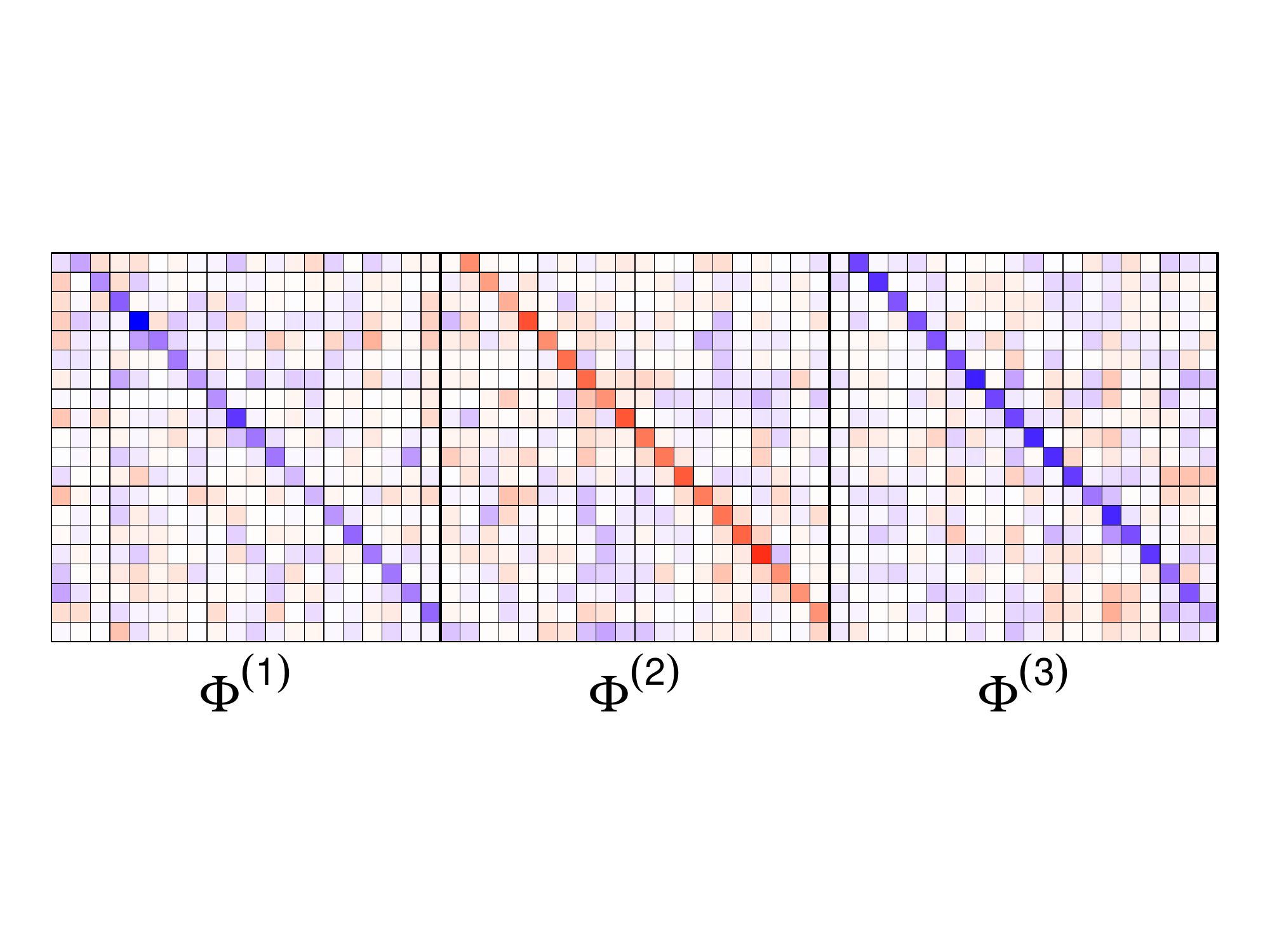}
    \caption{Left panel: results for detected change points at $\hat{t}_1=101$, and $\hat{t}_2=200$ (red lines); Right panel: plot of estimated model parameters.}
    \label{fig:lowrank-sparse-estimation}
\end{figure}

\subsection{An application to stocks return data}\label{sec:real-data-example}
We use the capabilities of the \pkg{VARDetect} package to examine \emph{weekly} log-return data of 20 US large stocks (components of the S\&P 100 index) for the 2001-2016 period. 

First, we install and import the \pkg{VARDetect} package, and then load the data set \emph{weekly} by using function \code{data}.
\begin{CodeChunk}
\begin{CodeInput}
R> install.package("VARDetect")
R> library("VARDetect")
R> data(weekly)
\end{CodeInput}
\end{CodeChunk}
Then, we set the tuning parameters and apply \code{tbss} function to detect change points, as well as estimate the model parameters. The following code demonstrates the results obtained by \emph{sparse} and \emph{group sparse} settings, separately, and the corresponding results are stored in variables \code{fit} and \code{fit_group}, respectively.
\begin{CodeChunk}
\begin{CodeInput}
R> set.seed(100)
R> lambda.1.max <- 1e-2; nlam <- 20 
R> lambda.1.min <- lambda.1.max * 1e-3
R> delata.lam <- (log(lambda.1.max) - log(lambda.1.min)) / (nlam - 1)
R> lambda.1.cv <-  sapply(1:(nlam), function(jjj) 
+                         lambda.1.min * exp(delata.lam * (nlam - jjj)))
R> fit <- tbss(weekly, method = "sparse", lambda.1.cv = lambda.1.cv, 
+              lambda.2.cv = 0.05, block.size = 8, an.grid = c(10, 15, 18))
R> print(fit)
R> plot(fit, display = "cp")
\end{CodeInput}
The detected change points based on a \emph{sparse} structure for the transition matrices are:
\begin{CodeOutput}
Estimated change points are: 92 363 417 477 568 765
\end{CodeOutput}
\end{CodeChunk}
Next, we apply the \emph{group sparse} structure for the transition matrices to detect change points:
\begin{CodeChunk}
\begin{CodeInput}
R> lambda.1.max <- 1e-2; nlam <- 20
R> lambda.1.min <- lambda.1.max * 1e-4
R> delata.lam <- (log(lambda.1.max) - log(lambda.1.min)) / (nlam - 1)
R> lambda.1.cv <- sapply(1:(nlam), function(jjj) 
+                        lambda.1.min * exp(delata.lam * (nlam - jjj)))
R> fit_group <- tbss(weekly, method = "group sparse", group.case = "columnwise", 
+                    max.iteration = 50, lambda.1.cv = lambda.1.cv, 
+                    lambda.2.cv = 0.1, block.size = 8, an.grid = c(10, 15, 18))
R> print(fit_group)
R> plot(fit_group, data, display = "cp")
\end{CodeInput}
Similarly, the detected change points for \emph{group sparse} transition matrices are
\begin{CodeOutput}
Estimated change points are: 93 363 426 481 571 602 755
\end{CodeOutput}
\end{CodeChunk}

\begin{figure}[!ht]
    \centering
    \includegraphics[width=.475\textwidth, trim={0 10 0 50}, clip]{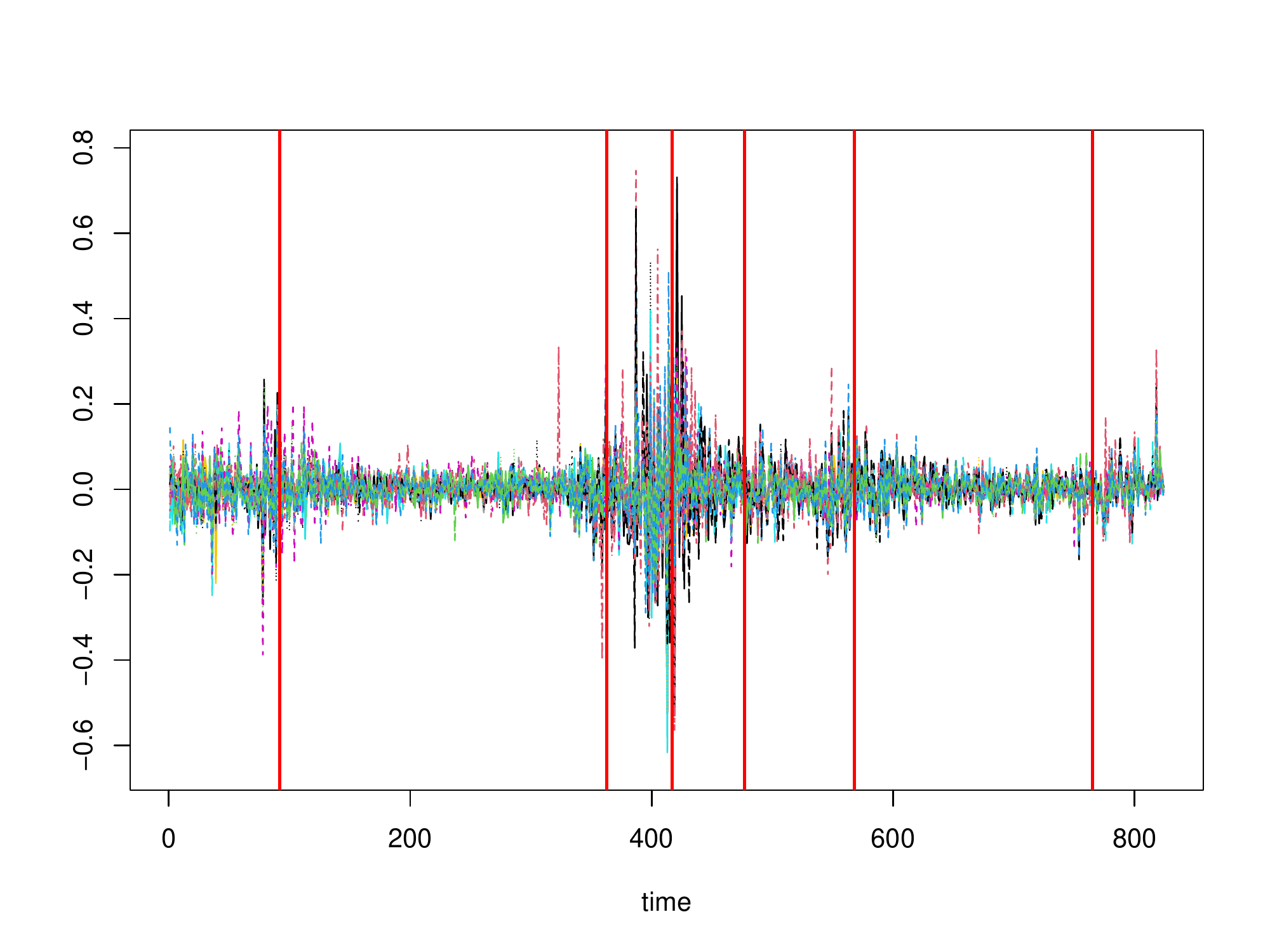}%
    \includegraphics[width=.475\textwidth, trim={0 10 0 50}, clip]{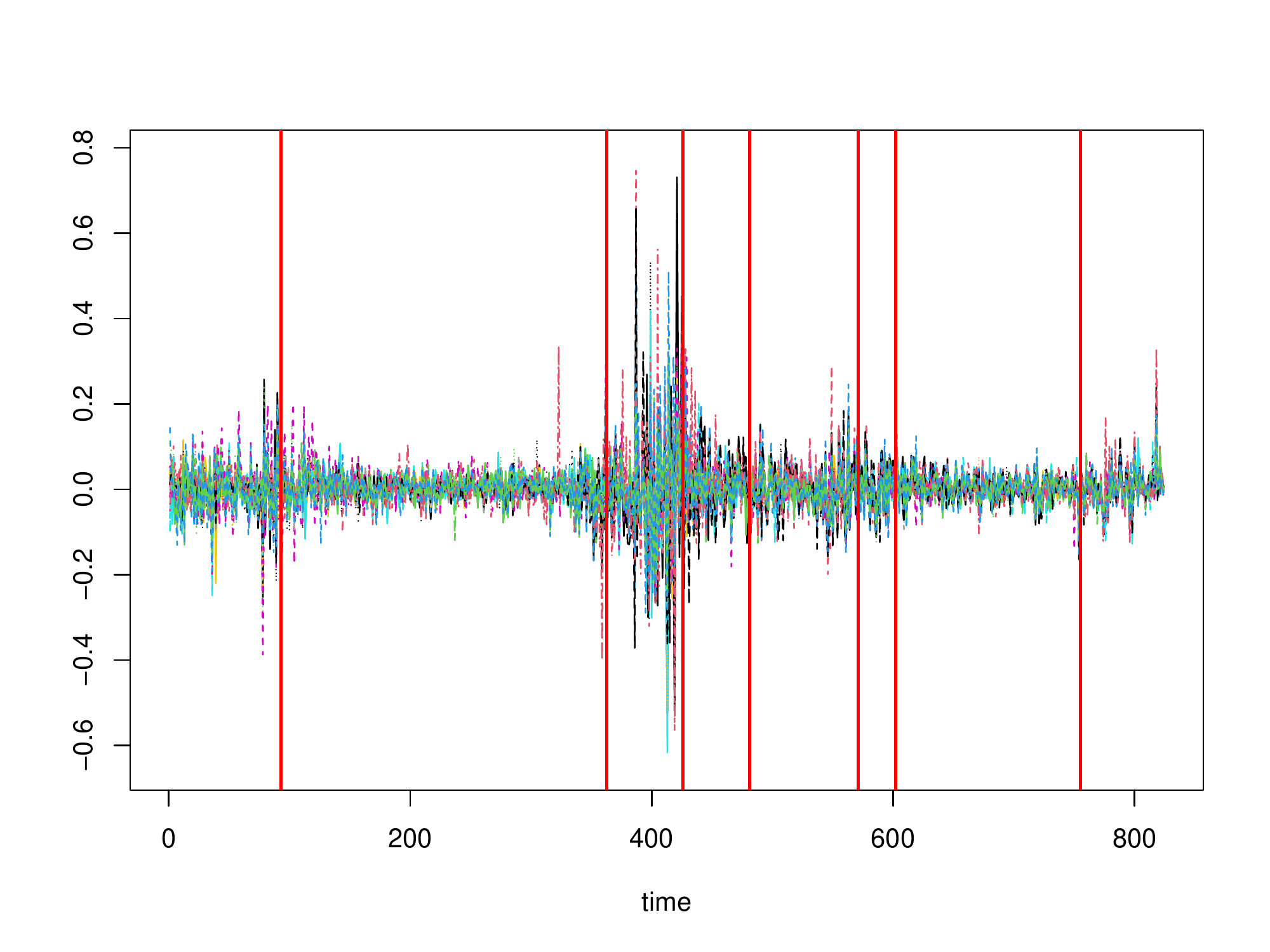}
    \caption{Detected change points in the stock data. Left panel: assuming \emph{sparse} transition matrices; Right panel: assuming \emph{group sparse} transition matrices.}
    \label{fig:real-examples}
\end{figure}

Figure \ref{fig:real-examples} presents the detected change points for both \emph{sparse} and \emph{group sparse} VAR models.

The following Table summarizes the change points detected by assuming \emph{sparse} and \emph{group sparse} transition matrices and associates them with relevant economic and financial events. 
\begin{table}[!ht]
    \centering
    \caption{Detected Change Points by Sparse and Group Sparse VAR Models}
    \label{tab:realexample-table}
    \resizebox{\textwidth}{!}{
    \begin{tabular}{c|c|c|l}
    \hline\hline
        No. of CPs & Sparse model & Group sparse model & Events \\
    \hline
        1 & 2002-10-29 & 2002-10-29 & Telecommunications bubble popped \\
        2 & 2008-02-05 & 2008-01-22 & Collapse of Bear Sterns \\
        3 & 2009-02-17 & 2009-03-24 & Sharp market downturn during the Great Financial Crisis  \\
        4 & 2010-04-13 & 2010-05-18 & European sovereign debt crisis \\
        5 & 2012-01-10 & 2012-01-10 & \multirow{2}{*}{Recovery from the Great Financial Crisis} \\
        6 & - & 2012-10-16 & \\
        7 & 2015-11-10 & 2015-09-01 & Sharp market correction\footnote{due to slowing growth in the GDP of China, a fall in petroleum prices, the Greek debt default in June 2015, and the effects of the end of quantitative easing in the US}  \\
    \hline\hline
    \end{tabular}}
\end{table}

\begin{figure}[!ht]
    \centering
    \includegraphics[width=.33\textwidth, trim={0 10 0 50}, clip]{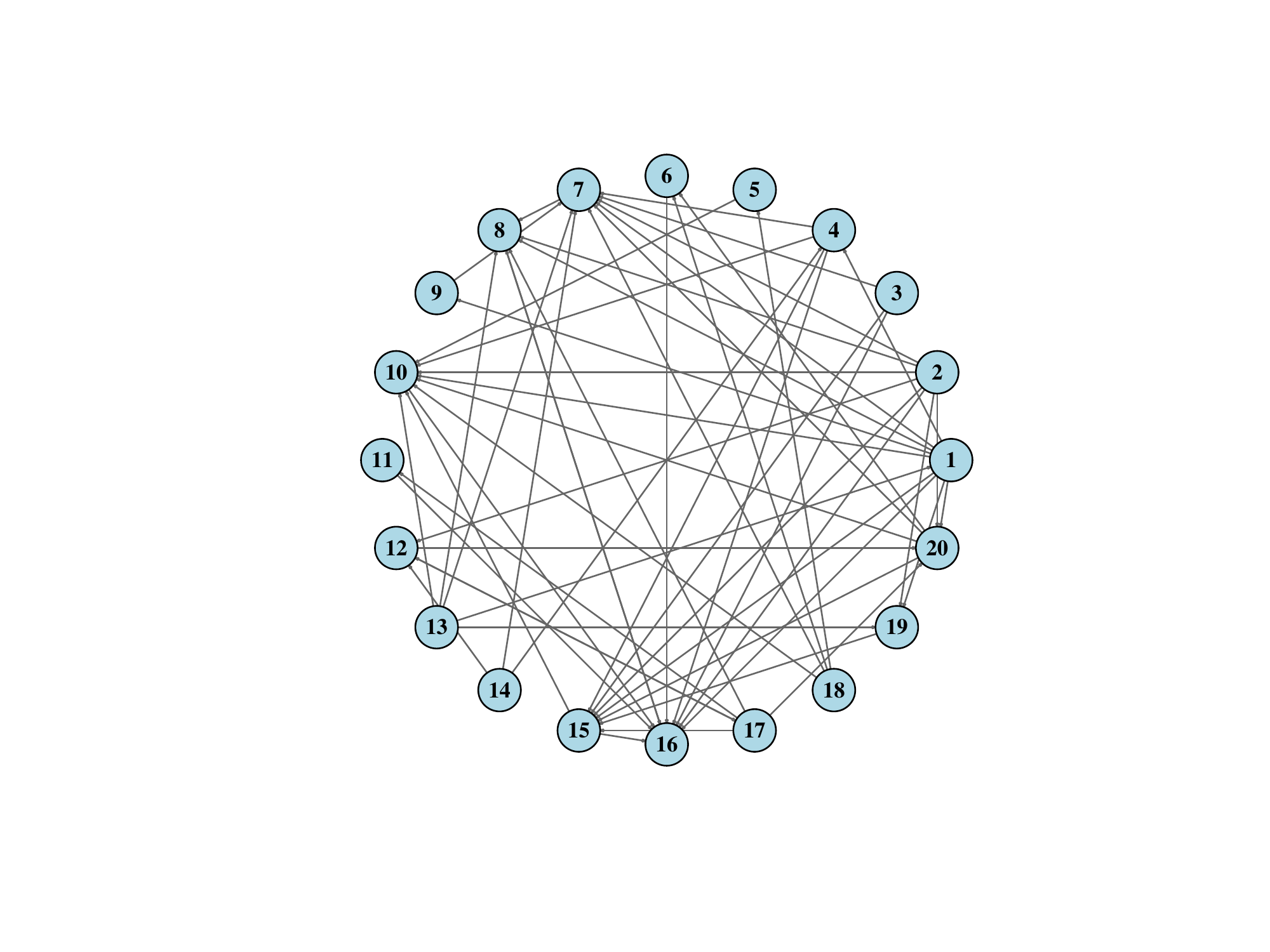}%
    \includegraphics[width=.33\textwidth, trim={0 10 0 50}, clip]{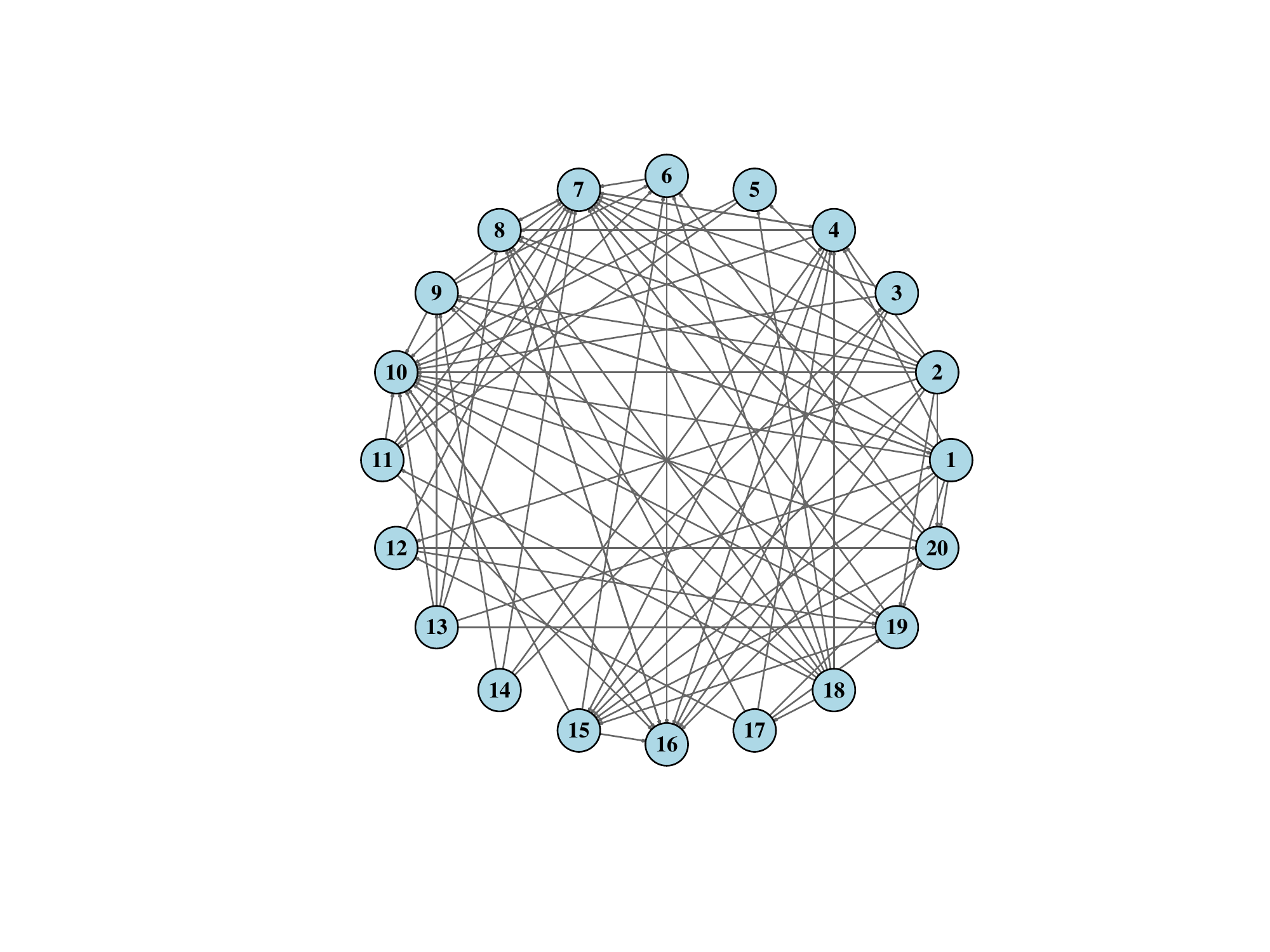}%
    \includegraphics[width=.33\textwidth, trim={0 10 0 50}, clip]{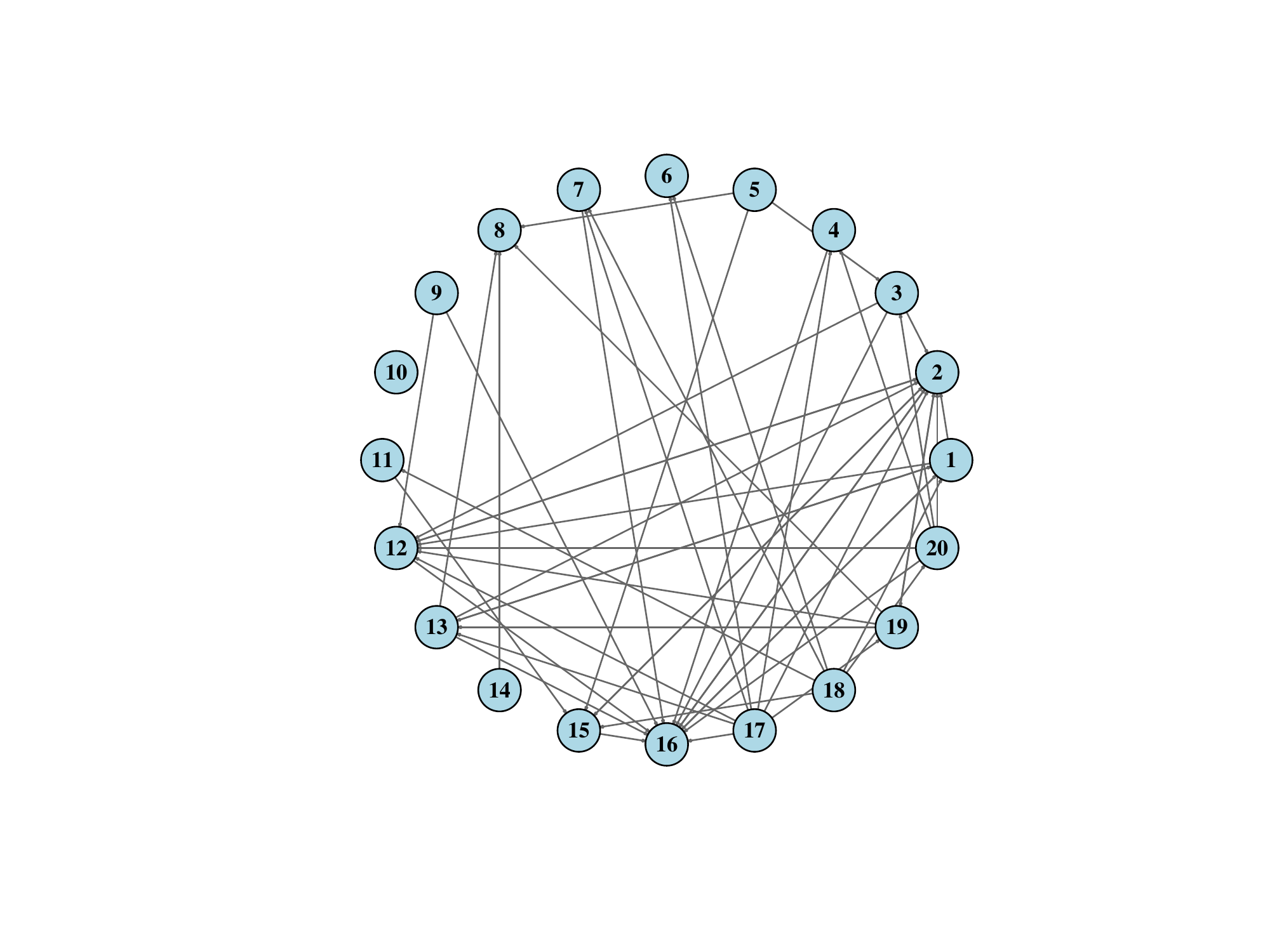}%
    \vfill
    \includegraphics[width=.33\textwidth, trim={0 10 0 50}, clip]{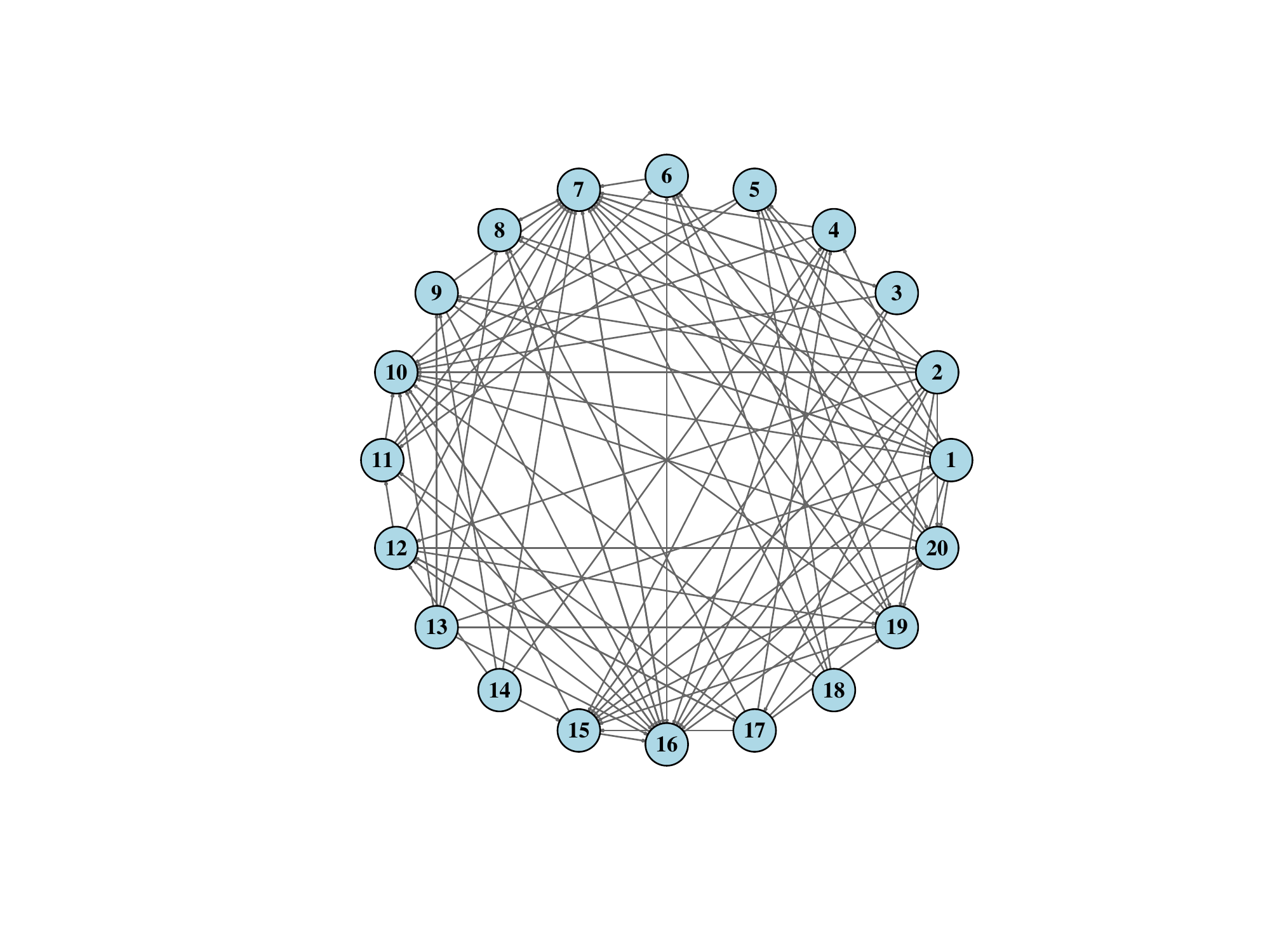}%
    \includegraphics[width=.33\textwidth, trim={0 10 0 50}, clip]{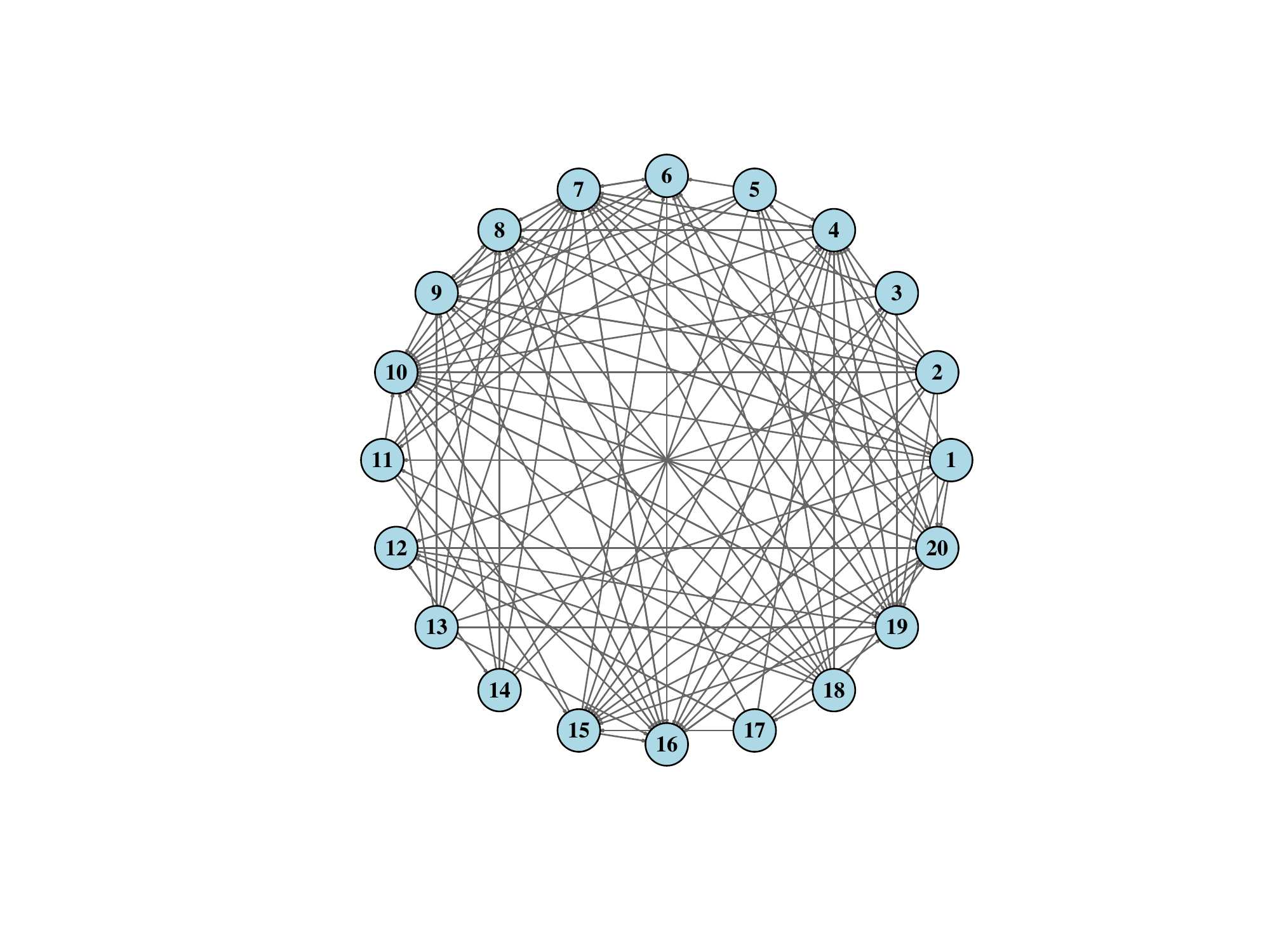}%
    \includegraphics[width=.33\textwidth, trim={0 10 0 50}, clip]{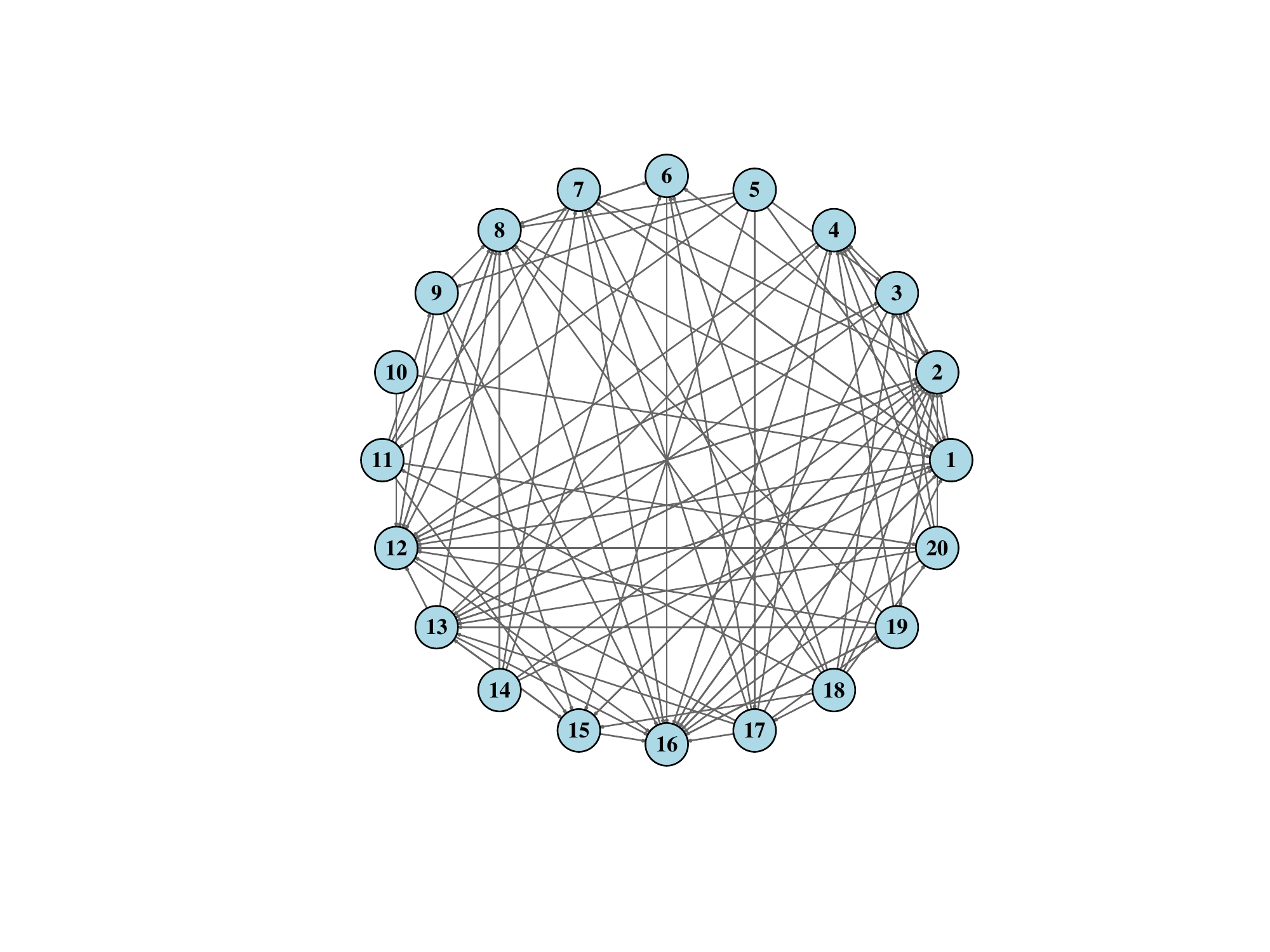}
    \caption{Estimated model parameters. Top panel (\emph{sparse} structure): pre-crisis (from 2002-10-29 to 2008-02-05), in-crisis (from 2008-02-05 to 2009-02-17), post-crisis (from 2012-01-10 to 2015-11-10); Bottom panel (\emph{group sparse} structure): pre-crisis (from 2002-10-29 to 2008-01-22), in-crisis (from 2008-01-22 to 2009-03-24), post-crisis (from 2012-10-16 to 2015-09-01).}
    \label{fig:real-example-modelparam}
\end{figure}

Next, we provide the code for visualizing the estimated model parameters for selected stationary segments. 
\begin{CodeChunk}
\begin{CodeInput}
R> plot(fit, display = "granger", threshold = 0.2)
R> plot(fit_group, display = "granger", threshold = 0.2)
\end{CodeInput}
\end{CodeChunk}
Figure \ref{fig:real-example-modelparam} presents \textit{selected} estimated Granger causal networks for the following time periods related to the Great Financial Crisis of 2008: pre-crisis (during 2002-2008), in-crisis (2008-2009), post-crisis (during 2012-2015), respectively. The specific time segments are given in the caption of Figure \ref{fig:real-example-modelparam}. It can be seen that the density of the transition matrix becomes much larger during the crisis period (2008-2009), since almost stocks exhibited strong negative returns during that period and hence their cross autocorrelations became larger in magnitude, as captured in the corresponding Granger causal networks.

\begin{figure}[!ht]
    \centering
    \includegraphics[width=.475\textwidth, trim={0 10 0 50}, clip]{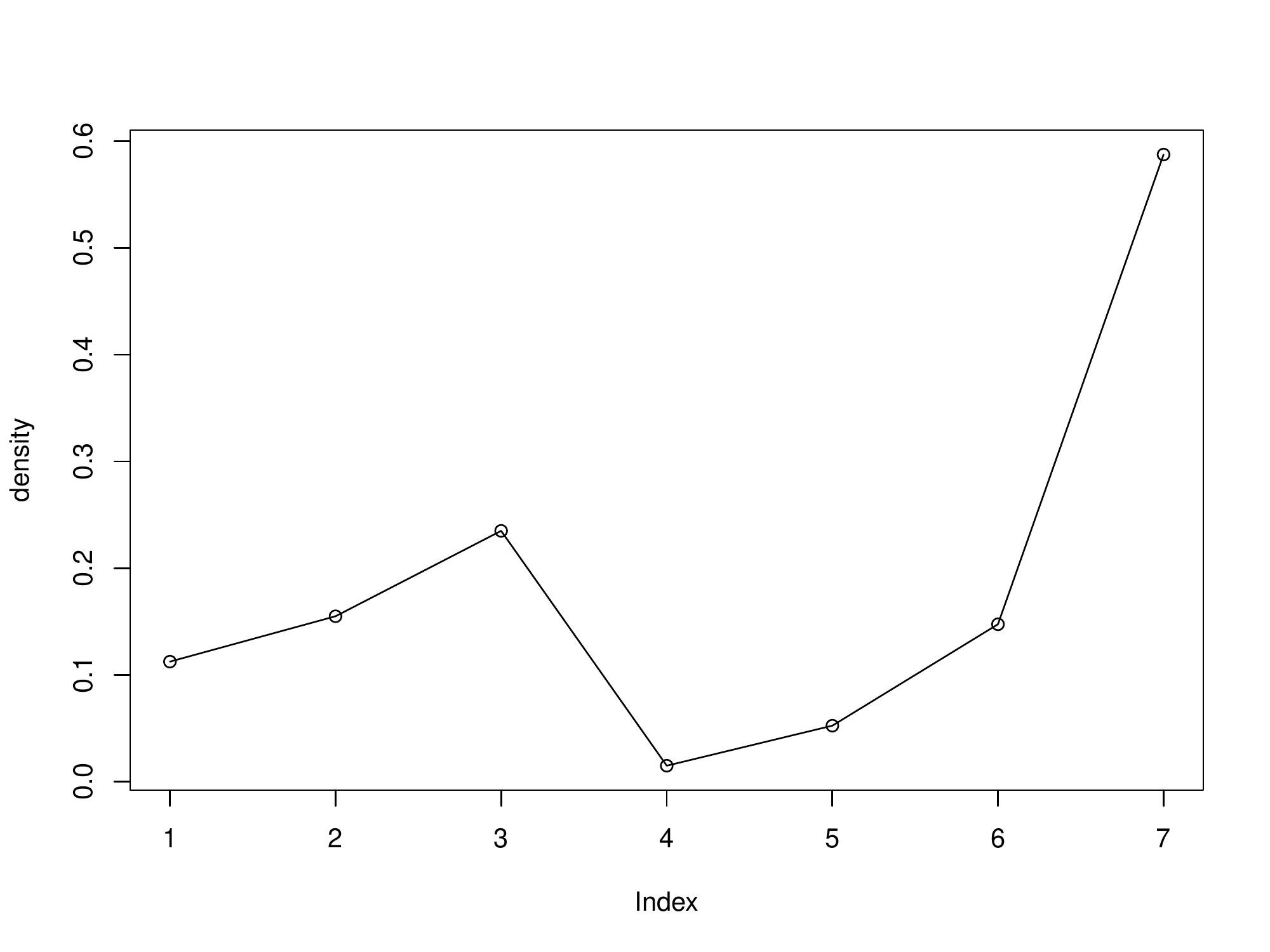}%
    \includegraphics[width=.475\textwidth, trim={0 10 0 50}, clip]{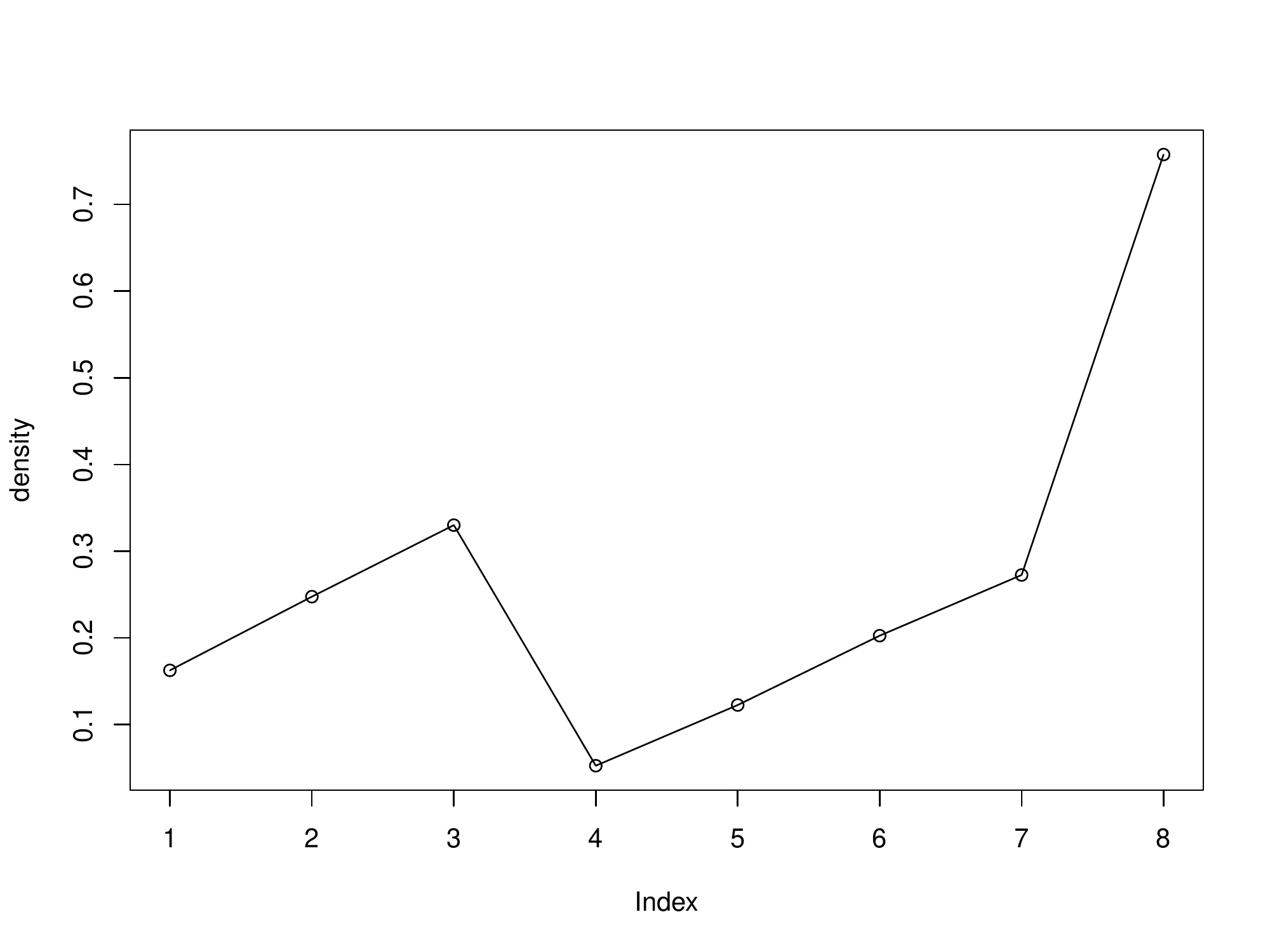}
    \caption{Density of the sparse component for each estimated segment based on a \emph{sparse} (left panel) and \emph{group sparse} transition matrix (right panel).}
    \label{fig:real-example-density}
\end{figure}

Further, we plot the density levels for the sparse component in each stationary segment in Figure \ref{fig:real-example-density}, using the \code{plot_density} function.
\begin{CodeChunk}
\begin{CodeInput}
R> plot(fit, display = "density", threshold = 0.2)
R> plot(fit_group, display = "density", threshold = 0.2)
\end{CodeInput}
\end{CodeChunk}
{
Further, we also employed a \emph{low rank plus sparse} structure for the transition matrices to identify the change points in the given data set. The code used is given next.
\begin{CodeChunk}
\begin{CodeInput}
R> n <- dim(weekly)[1]
R> k <- dim(weekly)[2]
R> lambda.1 <- c(0.015, 0.015); mu.1 <- c(0.05, 0.05)
R> lambda.2 <- c(0.01, 0.01); mu.2 <- c(0.05, 0.05)
R> lambda.3 <- c(0.008, 0.008); mu.3 <- c(0.0335, 0.0335)
R> N <- n - 1; omega <- (0.0015) * (((log(N))^1)*log(k))
R> fit_LpS <- lstsp(weekly, lambda.1 = lambda.1, mu.1 = mu.1, 
+              lambda.2 = lambda.2, mu.2 = mu.2, 
+              lambda.3 = lambda.3, mu.3 = mu.3, 
+              h = 80, step.size = 40, omega = omega,
+              niter = 20, skip = 5, verbose = TRUE)
R> print(fit_LpS)
R> plot(fit_LpS, display = 'cp')
R> plot(fit_LpS, display = 'density', threshold = 0.05)
\end{CodeInput}
\begin{CodeOutput}
Estimated change points are: 117 331 367 476 488 597
\end{CodeOutput}
\begin{figure}[!ht]
    \centering
    \includegraphics[width=.475\textwidth, trim={0 10 0 50}, clip]{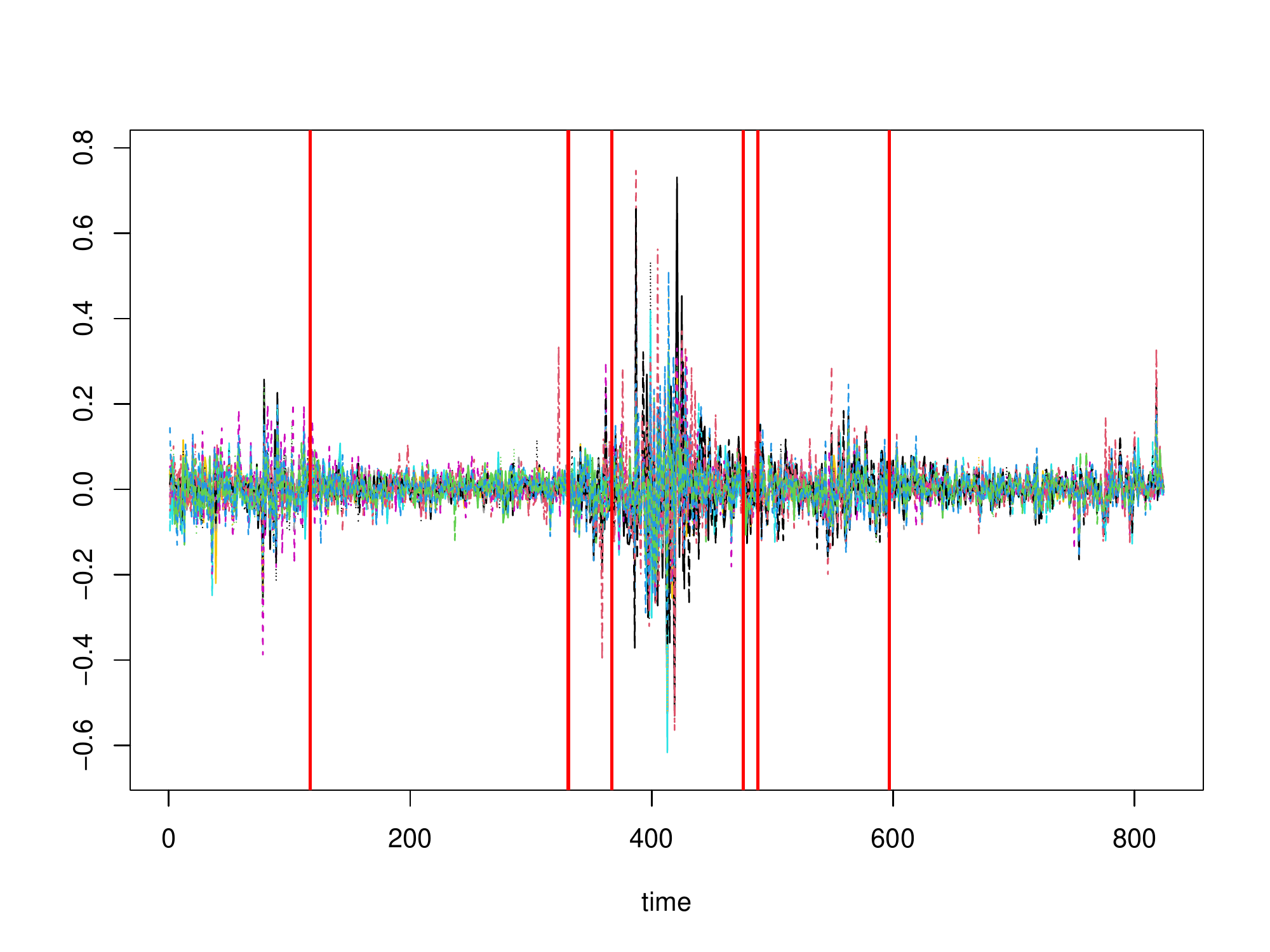}%
    \includegraphics[width=.475\textwidth, trim={0 10 0 50}, clip]{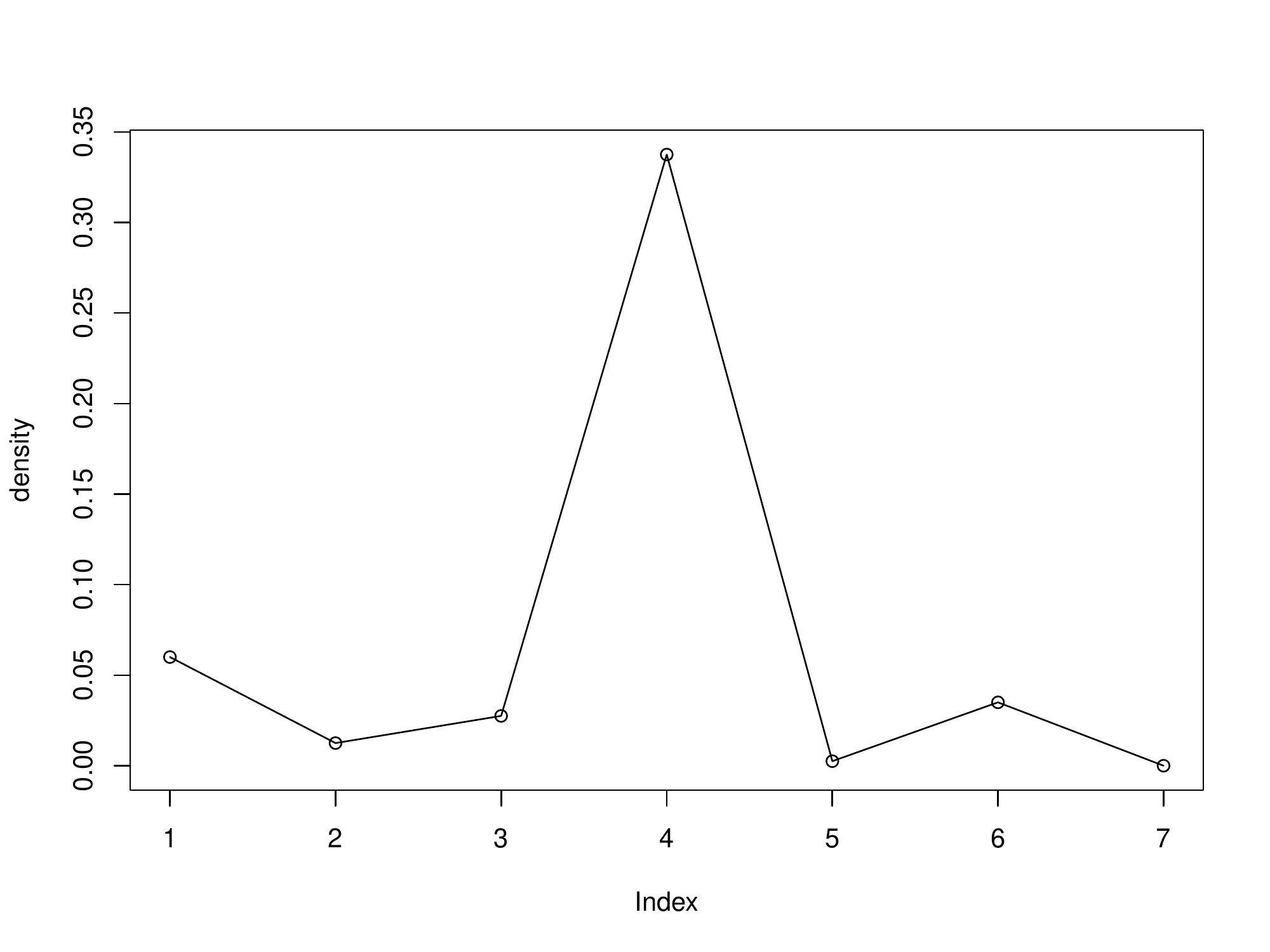}
    \caption{Left: Detected change points in the stock data by LSTSP algorithm; Right: Density plots for the sparse components}
    \label{fig:lps-example}
\end{figure}
\end{CodeChunk}
The following table summarizes the corresponding events to those identified change points by the LSTSP algorithm. 
\begin{table}[!ht]
    \centering
    \caption{Detected Change Points by Low rank plus Sparse VAR Models}
    \label{tab:realexample-table-lstsp}
    \resizebox{\textwidth}{!}{
    \begin{tabular}{c|c|l}
    \hline\hline
        No. of CPs & Date & Events \\
    \hline
        1 & 2003-04-22 & Market recovered from Telecommunications bubble popped \\
        2 & 2007-06-12 & Collapse of Bear Sterns \\
        3 & 2008-03-04 & Sharp market downturn during the Great Financial Crisis  \\
        4 & 2010-04-06 & \multirow{2}*{European sovereign debt crisis} \\
        5 & 2010-06-29 & \\
        6 & 2012-07-31 & Recovery from the Great Financial Crisis\\
    \hline\hline
    \end{tabular}}
\end{table}

In summary, the LSTSP algorithm performs similar to the TBSS algorithm (see Table \ref{tab:realexample-table}) up to 2012. However, the LSTSP algorithm missed the last change point located between September and November 2015. Further, the density plot for the sparse components obtained by the LSTSP algorithm is different to that by the TBSS algorithm, since the 4th segment in the LSTSP result is the 3rd segment in TBSS result, which is the highest in the first seven segments. 
}

{
We also compare the performance and results obtained by the TBSS algorithm with the results derived by another algorithm that does not consider specifically VAR models, but is suited for multivariate time series data. The change points detected by the Sparified Binary Segmentation (SBS) algorithm proposed by \cite{cho2015multiple} are obtained by the following code snippet.
\begin{CodeChunk}
\begin{CodeInput}
R> library("hdbinseg")
R> data <- read.csv("../Weekly.csv")
R> fit <- sbs.alg(t(data), cp.type = 2)
R> print(paste("Estimated change points are:, fit$ecp, sep = " "))
\end{CodeInput}
\begin{CodeOutput}
Estimated change points are: 124 384 432 506
\end{CodeOutput}
\end{CodeChunk}
The estimated change points are presented in the following Figure \ref{fig:sbs-example}, and listed with the approximate corresponding events in the Table \ref{tab:sbs-events}. 
\begin{figure}[!ht]
    \centering
    \includegraphics[width=.475\textwidth, trim={0 10 0 50}, clip]{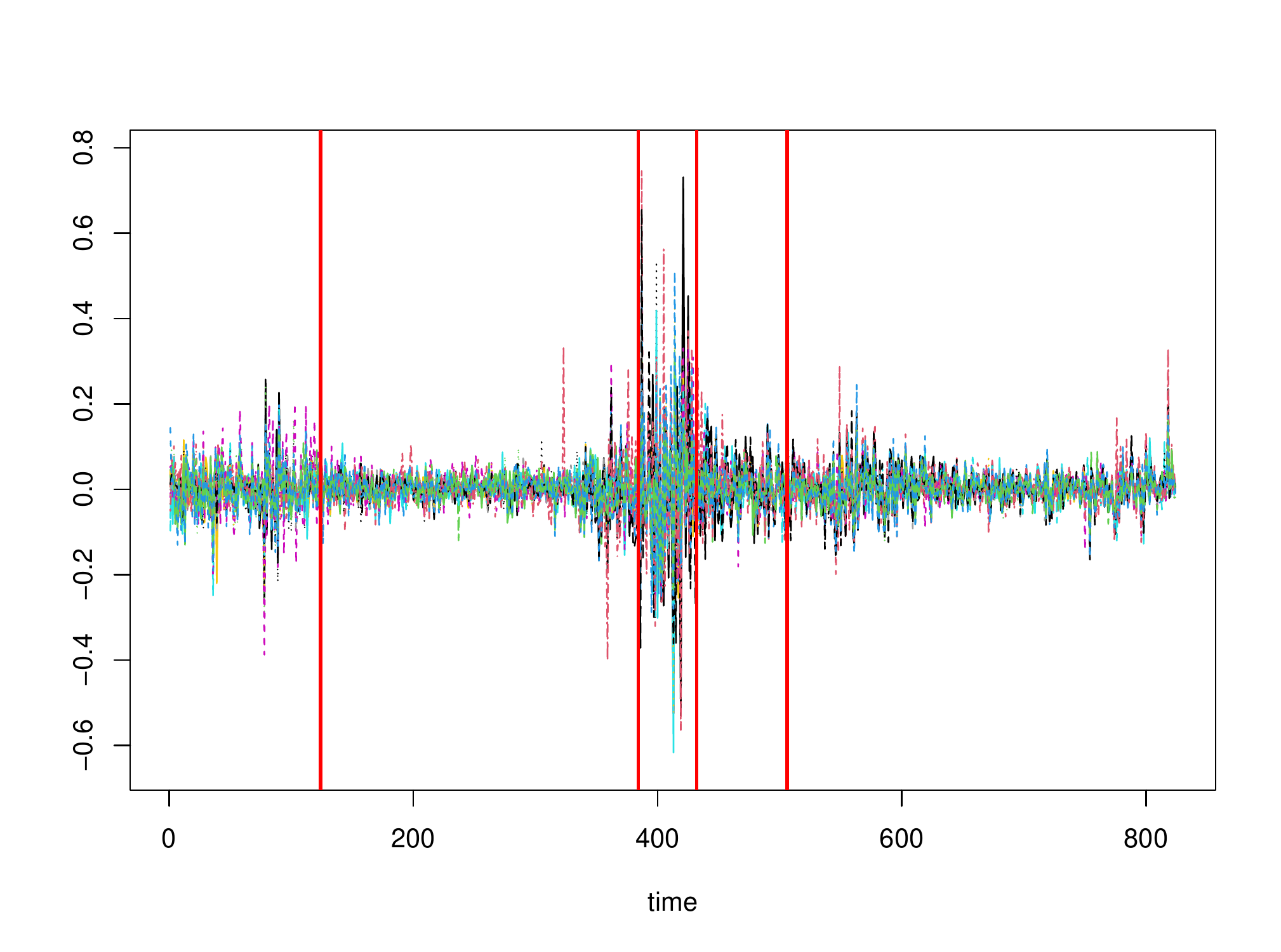}
    \caption{Detected change points in the stock data by SBS algorithm.}
    \label{fig:sbs-example}
\end{figure}
\begin{table}[!ht]
    \centering
    \caption{Detected Change Points and Corresponding Events by SBS Algorithm.}
    \label{tab:sbs-events}
    \begin{tabular}{c|c|c}
    \hline\hline
       No. of CPs & Date & Events \\
    \hline
       1 & 2003-06-10 & Aftermath of telecommunications bubble popped \\
       2 & 2008-07-01 & The middle of the Great Financial Crisis \\
       3 & 2009-06-02 & Sharp market downturn during the Great Financial Crisis\\
       4 & 2010-11-02 & European sovereign debt crisis \\
    \hline\hline
    \end{tabular}
\end{table}

It can be seen that the TBSS algorithm based on a sparse/group sparse model identifies seven change points corresponding to major economic/financial shocks that occurred during the period under consideration. In contrast, the SBS algorithm detects only four change points, and is not able to identify any change points after 2010. 
}

\subsection{An application to EEG signals data}
{
We consider a data set that contains EEG signals from 72 electrodes positioned on the scalp of a subject (see details in \cite{trujillo2017effect}). The recording time is 480 secs at a sampling frequency of 256Hz, resulting in time series of length 122,880. The subject was asked to keep her/his eyes open for 1 min and then close them for 1 min and repeat this pattern four times. In the ensuing analysis, the selected data correspond to the segment with time index from 49000 to 113999, that were subsequently downsampled to 4063 observations by choosing every 16 time points and selected 20 channels at random from the data set. There are three change points in the selected time segment according to the design of the experiment that are located around $t_1=1000, t_2=2000$, and $t_3=3000$, respectively.

First, we used a \emph{sparse} structure for the VAR transition matrices, and the lag of the VAR model was determined by using the BIC procedure introduced in Section \ref{sec:bic}. The results for the lag selection are given in Table \ref{tab:bic-table}.
\begin{table}[!ht]
    \centering
    \caption{BIC values for different time lag VAR models. }
    \label{tab:bic-table}
    \begin{tabular}{c|c|c|c|c|c}
    \hline\hline
        Lags & 1 & 2 & 3 & 4 & 5 \\
    \hline
        Total BIC value $(\times 10^3)$ & 2.331 & 4.277 & 10.021 & 22.417 & 32.592 \\
    \hline\hline
    \end{tabular}
\end{table}
It can be seen that a VAR(1) is selected by the BIC criterion.
Then, based on the optimal lag ($q=1$) supplied to the TBSS algorithm we obtained the change points.
\begin{CodeChunk}
\begin{CodeInput}
R> library(VARDetect)
R> lambda.1.cv <- c(0.1)
R> lambda.2.cv <- c(0.001)
R> fit <- tbss(as.matrix(data), method = "sparse", q = 1, 
+             lambda.1.cv = lambda.1.cv, 
+             lambda.2.cv = lambda.2.cv, 
+             block.size = floor(0.8*sqrt(n)), 
+             an.grid = c(150, 300), refit = TRUE)
R> print(fit)
R> plot(fit, display = 'cp')
R> plot(fit, display = 'granger', threshold = 0.75)
\end{CodeInput}
\begin{CodeOutput}
Estimated change points are: 751 2279 2951
\end{CodeOutput}
\end{CodeChunk}
Note that according to the design of the experiment, a transition between eyes closed to open and vice versa,
occurs around $t_1 = 1000, t_2 = 2000$, and $t_3= 3000$.
It can be seen that the estimated change points are located fairly close to the ones implied by the experimental design. Nevertheless, the first two exhibit a larger deviation that can be a consequence of selecting 20 (out of 72) EEG channels, noise in the data, and also of the reaction time of the subject. The left panel of Figure \ref{fig:eeg-example} presents the locations of detected change points.

Next, we assumed a low rank plus sparse structure for the transition matrix and employed the LSTSP algorithm to detect the change points. 
\begin{CodeChunk}
\begin{CodeInput}
R> lambda.1 <- c(0.5, 0.5); mu.1 <- c(5, 5)
R> lambda.3 <- c(0.35, 0.35); mu.3 <- c(200, 200)
R> N <- n-1
R> lambda.2 <- rep((1/1)*(log(N)*log(p))/N, 2); mu.2 <- c(1, 1)
R> omega <- (250)*(((log(N))^1)*log(p))
R> h <- 8*floor(sqrt(n))+1; steps <- floor(0.45*h)
R> fit <- lstsp(as.matrix(data), lambda.1 = lambda.1, mu.1 = mu.1, 
+              lambda.2 = lambda.2, mu.2 = mu.2, 
+              lambda.3 = lambda.3, mu.3 = mu.3, 
+              omega = omega, h = h, step.size = steps, skip = 125)
R> print(fit)
R> plot(fit, display = 'cp')
R> ranks <- rep(0, length(fit$cp)+1)
R> for(i in 1:(length(fit$cp)+1)){
+      ranks[i] <- qr(fit$lowrank_mats[[i]])$rank
+      print(ranks[i])
+ }
R> plot(ranks, type = 'o', ylab = 'rank')
\end{CodeInput}
\begin{CodeOutput}
Estimated change points are: 609 1943 3106
\end{CodeOutput}
\end{CodeChunk}
\begin{figure}[!ht]
    \centering
    \includegraphics[width=.475\textwidth, trim={0 10 0 50}, clip]{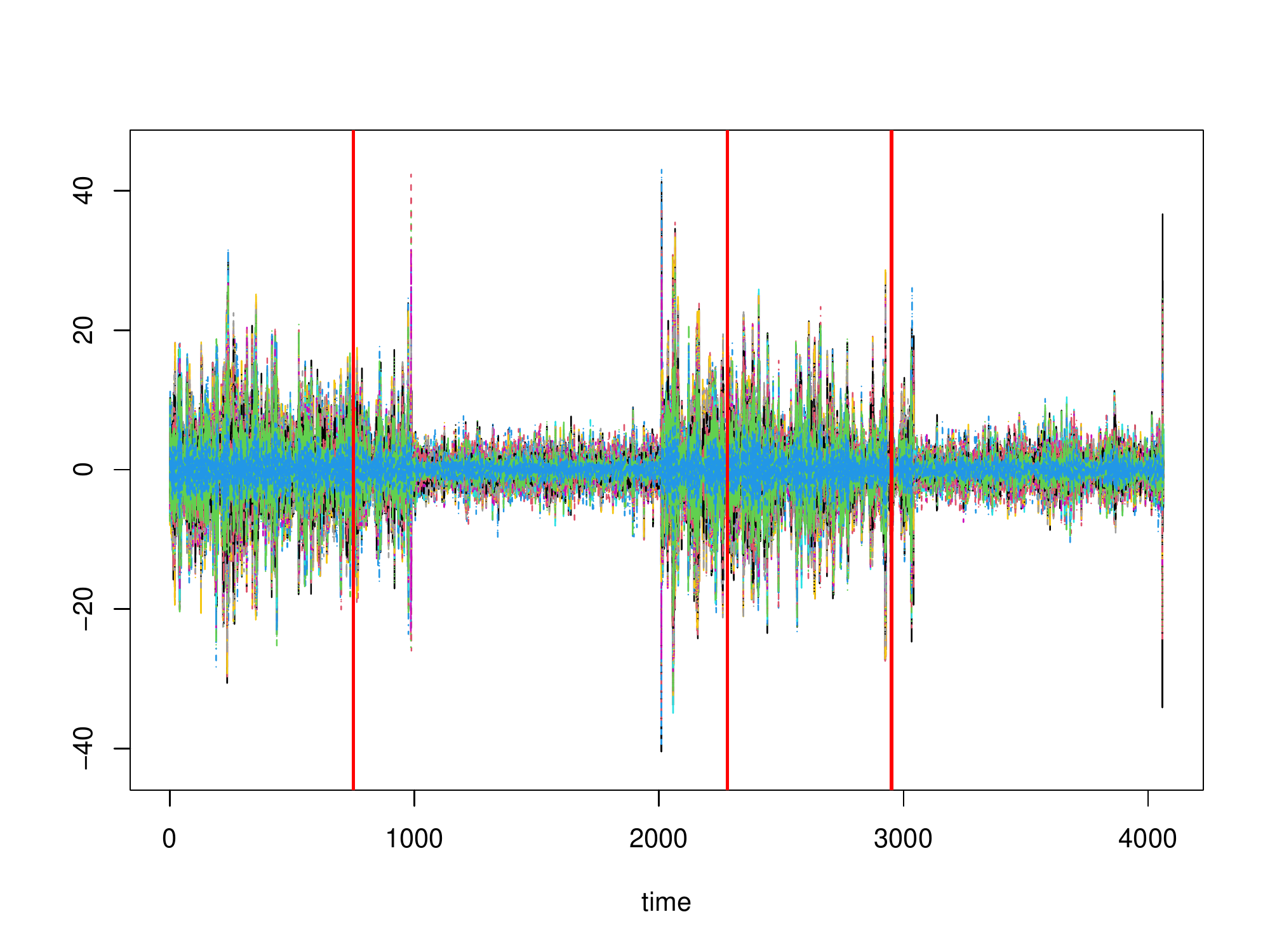}%
    \includegraphics[width=.475\textwidth, trim={0 10 0 50}, clip]{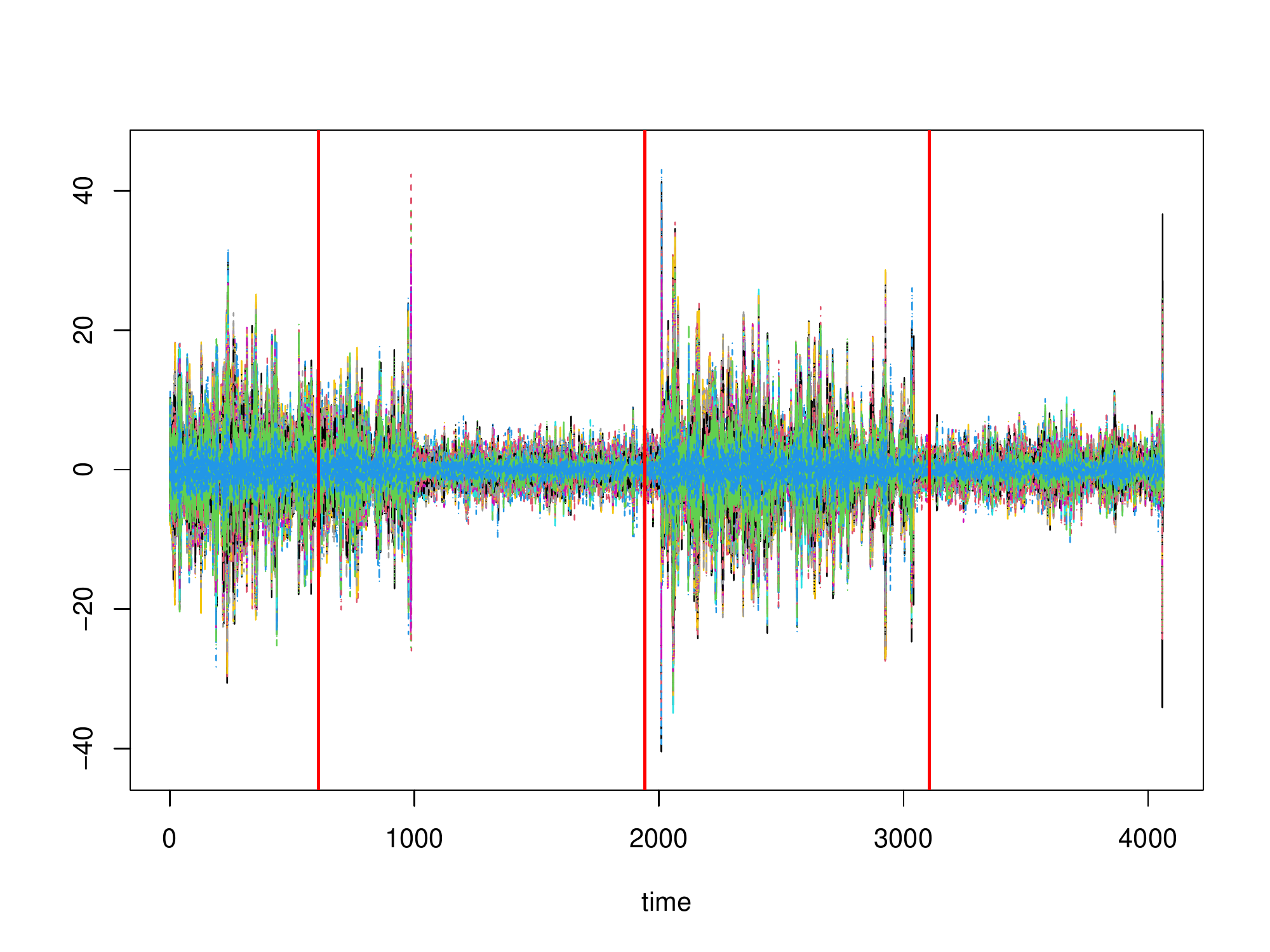}
    \caption{Left: Identified change points by TBSS algorithm; Right: Identified change points by LSTSP algorithm.}
    \label{fig:eeg-example}
\end{figure}

The detected change points are presented in the right panel of Figure \ref{fig:eeg-example}. It can be seen that the first detected change point by LSTSP is a little bit off, but the second and the third ones align better with those suggested by the experimental design.

We also depict the Granger causal networks for the estimated segments obtained by the TBSS algorithm in Figure \ref{fig:eeg-networks}. 
\begin{figure}[!ht]
    \centering
    \includegraphics[width=.475\textwidth, trim={0 10 0 50}, clip]{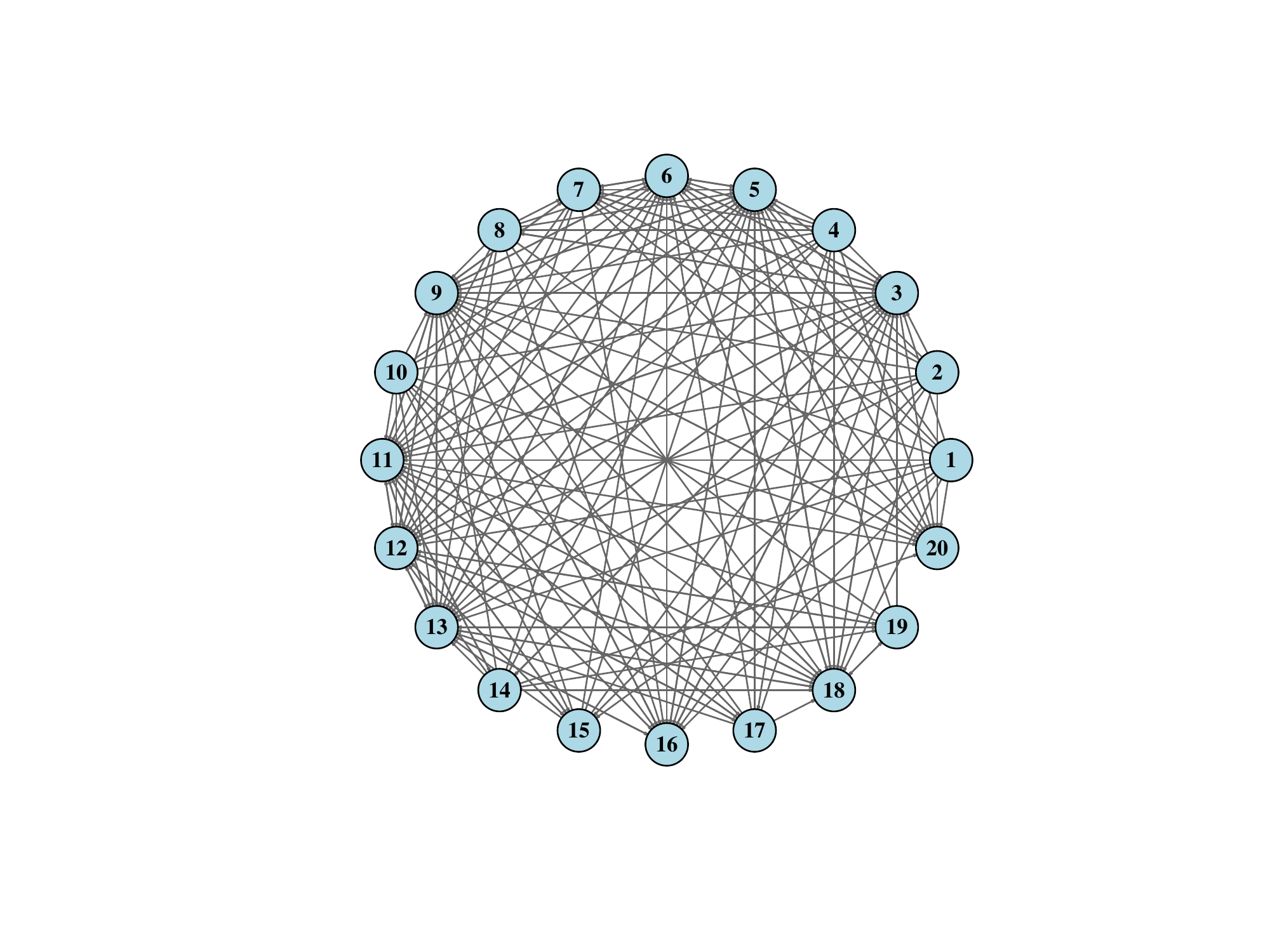}%
    \includegraphics[width=.475\textwidth, trim={0 10 0 50}, clip]{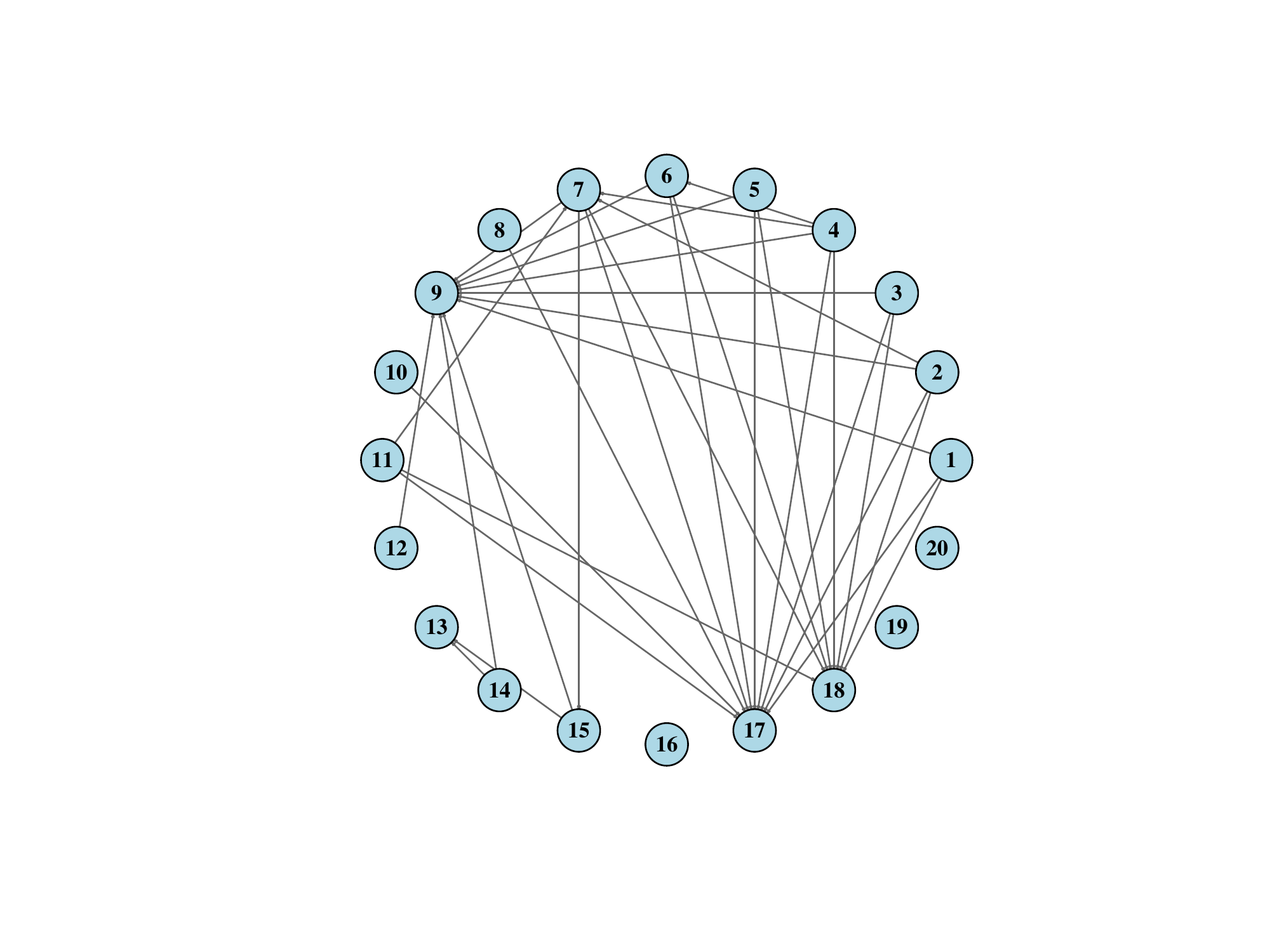}%
    \vfill
    \includegraphics[width=.475\textwidth, trim={0 10 0 50}, clip]{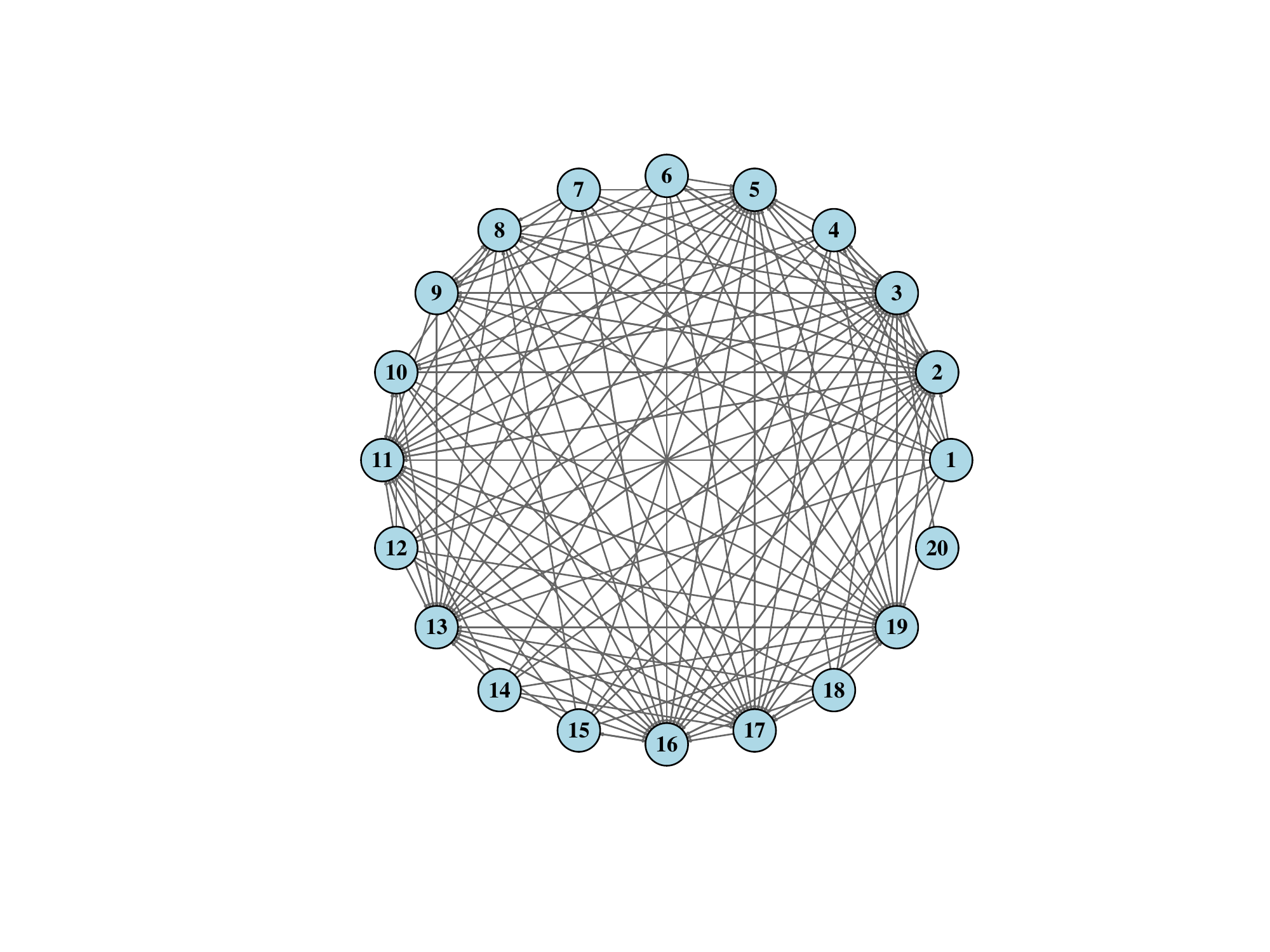}%
    \includegraphics[width=.475\textwidth, trim={0 10 0 50}, clip]{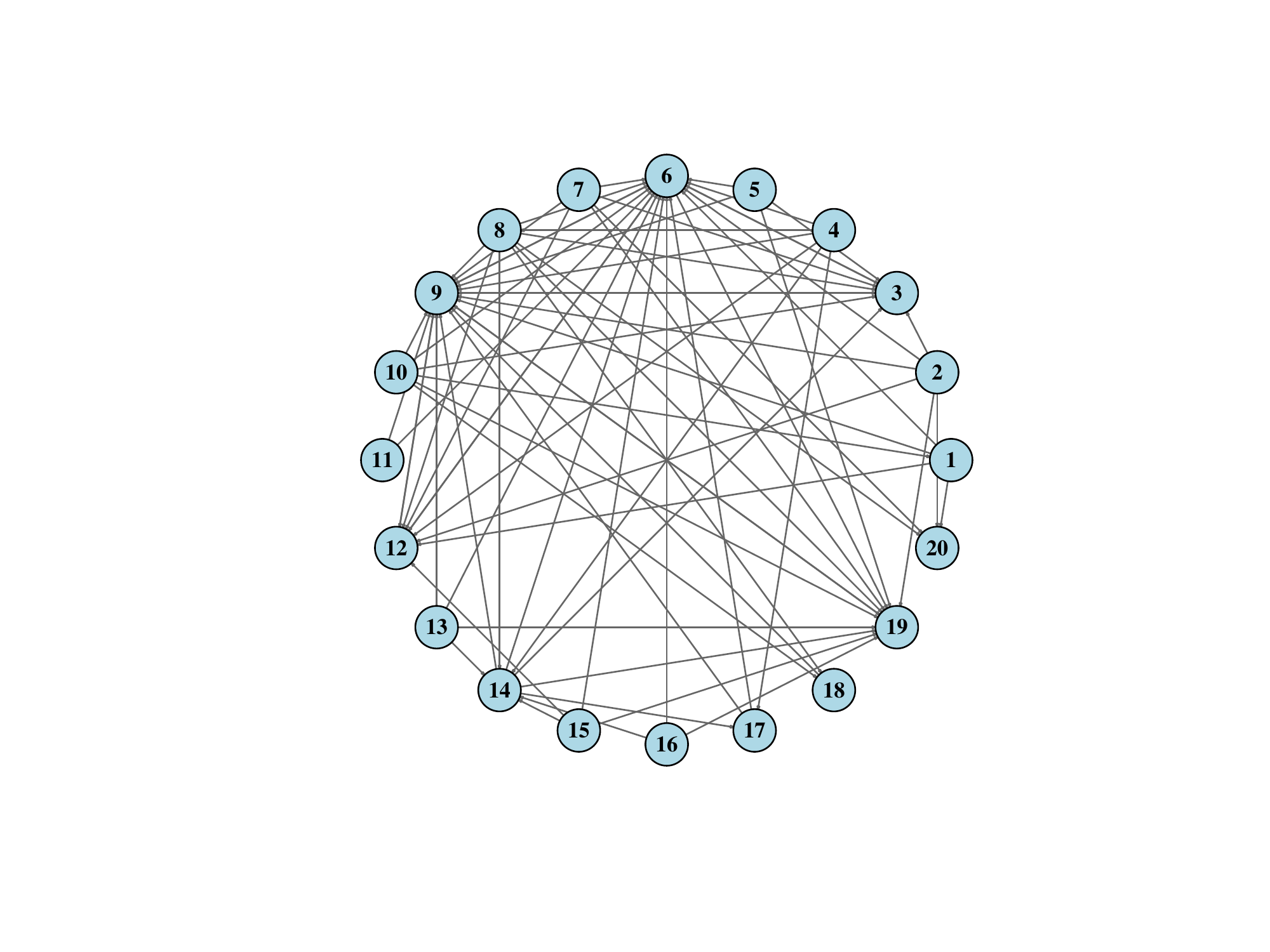}
    \caption{The Granger causality networks for all four estimated segments. Left column: two open segments (1st and 3rd); Right column: two closed segments (2nd and 4th).}
    \label{fig:eeg-networks}
\end{figure}
The results indicate that the open yes segments have much denser networks than the closed eyes segments. 

By using the LSTSP algorithm, we provided the sparsity levels plot as well as the estimated ranks plot for each segment. Specifically, the estimated ranks are 16, 12, 17, and 2, respectively. It becomes apparent that the open eyes segments associate with higher ranks (almost full rank) than the closed eyes segments. 
\begin{figure}[!ht]
    \centering
    \includegraphics[width=.475\textwidth, trim={0 10 0 50}, clip]{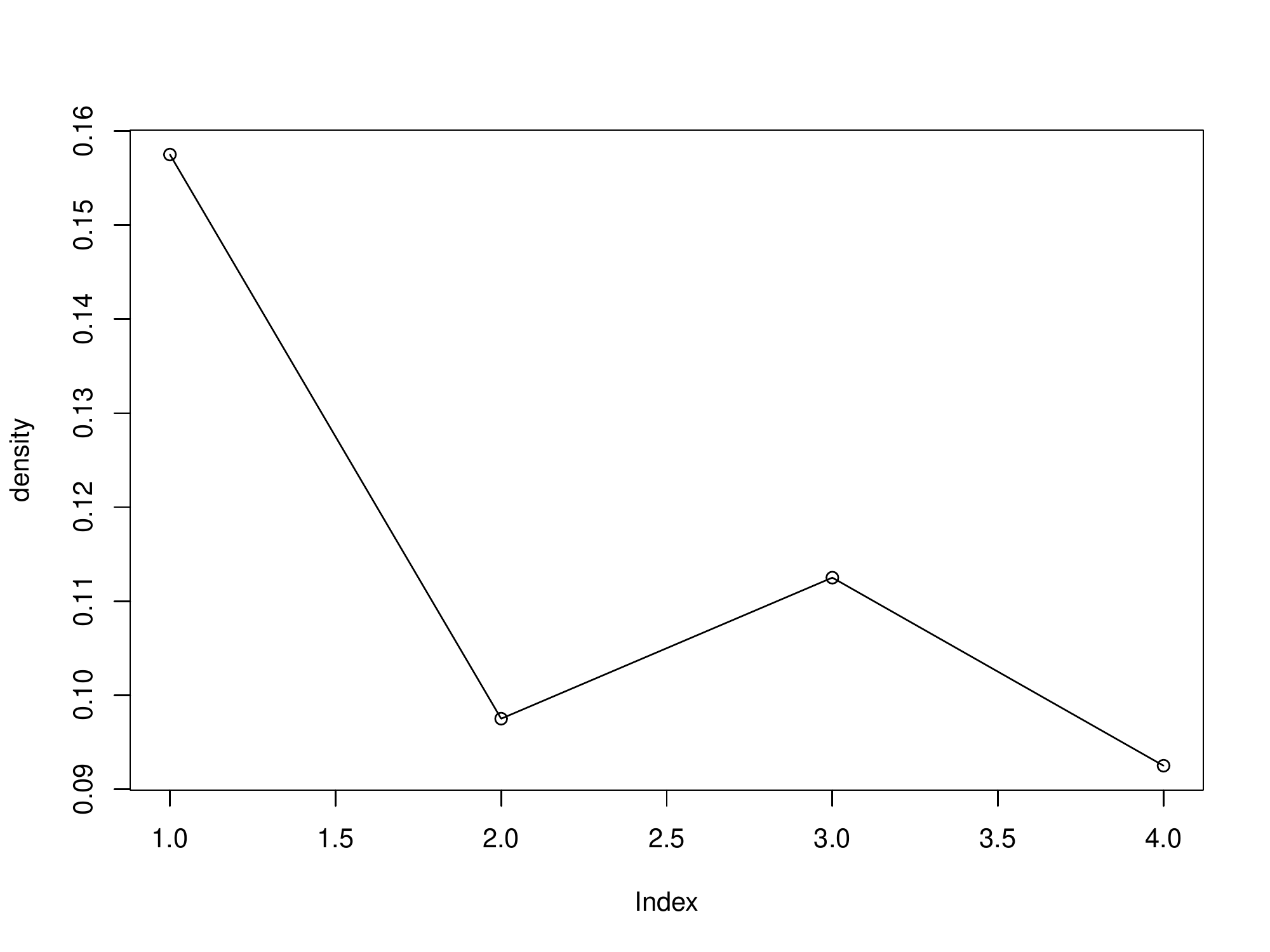}%
    \includegraphics[width=.475\textwidth, trim={0 10 0 50}, clip]{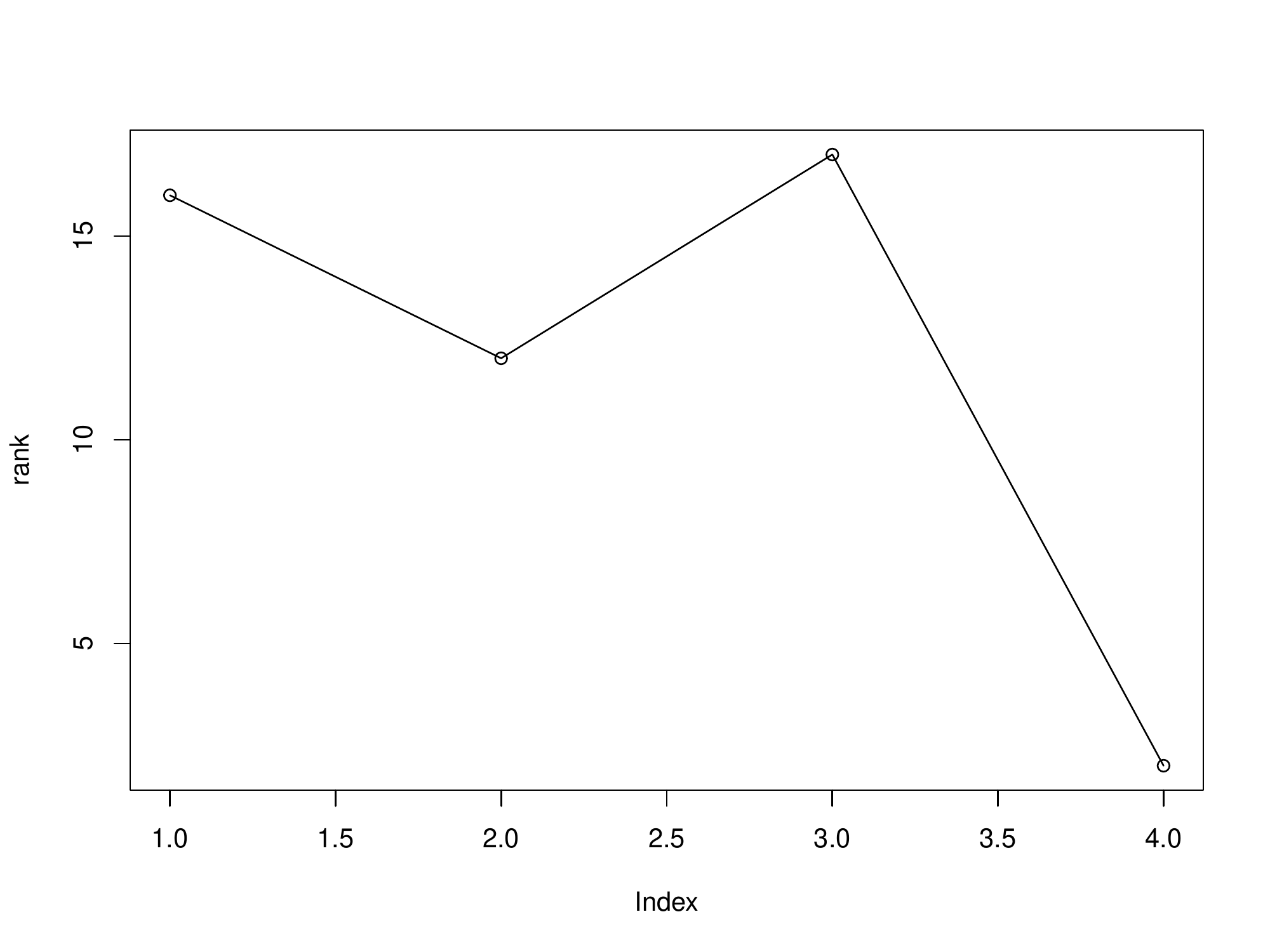}
    \caption{Left: sparsity levels for each detected segments; Right: ranks for the low rank components for each detected segments.}
    \label{fig:eeg-density-rank}
\end{figure}
It can also bee seen that the sparsity level and rank for the open eyes segments (1st and 3rd) are higher than te closed eyes segments, in accordance with the Granger causal network plots in Figure \ref{fig:eeg-networks}.
}

{
We also applied the SBS detection method \citep{cho2015multiple} to this data set. The method estimated too many change points, many of them far from those implied by the experimental design. Hence, we do not pursue any further this comparison.}

\section{Conclusion and Outlook}
The developed \proglang{R} package \pkg{VARDetect} \citep{bai2021vardetect} is designed to detect multiple change points in high dimensional VAR models, whose transition matrices exhibit structured sparse or low rank structure. Its main detection functions are based on the following algorithms: TBSS and LSTSP described in Section \ref{sec:tbss} and Section \ref{sec:lstsp}, respectively. The results obtained by functions \code{tbss} and \code{lstsp} can be visualized and summarized by the corresponding functions in the package. The package also includes options to generate data from the VAR models under consideration and provides summary statistics of the performance of the detection algorithms across replicates, which is a useful option when conducting simulation studies. Furthermore, the \pkg{VARDetect} package provides a data-driven method to automatically determine the tuning parameters in the functions and we also provide a step-by-step guideline on how to select parameters.

We note that detection of multiple change points in VAR models can also be achieved by a dynamic programming algorithm \citep{wang2019localizing} that exhibits \textit{quadratic} time complexity in the number of time points $T$. For that reason, it was not included in the \pkg{VARDetect} package. Nevertheless, an \proglang{R} function implementing such an algorithm for a sparse VAR model is available at \url{https://github.com/peiliangbai92/VARDetect}.

Finally, additional enhancements in future versions of the \pkg{VARDetect} package would include support for \textit{weakly sparse} \citep{negahban2012unified} transition matrices. 

\bibliography{reference}
\end{document}